\documentclass{jfm}

\usepackage[english]{babel}
\usepackage[utf8x]{inputenc}
\usepackage[T1]{fontenc}

\usepackage{amsmath}
\usepackage{graphicx}
\usepackage{subcaption}
\usepackage{float} 

\graphicspath{{figures/}}
\usepackage[colorlinks=true, allcolors=blue]{hyperref}
\usepackage{xcolor}
\usepackage{pgfplots} 
\usepackage{bm}
\usepackage{todonotes}
\usepackage{soul}

\newcommand{\ethan}[1]{\textcolor{blue}{#1}}

\title{Lift-up, Kelvin-Helmholtz and Orr mechanisms in turbulent jets}

\author{Ethan Pickering\aff{1}, Georgios Rigas\aff{1}, Petr{\^o}nio A. S. Nogueira\aff{2}, Andr\'{e} V. G. Cavalieri\aff{2}, Oliver T. Schmidt\aff{3}, Tim Colonius\aff{1}}

\affiliation{
\aff{1} Mechanical Engineering, California Institute of Technology, Pasadena, CA 91125, USA
\aff{2} Instituto Tecnol\'ogico de Aeron\'autica, S\~ao Jos\'e dos Campos, SP 12228-900, Brazil
\aff{3} Department of Mechanical and Aerospace Engineering, University of California, San Diego, La Jolla, CA 92093, USA
}

\begin{document}
    \maketitle
    \begin{abstract}
    Three amplification mechanisms present in turbulent jets, namely lift-up, Kelvin-Helmholtz, and Orr, are characterized via global resolvent analysis and spectral proper orthogonal decomposition (SPOD) over a range of Mach numbers. The lift-up mechanism was recently identified in turbulent jets  via local analysis by Nogueira \textit{et al.} (\textit{J. Fluid Mech.}, vol. 873, 2019, pp. 211–237) at low Strouhal number ($St$) and non-zero azimuthal wavenumbers ($m$). In these limits, a global SPOD analysis of data from high-fidelity simulations reveals streamwise vortices and streaks similar to those found in turbulent wall-bounded flows. These structures are in qualitative agreement with the global resolvent analysis, which shows that they are a response to upstream forcing of streamwise vorticity near the nozzle exit.  Analysis of mode shapes, component-wise amplitudes, and sensitivity analysis distinguishes the three mechanisms and the regions of frequency-wavenumber space where each dominates, finding lift-up to be dominant as $St/m \rightarrow 0$. Finally, SPOD and resolvent analyses of localized regions show that the lift-up mechanism is present throughout the jet, with a dominant azimuthal wavenumber inversely proportional to streamwise distance from the nozzle, with streaks of azimuthal wavenumber exceeding five near the nozzle, and wavenumbers one and two most energetic far downstream of the potential core.
    \end{abstract}

\section{Introduction}

Coherent structures in turbulence are responsible for the transport of mass, momentum, and energy, and the radiation of acoustic waves.  Early observations of \cite{mollo1967jet} and \cite{crow1971orderly} found orderly structure in turbulent jets and \cite{brown1974density} in planar mixing layers. Almost 50 years later, a full understanding of the underlying mechanisms driving the generation and sustenance of coherent structures remains elusive, yet their connection to longstanding engineering problems such as jet acoustics \citep{jordan2013wave} and drag in wall-bounded flows \citep{jimenez2018coherent} has become increasingly clear.

The interpretation and modelling of coherent structures in turbulence have historically taken the form of instabilities of (typically steady) basic flows. In jet turbulence, \cite{crighton1976stability} hypothesized that coherent structures could be described as linear instabilities of the mean flow via a modal analysis. Since that time, researchers have computed modal and non-modal mechanisms in jets with varying degrees of generality. Earlier work focused on parabolized stability equations (PSE), and \cite{gudmundsson2011instability} showed that PSE solutions for an experimentally measured jet mean flow yielded good predictions for the dominant frequency/azimuthal mode structures educed from a near-field caged microphone array.  However, in fully global studies, \cite{garnaud2013modal} yield a stable spectrum (for jets that are not too highly heated), and the characterization of instability mechanisms required transient (non-modal) growth.  Resolvent analysis, by contrast, characterizes linear amplification of disturbances in the frequency domain, and is therefore easier to relate experimental mechanisms in both transitional and turbulent flows. \cite{mckeon2010critical} proposed a resolvent interpretation for the turbulent case, where nonlinear interactions are regarded as forcing terms to a linearized operator that amplifies them according to the turbulent mean flow.  Resolvent analysis of a variety of jet mean flows has been reported in the literature \citep{garnaud2013preferred,jeun2016input,semeraro2016modeling,SchmidtJFM2018,lesshafft2018resolvent} and has shown two essentially different linear amplification mechanisms, one associated with the traditional, modal (parallel-flow) Kelvin-Helmholtz (KH) instability and the other as an Orr-type mechanism, also identified through PSE by \cite{tissot2017wave}.

The KH mechanism has long been invoked to describe coherent structures in both transitional and turbulent planar shear layers and jets. Strictly speaking, the mechanism itself is not defined outside of the context of parallel/quasi-parallel laminar shear layers, where, in the spatial stability theory, KH is an unstable modal solution with an associated spatial growth rate. For jets, the solution is typically a convective instability, and, under spreading of the flow, an initially growing wave (at a fixed frequency) will eventually become neutral and decay \citep{crighton1976stability}. For the axisymmetric azimuthal wavenumber, $m=0$, the resulting wavepacket has a nearly constant phase speed $c_{ph} \approx 0.8$ (for Strouhal number, $St \in [0.3,1]$) \citep{michalke1984survey,malik1997pse,SchmidtJFM2018}, a phase change in streamwise velocity over the critical layer \citep{cavalieri2013wavepackets}, and is coincident with what has been termed the preferred mode of the jet (most amplified mode to external forcing) between Strouhal numbers $\approx 0.3-1$ \citep{morris1976spatial,ho1984perturbed,tam1989three,gudmundsson2011instability,garnaud2013preferred,rodriguez2015study,semeraro2016modeling,SchmidtJFM2018,lesshafft2018resolvent}.  According to the parallel theory and its quasi-parallel extensions, the mode is unstable at low frequencies but with a growth rate that goes to zero faster than the frequency.  In more recent global stability \citep{nichols2011global} and resolvent analyses \citep{garnaud2013preferred,semeraro2016modeling,jeun2016input,SchmidtJFM2018,lesshafft2018resolvent}, the optimal resolvent response has been related to the KH mechanism from $St \in [0.3,1]$, possessing similar characteristics described above through quasi-parallel analysis. Each of these resolvent studies found the KH response, triggered by optimal forcing localized to the initial shear layer (i.e. the lip line ($r/D=0.5$) at the nozzle exit), to exhibit high gain, as well as significant gain separation between the optimal and sub-optimal modes, suggesting an intrinsic mechanism reminiscent of a parallel-flow modal instability. At low frequencies, the gain separation diminished and \cite{SchmidtJFM2018} explicitly tracked the KH mode for $St < 0.3$, finding the KH mode fell into the sub-optimal gains of the resolvent spectrum and was overtaken by another family of resolvent response modes governed by the Orr mechanism. 

What we term the Orr mechanism, by contrast with KH, is, in the context of the response of jets, a relatively recent observation based on either non-modal parallel-flow or global analysis \citep{garnaud2013preferred,semeraro2016modeling,tissot2017wave,tissot2017sensitivity,SchmidtJFM2018,lesshafft2018resolvent}. The characteristics of the Orr mechanism are similar to what has been observed in non-modal analyses of wall-bounded flows \citep{farrell1988optimal,dergham2013stochastic,sipp2013characterization,jimenez2018coherent}.  The Orr wavepacket has a phase speed of $c_{ph} \approx 0.4$  \citep{SchmidtJFM2018} (at low frequencies, $St < 0.3$) and is forced (or most efficiently initialized) by structures oriented at 45 degrees against the direction of mean shear, while the response structure that appears downstream is oriented at 45 degrees along the direction of mean shear. Structures are thus tilted by the mean shear, with algebraic amplitude growth of velocity fluctuations in this process \citep{jimenez2013linear}.  \cite{SchmidtJFM2018} showed that modes with these characteristics were the dominant global resolvent modes in the jet at high frequencies (i.e. $St > 2$), for all $m$.  For $m=0$, at low frequencies (i.e. $St<0.3$ and mentioned above) the Orr mechanism appears again as the dominant mode as the gain associated with the KH mode becomes small and falls below the gains of Orr responses.

However, these previous global resolvent computations neglected the lowest-frequency region of the spectrum, owing largely to computational difficulties, i.e. the large spatial domains required.  Recently, \cite{Nogueira2019Streaks} presented evidence, both through spectral proper orthogonal decomposition (SPOD) of particle image velocimetry data and a locally parallel resolvent analysis, that a third and uniquely distinct mechanism is at play in the low-frequency region of the spectrum, namely the lift-up mechanism. 

The lift-up mechanism \citep{ellingsen1975stability} and the associated coherent structures, streaks, have long been understood as an important mechanism in wall-bounded flows (see review by \cite{brandt2014lift} and references therein). Parallel to the history of mechanism identification in jet flows, the lift-up mechanism and its interplay with the Tollmien-Schlichting and Orr mechanisms has been systematically described, first through observations \ethan{\citep{kline1967structure,klebanoff1971effect,kim1971production}}, then local analyses \citep{moffatt1965interaction,ellingsen1975stability,landahl1980note}, followed by transient growth  \citep{butler1992three,farrell1993optimal}, and most recently resolvent analysis \citep{hwang2010amplification,marant2018influence,abreureduced2019}. The salient properties of the lift-up mechanism include streamwise vortices (rolls with streamwise vorticity) that lift up low-speed fluid from the wall (and push high-speed fluid toward the wall) until viscous dissipation becomes important. The associated optimal forcing takes the form of streamwise rolls (cross-stream forcing components) which result in growth of both streamwise response rolls and the streamwise velocity component (streaks). Although research surrounding streaks is most closely tied with wall-bounded flows, it is known that the wall is not necessary for the lift-up mechanism to persist \citep{jimenez1999autonomous,mizuno2013wall,chantry2016turbulent}.

Around the same time that the identification and recognition of the lift-up mechanism was being studied in wall-bounded flows, so too were the implications of streamwise vortices in free shear flows, in particular, the plane free shear layer. The earliest theoretical evidence of streaks in the plane free shear layer likely belongs to \cite{Benney1960} and \cite{benney1961non}, who studied secondary instabilities of the flow in the form of weakly nonlinear interactions between two-dimensional and three-dimensional waves. They found streamwise vortices to be responsible for alternate steepening and flattening of the velocity profile, associated with the presence of streaks. \cite{miksad1972experiments} first observed this behaviour, followed by the observations of \cite{konrad_1976}, \cite{breidenthal1978chemically}, \cite{bernal1979development}, \cite{bernal1981coherent}, \cite{breidenthal1981structure}, and \cite{bernal1986streamwise}, along with numerical visualizations of \cite{jimenez1985perspective}, \cite{metcalfe1987secondary}, and \cite{rogers1992three}. 

Advancing the original theoretical work of Benney and Lin, \cite{widnall1974instability} and \cite{pierrehumbert1982two} made explicit the connection of streamwise vortices and streaks through secondary instability of the plane free shear layer. Introducing KH-like vortices into the flow and increasing their amplitude produced three-dimensional modes that became unstable, leading to streamwise vortex generation along with streaks. Further three-dimensional, secondary-flow theory came via the approximate dynamical models of \cite{lin1981}, \cite{lin1984mixing}, and \cite{neu1984dynamics} who assumed generation of streamwise vorticity occurs in the braid regions by the strain field of two adjacent spanwise structures, resulting in streamwise streaks. A comprehensive review of the early works on secondary instability and streaks in plane shear flow can be found in \cite{ho1984perturbed}.

Although the plane shear layer is a general case of the free shear layer, explicit connections of jet flow to streamwise vortices and streaks did not receive such detailed attention during the same time period. \cite{bradshaw1964turbulence} identified elongated structures, albeit in the transitional region. Their visualizations provided evidence of ``mixing jet'' structures present in the flow, inclined in the $x-r$ plane, and forced through radial/azimuthal components leading to streamwise momentum transfer. They also postulated, similar to boundary-layer transition, that the highly energetic structures were essential for breakdown to turbulence. Later works by \cite{becker1968vortex}, \cite{browand1975roles}, \cite{yule1978large}, \cite{dimotakis1983structure}, and \cite{agui1988flow} all observed streamwise vortices in jet flow, but questions remained -- where does the instability originate and by what means does it grow?

Two works, one numerical \citep{martin1991numerical} and one experimental \citep{liepmann1992role}, sought to address the above questions. \cite{martin1991numerical} showed how the strain field and radial shear was instrumental in the creation and evolution, of streamwise vortices in the braid region. While \cite{liepmann1992role}, through the use of digital image particle velocimetry on a transitional round jet, identified streamwise vortices and again connected them to highly strained regions between vortex rings (i.e. the braid region) -- finding that the streamwise vortices have a fundamental effect on the dynamics and statistical properties of the flow. Of particular note, they also found strong streamwise vortices persisted in the fully turbulent region of their flow, even when the energy of azimuthal rollers, necessary for further vortex generation, diminished. Regardless of this final observation, both studies supported the stability theory of \cite{pierrehumbert1982two} and the collapse of the braid region into vortices by \cite{lin1984mixing} and \cite{neu1984dynamics}, however, they each brought forth questions regarding additional tertiary instabilities resulting from their observations, increasing analysis complexity. The resolvent framework presented here seeks to eliminate the need for continued secondary and tertiary stability analyses.

In more recent work, \cite{citriniti2000reconstruction} identified streamwise vortices using a 138 hot-wire array located 3 diameters downstream of a turbulent round jet to identify, through SPOD, structures. Their structures contained both streamwise vorticity (i.e. rolls) and streamwise velocity (i.e. streaks), with peak energies at $m=4$.  \cite{jung2004downstream}, expanding the above experiment, found that the most energetic, non-zero azimuthal wavenumber scaled inversely to downstream location. Other experiments investigating streamwise vortices in high Reynolds number jets, exiting from both transitional and turbulent boundary layers, found that interactions between strong streamwise vortices and weak axisymmetric vortices may generate further streamwise vortices \citep{davoust2012Dynamics,kantharaju2020interactions}. As the importance of streamwise vortices to the overall dynamics became increasing apparent, various techniques such as chevrons \citep{bridges2003control,saiyed2003acoustics, bridges2004parametric,violato2011three} and microjets \citep{arakeri2003use, greska2005effects, alkislar2007effect, yang2016turbulent} were implemented to impact the near-field turbulence for far-field noise reduction. In these cases, the addition of chevrons and microjets directly imprint strong streamwise vortices and velocity variations upon the mean of each flow case. However, not all configurations led to reductions in noise, motivating further theoretical understanding of streamwise vortices in jets.

Only recent work on non-modal instability provided direct theoretical evidence of the lift-up mechanism in round jets. \cite{boronin2013non} performed local transient growth analyses on liquid, laminar jets of Reynolds number 1,000 and found lift-up to be responsible for optimal growth. \cite{jimenez2017transient} extended this approach considering the influence of jet aspect ratio, Reynolds number, and azimuthal wavenumber on lift-up. Similar results were found by \cite{garnaud2013modal} using a global transient analysis, for both incompressible laminar and turbulent jets, however, they noted there was no firm evidence of lift-up effects at the present time and did not investigate the zero-frequency limit in their companion resolvent study \citep{garnaud2013preferred}.

Although, it is clear that lift-up is present in laminar, transitional, and, in a more limited sense, turbulent jets at high Reynolds number, a direct quantitative comparison between globally observed and computed structures is lacking. Resolvent analysis and SPOD provide a  statistical connection between structures observed in turbulent flow to those computed through linear-amplification theory \citep{towne2018spectral}.  

In this paper, we use these tools to study characteristics of the lift-up mechanism, expanding upon the work of \cite{Nogueira2019Streaks}. The jets we consider are fully turbulent at high Reynolds numbers and are issued from nozzles consisting of a fully turbulent boundary layer. We demonstrate the presence and energetic importance of the lift-up mechanism over subsonic to supersonic regimes, and characterize, as a function of frequency and azimuthal mode number, the interplay between the (now three) mechanisms (KH, Orr, lift-up). We show that previously described experimental \citep{liepmann1991streamwise,paschereit1992flow,liepmann1992role,arnette1993streamwise,citriniti2000reconstruction,jung2004downstream,alkislar2007effect,cavalieri2013wavepackets} and numerical \citep{martin1991numerical,caraballo2003application,freund2009turbulence} observations result from the lift-up mechanism, yet do not reject the potential origin of streamwise vortices through secondary instabilities of the flow \citep{widnall1974instability,pierrehumbert1982two,lin1984mixing,neu1984dynamics}. In this work, resolvent analysis provides the key statistical link between response modes observed in the SPOD spectrum and their underlying resolvent forcing modes, shedding light on the most active linear-amplification mechanisms throughout the frequency-wavenumber space.

The manuscript is organized as follows. In \S~\ref{sec:methods} we describe the resolvent and SPOD methodology and discuss the large eddy simulation (LES) databases used to educe coherent structures. In \S~\ref{sec:LES} we introduce a collection of snapshots from the LES of a Mach 0.4 jet showing streaky characteristics. In \S~\ref{sec:SPOD_Res} we present the dominant energies and gains computed using SPOD and resolvent analysis respectively. In \S~\ref{sec:Zero_St} SPOD results approaching Strouhal number $St \rightarrow 0$ and resolvent analysis at $St = 0$ are shown, identifying streaks and the direct presence of the lift-up mechanism in turbulent jets. We then compare SPOD and resolvent modes in \S~\ref{sec:Mod-Low} at non-zero azimuthal wavenumbers and frequencies from $St=0.6$ to $St \rightarrow 0$ and present a sensitivity analysis delineating the various mechanisms pertaining to KH, Orr, and lift-up, while also suggesting a mechanism map of the most amplified response mechanism throughout the frequency-wavenumber space of turbulent jets. We then conclude the manuscript in \S~\ref{sec:Domain} addressing the presence of streaks throughout the domain, both near and far from the nozzle.

\section{Methods} \label{sec:methods}

The LES database, resolvent analysis and SPOD were described in \cite{SchmidtJFM2018} and \cite{towne2018spectral}. For brevity, we recall the main details here.

\subsection{Large Eddy Simulation database}

\begin{table}
    \centering
\begin{tabular}{c c c c c c c c}
 case & $M_j$ & $Re_j$ & $\frac{p_0}{p_\infty}$ & $\frac{T_0}{T_\infty}$ & $n_{\text{cells}}$ & $\Delta t a_\infty/ D$ & $ \Delta St$\\  
 subsonic & $0.4$ & $4.5 \times 10^5$ & 1.117 & 1.03 & $15.9 \times 10^6$ & 0.2 & $0.049$ \\ 
 transonic & $0.9$ & $1.01 \times 10^6$ & 1.7 & 1.15 &$15.9 \times 10^6$ &0.2 & $0.022$ \\  
 supersonic & $1.5$ & $1.76 \times 10^6$ & 3.67 & 1.45 &$31 \times 10^6 $ &0.1 & $0.026$ \\
\end{tabular}
\caption{Parameters, sampling rate, and frequency resolution for the LES.}
    \label{tab:LES}
\end{table}


The LES databases, including subsonic (Mach 0.4), transonic (Mach 0.9), and supersonic (Mach 1.5) cases, were computed using the flow solver Charles; details on the numerical method, meshing, and subgrid models can be found in \cite{bres2017unstructured}.  The simulations were validated with experiments conducted at PPRIME Institute, Poitiers, France for the Mach 0.4 and 0.9 jets \citep{bres2018importance}. A summary of parameters for the three jets considered is provided in table \ref{tab:LES}.  These include the Reynolds number based on diameter $Re_j = \rho_j U_j D / \mu_j$ (where subscript $j$ indicates the value at the centre of the jet, $\rho$ is density, $\mu$ is viscosity) and the Mach number, $M_j = U_j/a_j$, where $a_j$ is the speed of sound at the nozzle exit. The simulated $M_j = 0.4$ jet corresponds to the experiments in \cite{cavalieri2013wavepackets,jaunet2017two,Nogueira2019Streaks} with the same nozzle geometry and similar boundary-layer properties at the nozzle exit.  Throughout the manuscript, reported results are non-dimensionalized by the mean jet velocity $U_j$ and jet diameter $D$; pressure is made non-dimensional using $\rho_j U_j^2$. Frequencies are reported in Strouhal number, $St = f D / U_j$, where $f$ is the frequency. 

\begin{figure}
\centering
\vspace{0.5cm}
\includegraphics[width=1\textwidth]{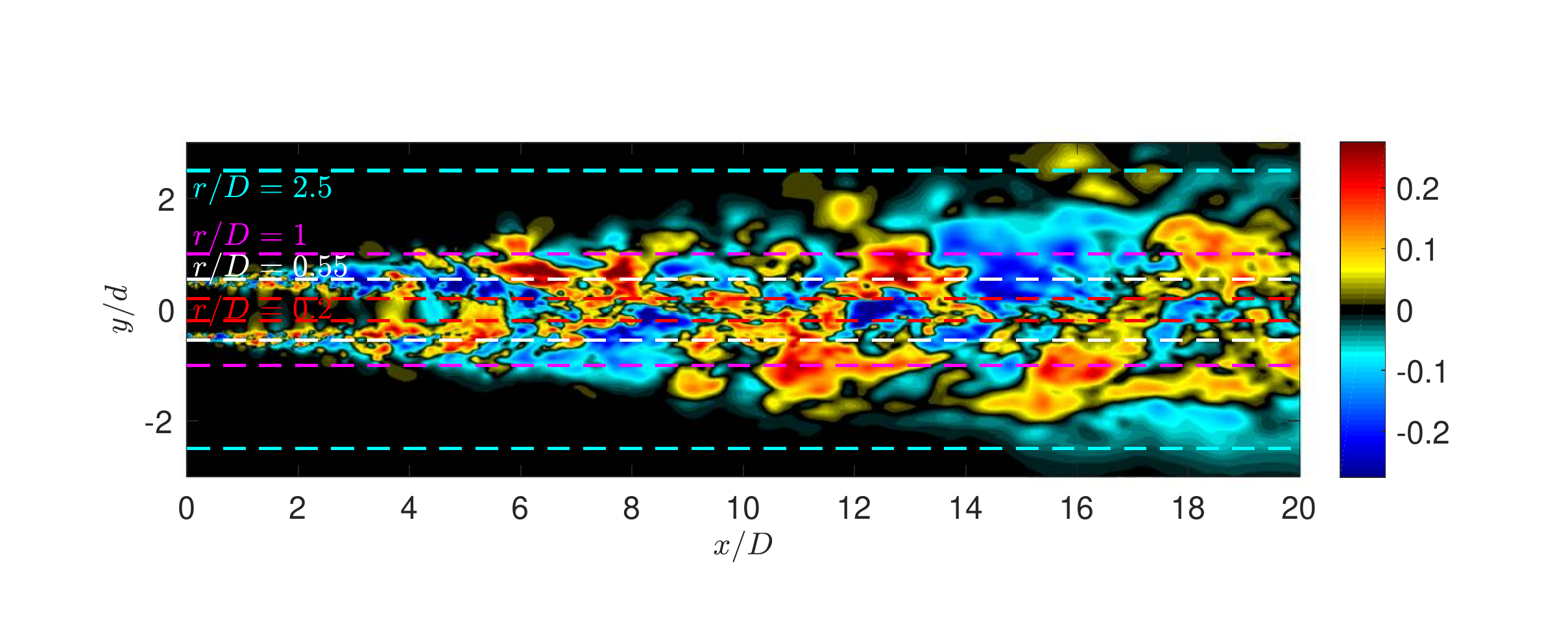}
\caption{LES snapshot of fluctuating streamwise velocity in a plane through the $M_j = 0.4$ jet centreline. The lines show the areas of interest plotted in figure \ref{fig:LES_rD}.}
\label{fig:LES_annotated}
\end{figure}

Each database consists of 10,000 snapshots separated by $\Delta t a_\infty/ D$, where $a_\infty$ is the ambient speed of sound, and interpolated onto a structured cylindrical grid $x,r,\theta \in [0,30] \times [0,6] \times [0, 2\pi]$, where $x$, $r$, $\theta$ are streamwise, radial, and azimuthal coordinates, respectively. Variables are reported by the vector
\begin{align}
\bm{q} = [\rho, u_x, u_r, u_\theta, T]^T,    
\end{align}
where $u_x$, $u_r$, $u_\theta$ are the three velocity components, $T$ is temperature, and a standard Reynolds decomposition separates the vector into mean, $\bar{\bm{q}}$, and fluctuating, $\bm{q}'$, components
\begin{align}
   \bm{q}(x,r,\theta,t) = \bar{\bm{q}}(x,r) + \bm{q}'(x,r,\theta,t).
\end{align}
Figure \ref{fig:LES_annotated} displays one snapshot of the streamwise velocity component in a plane through the $M_j = 0.4$ jet centreline.

\subsection{Spectral Proper Orthogonal Decomposition}

SPOD is implemented to determine a set of orthogonal space-time correlated modes that optimally describe the turbulent flow statistics \citep{towne2018spectral}. To perform SPOD we decompose each LES database, $\bm{Q}$, where columns of $\bm{Q}$ are temporal snapshots of the state vector $\bm{q}$, in the azimuthal and temporal dimensions via discrete Fourier transforms to give, $\bm{Q}_{m,\omega}$. The azimuthal decomposition is performed over 128 azimuthal grid points, allowing for valid transforms of up to $m=64$, however we only report results using up to $m \approx 30$ with most of the analysis at low azimuthal wavenumbers, $m \leq 5$. The temporal transforms require the dataset to be segmented into sequences $\bm{Q} = [\bm{q}^{(1)} \bm{q}^{(2)} ... \bm{q}^{(n_{\text{freq}})}]$ where each segment (or block) contains $n_{\text{freq}}$ instantaneous snapshots (including overlap), with a periodic Hanning window employed over each block to prevent spectral leakage. Applying the temporal and azimuthal Fourier transforms gives $\hat{\bm{Q}}_{m,\omega_k}^{(l)} = [ \hat{\bm{q}}_{m,\omega_1}^{(l)}  \hat{\bm{q}}_{m,\omega_2}^{(l)} ...  \hat{\bm{q}}_{m,\omega_{n_{\text{freq}}}}^{l}]$, where $\hat{\bm{q}}_{m,\omega_k}^{l}$ is the $l$-th realization of the transform at the $k$-th frequency. The cross-spectral density tensor at a given frequency $\omega_k = 2 \pi St$ and azimuthal wavenumber $m$ is then given by
\begin{equation}
\textbf{S}_{m, \omega_k} = \hat{\bm{Q}}_{m,\omega_k} \hat{\bm{Q}}_{m,\omega_k}^*
\end{equation}
and the SPOD eigenvalue problem presented by \cite{lumley1967, lumley1970} can be solved
\begin{equation}
\textbf{S}_{m, \omega_k}\textbf{W}\bm{ \Psi }_{m, \omega_k} = \bm{\Psi}_{m, \omega_k} \bm{\Lambda}_{m, \omega_k}. 
\end{equation}
The SPOD modes are represented by the columns of $\boldsymbol{\Psi}_{m, \omega_k}$ and are ranked by the diagonal matrix of eigenvalues $\bm{\Lambda}_{m, \omega_k}= \text{diag}(\lambda_1, \lambda_2, ... , \lambda_N)$. The modes are orthonormal in the norm $\bm{W}$, representing the compressible energy norm of \cite{chu1965energy}
\begin{equation}
    \langle \bm{q}_1, \bm{q}_2 \rangle_E = \int \int \int \bm{q}_1^* \text{diag} \bigg( \frac{\bar{T}}{\gamma \bar{\rho} M^2}, \bar{\rho},  \bar{\rho}, \bar{\rho}, \frac{\bar{\rho}}{\gamma (\gamma - 1) \bar{T} M^2} \bigg) \bm{q}_2 r dr dx d\theta
\end{equation}
and satisfy $\bm{\Psi}_{m, \omega_k}^*\bm{W}\bm{\Psi}_{m, \omega_k}= \bm{I}$.

Given that SPOD is a discrete method for educing turbulent coherent structures, challenges persist for approaching the limit of $St \rightarrow 0$. We must strike a balance between snapshots per block (frequency resolution) and the number of blocks (convergence) when performing SPOD. When considering $St \in [0.2,1]$ for the $M_j = 0.4$ jet, \cite{SchmidtJFM2018} used block sizes of 256 snapshots with 50 \% overlap (resulting in 78 blocks) and we use these same parameters in \S~\ref{sec:SPOD_Res}. However, to uncover stationary-in-time structures such as streaks, increases in frequency resolution, and therefore block sizes, are necessary in the limit of $St \rightarrow 0$. For the global SPOD analysis presented, and after experimentation with block sizes (256, 512, 1024, 2048) and overlap (50\%, 75\%), we use 1024 snapshots with 75\% overlap (36 blocks) to attain the required frequency resolution, yet maintain convergence with sufficient blocks as $St \rightarrow 0$ in \S~\ref{sec:Zero_St}. We also note that the use of 2048 snapshots per block produced almost identical SPOD modes as using 1024 snapshots, albeit with a greater uncertainty as we reduced the block ensemble to 16 blocks.

\subsection{Resolvent analysis}
For the round, statistically stationary, turbulent jets we consider, the compressible Navier-Stokes, energy, and continuity equations are linearized through a standard Reynolds decomposition, and Fourier transformed both temporally and azimuthally to the compact expression
\begin{equation}
(i\omega\textbf{I} - \textbf{A}_m) \bm{q}_{m, \omega} = \bm{f}_{m, \omega}.
\label{LNS_eq}
\end{equation}
Here, $ \textbf{A}_m $ is the discretized linear operator considering the mean field as the base flow, $\bm{q} = [\rho', u_x', u_r', u_\theta', T']$ is the state vector, and $\bm{f}$ constitutes the forcing in each variable.

\begin{figure}
\centering
\vspace{0.5cm}
\includegraphics[width=1\textwidth,trim={0cm 0cm 0cm 0.8cm},clip]{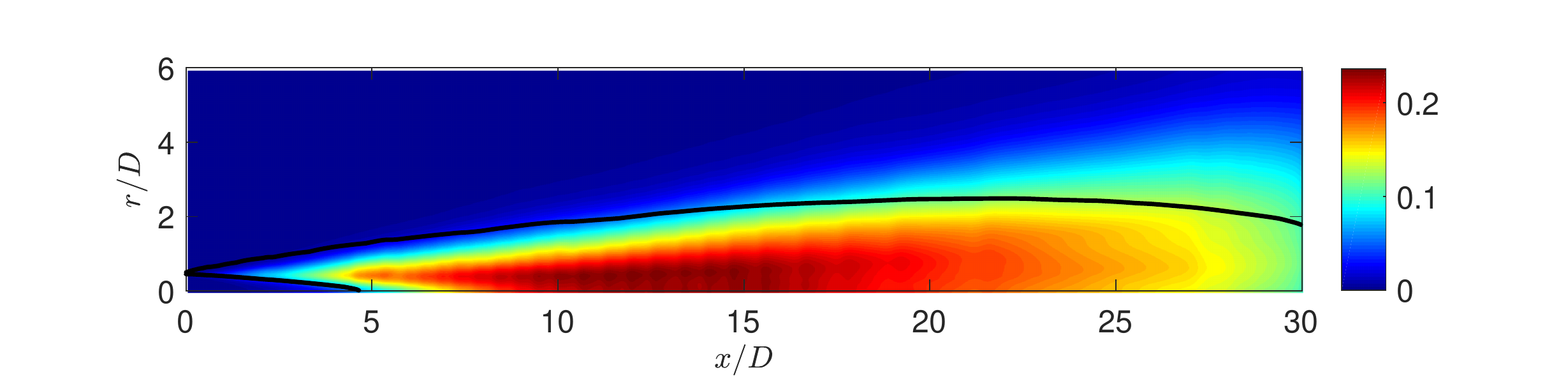}
\caption{Resolvent eddy-viscosity field using the turbulent kinetic energy model,  $\mu_{T} =  \bar{\rho} c k^{1/2} l_m $, with $c=1$. The black lines denote the region of $ > 10$\% maximum TKE which defines the length scale, $l_m$.}
\label{fig:TKE_field}
\end{figure}

The influence of viscosity on the linear operator, $\bm{A}_m$, was previously \citep{SchmidtJFM2018} based on a spatially uniform Reynolds number using an altered molecular viscosity.  However, recent resolvent analyses for turbulent jets \citep{pickering2019eddy} and  channel flows \citep{morra2019relevance}, have shown substantial improvement in SPOD-resolvent correspondence by using a mean-flow eddy-viscosity model on the fluctuations. We proceed here by including an eddy-viscosity model based upon the turbulent kinetic energy (TKE) suggested for turbulent jets by \cite{pickering2019eddy}. This model was primarily chosen due to its simplicity and availability of the corresponding quantities from the LES database.  The model takes the form
\begin{equation}
    \mu_{T} =  \bar{\rho} c k^{1/2} l_m, 
\end{equation}
where $c$ is a scaling constant, $k$ is the mean-flow turbulent kinetic energy, and $l_m$ is a chosen length scale representative of the mean shear layer thickness; $l_m$ is chosen as the width of the shear layer where the turbulent kinetic energy is more than $10\%$ of its maximum value at each streamwise location, and the scaling constant $c = 0.0065$ is used considering the favourable SPOD-resolvent alignments previously shown for $m=0$ and $St = 0.05-1$. The eddy-viscosity field implemented in this analysis is shown in figure \ref{fig:TKE_field}, note that the field is shown when $c=1$. 

With the inclusion of an eddy-viscosity model the forward operator becomes $(i\omega \bm{I} - \bm{A}_m - \bm{A}_{m,T}(\mu_T))$, where $ \bm{A}_{m,T}(\mu_T)$ only possesses terms that include $\mu_T$ (equations for $\bm{A}_{m,T}(\mu_T)$ are included in \cite{pickering2019eddy}). The streamwise plane is discretized using fourth-order finite difference operators satisfying a summation by parts rule developed by \cite{mattsson2004summation}, the polar singularity is treated as in \cite{mohseni2000numerical}, and non-reflecting boundary conditions are implemented at the domain boundaries.

We can rewrite equation \ref{LNS_eq} by defining the resolvent operator, $\bm{R}_{\omega,m} = (i\omega \bm{I} - \bm{A}_m - \bm{A}_{m,T}(\mu_T))^{-1}$,
\begin{equation}
\bm{q}_{m, \omega}  = \bm{R}_{m, \omega} \bm{f}_{m, \omega},
\label{eqn:resolvent}
\end{equation}
and introducing the compressible energy norm \citep{chu1965energy} via the matrix $\bm{W}$ to the forcing and response, where $\bm{W} = \bm{W}_f = \bm{W}_q$, gives the weighted resolvent operator, $\hat{\bm{R}}_{m, \omega}$,
\begin{equation}
\hat{\bm{R}}_{m, \omega}  = \boldsymbol{W}_q^{1/2} \bm{R}_{m, \omega} \boldsymbol{W}_f^{-1/2}.
\end{equation}
Taking the singular value decomposition of the weighted resolvent operator gives
\begin{equation}
\hat{\bm{R}}_{m, \omega}  = \hat{\bm{U}}_{m,\omega} \bm{\Sigma} \hat{\bm{V}}_{m, \omega}^*
\end{equation}
where the optimal response and forcing modes are contained in the columns of $\bm{U}_{m,\omega}  = \bm{W}_y^{-1/2} \hat{\bm{U}}_{m,\omega}$, with $\bm{U}_{m, \omega} = [\bm{u}_{m, \omega}^1, \bm{u}_{m, \omega}^2, ... , \bm{u}_{m, \omega}^N]$, $\bm{V}_{m,\omega}  = \bm{W}_f^{-1/2} \hat{\bm{V}}_{m,\omega}$, with $\bm{V}_{m, \omega} = [\bm{v}_{m, \omega}^1, \bm{v}_{m, \omega}^2, ... , \bm{v}_{m, \omega}^N]$, respectively, while $\bm{\Sigma} = \text{diag}(\sigma_1, \sigma_2, ... , \sigma_N)$ are the ranked gains. 

To avoid ambiguity in referring to computed SPOD and resolvent modes, the following notation is used in the remainder of the manuscript: $\bm{\psi}_{n}$ represents the $n$-th most energetic SPOD mode, while $\bm{v}_n$ and $\bm{u}_n$ denote the resolvent forcing and response, respectively, that provide the $n$-th largest linear-amplification gain between $\bm{v}_n$ and $\bm{u}_n$. In this paper, we only consider the optimal, or dominant, modes such that $n$ is unity. Additionally, when referring to specific components of each mode, such as streamwise velocity, the notation $\bm{\psi}_{1}: u_x$ is used. Finally, for all computed modes subscripts $m,\omega$ are dropped, but referenced when necessary in the text. 

\subsection{Resolvent-based sensitivity analysis}


A structural sensitivity analysis \citep{qadri2017frequency} is applied to reveal regions of the flow where small perturbations to the operator have the largest effect on the resolvent gain. The component-wise sensitivity tensor is defined as the component-wise product of the forcing and response modes,
\begin{align}
S_{i,j} = \sigma^2 Re((\bm{v}_{1}:i) \circ (\bm{u}_{1}:j)^*),
\end{align}
where subscripts $i,j$ denote the $i$-th or $j$-th component of the forcing or response, and $\circ$ denotes component-wise multiplication, giving the spatial sensitivity for each $i,j$ forcing-response combination to the gain, $\sigma$. The sensitivity is calculated for the three forcing and response velocities of the momentum equation, presenting a $3\times 3$ sensitivity tensor.

\section{Lift-up mechanism \& streaks} \label{sec:lift-up}

The lift-up mechanism, and the associated streaks, were first described by \cite{moffatt1965interaction} and \cite{ellingsen1975stability}, where low-energy vortices `lift-up' streamwise momentum of the basic flow and generate high-energy streaks. This simple description continues to provide an intuitive and accurate description in line with modern analyses. Here, we extend findings on non-modal growth of streaks in jets \citep{boronin2013non,garnaud2013modal,jimenez2017transient,Nogueira2019Streaks} and compare our findings to wall-bounded optimal forcing studies of \cite{hwang2010amplification} and \cite{monokrousos2010global}, as well as the transient growth analysis of planar mixing layers by \cite{arratia2013transient}. 

There are three salient characteristics that we use to identify the lift-up mechanism in the SPOD and resolvent results. First, the optimal forcing structures take the form of cross-stream vortices while the response assumes a highly amplified streamwise velocity structure.  Second, the structures are elongated in the streamwise coordinate and are most amplified as their wavelength lengthens (i.e. represented in wall-bounded flows as a streamwise wavenumber of 0, and in the jet as $St=0$).  Third, streaks are azimuthally non-uniform (spanwise non-uniform in planar flows), i.e. they do not occur at $m=0$. 

For the remainder of this manuscript we report results for the $M_j = 0.4$ turbulent jet. Similar analyses are provided for both the $M_j=0.9,1.5$ jets in appendix \ref{sec:TransSuper}.

\subsection{LES Visualization} \label{sec:LES}

\begin{figure}
\centering
\includegraphics[width=0.9\textwidth,trim={0cm 1cm 0cm 1cm},clip]{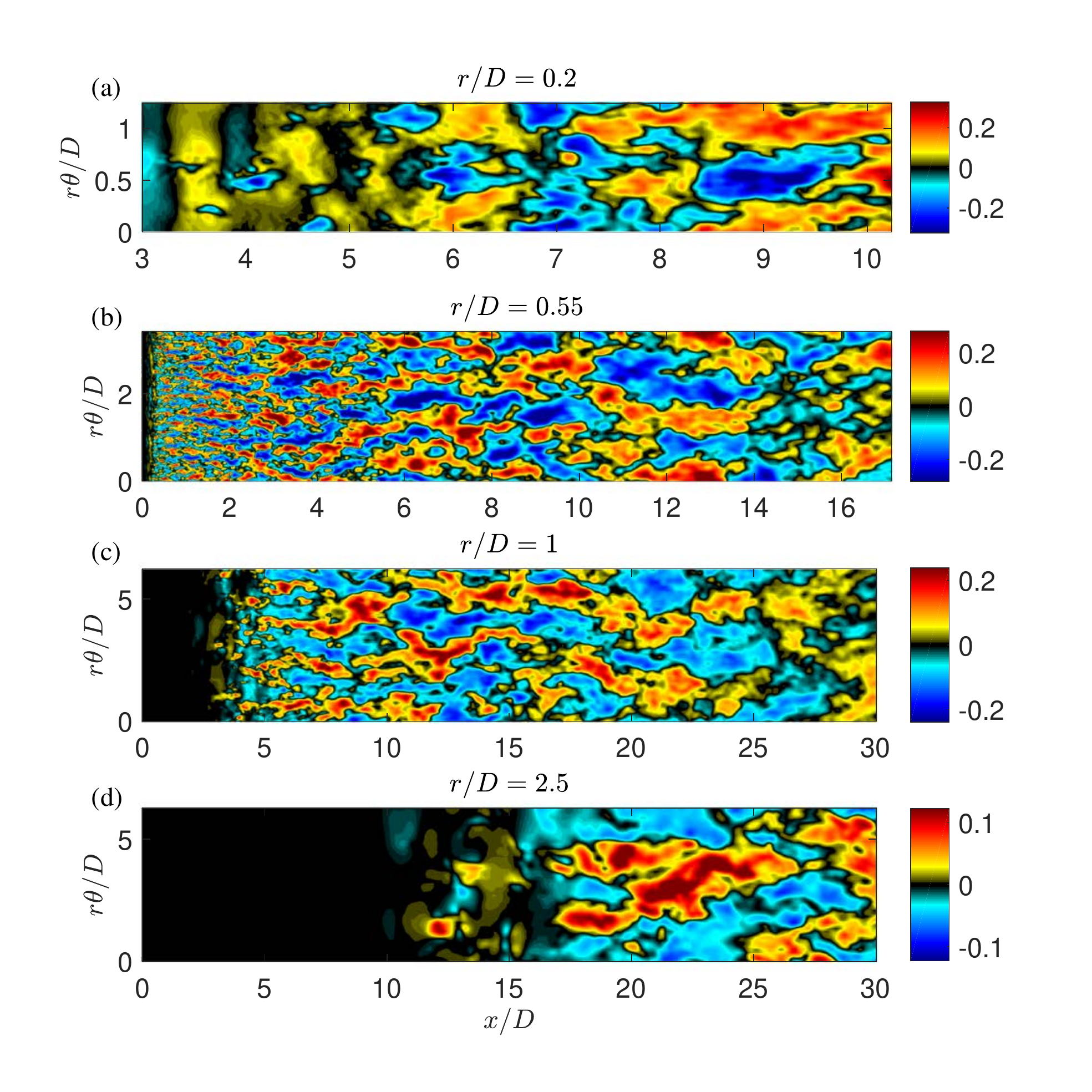}
\caption{Instantaneous LES snapshots of streamwise fluctuating velocity, $u_x'$, taken at radial locations $r/D=0.2,0.55,1,2.5$ (a-d) with the y-axis representing the unwrapped surface $r \theta /D$. Only $2/5$ of the $r/D=2.5$ surface is shown in the last plot.}
\label{fig:LES_rD}
\end{figure}

Previous SPOD and resolvent analyses of turbulent jets \citep{SchmidtJFM2018} showed that the energy of the most amplified mode increases for non-zero azimuthal wavenumber as $St \rightarrow 0$. Considering significant energy is found at low frequencies, we expect to find elongated, streak-like structures to be visually present in the LES snapshots. Instantaneous snapshots of fluctuating streamwise velocity, $u_x'$, are shown on four unwrapped cylindrical surfaces at $r/D = 0.2,0.55,1,$ and $2.5$  (denoted in figure \ref{fig:LES_annotated}) in figure \ref{fig:LES_rD}. The figures are plotted with the circumferential location of the unwrapped surface, $r\theta/D$, as the y-axis to maintain the physical aspect ratio of the data. Note that the plots are therefore increasingly zoomed into the near nozzle region as the radial position is decreased.

These instantaneous visualizations bear a resemblance to streaky structures found in plane shear flows, wall-bounded flows, and most recently the visualization of the present flow by \cite{Nogueira2019Streaks} invoking Taylor’s hypothesis.  We see elongated structures at each radial location taking forms analogous to \cite{konrad_1976} and \cite{bernal1986streamwise} in plane shear flow, turbulent boundary layers \citep{swearingen1987growth,hutchins2007evidence,eitel2014simulation}, channel flow \citep{monty2007large}, and pipe flow \citep{hellstrom2011visualizing}. The figures also show, more subtly, the presence of KH rollers. Beginning with radial surface $r/D=0.2$  in figure \ref{fig:LES_rD} (a) we attain a viewpoint from within the potential core, showing both KH rollers and streaky structures. From $x/D = [3,6]$ structures appear to have a dominant $r \theta/D$ dimension and are likely to be associated with KH $m=0,1$ instabilities (i.e. KH rollers and non-streaky) extending throughout the potential core. However, at $x/D \approx 6$, where the potential core ends, the turbulent kinetic energy increases and streamwise-elongated (i.e. $x/D$ dominated) structures begin to appear; they become more pronounced further downstream. For the radial surface $r/D=0.55$ (b), we see streaky structures of smaller scale close to the nozzle and larger scale further downstream. These are similar to the structures visualized by \cite{Nogueira2019Streaks} at this radial location. The structures also meander as they propagate downstream, a quality also observed in turbulent boundary-layer flows by \cite{hutchins2007evidence}. Comparable behaviour is seen for both $r/D=1$ (c) and $2.5$ (d). 

\subsection{SPOD and Resolvent Analysis} \label{sec:SPOD_Res}

\begin{figure}
\centering
\vspace{0.5cm}
\includegraphics[width=1.0\textwidth,trim={0.5cm 0cm 0 0},clip]{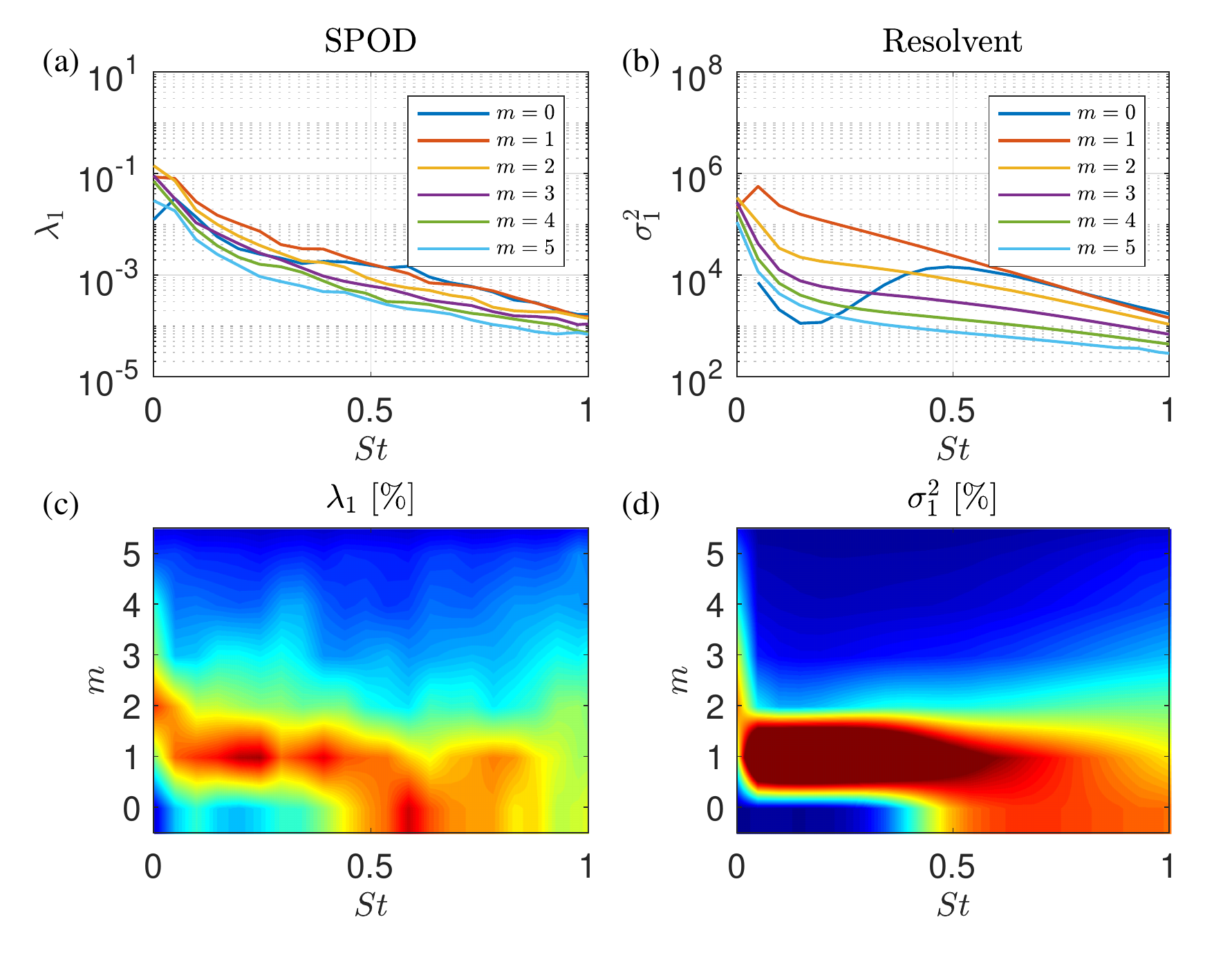}
\caption{Energy of the most amplified mode from SPOD (left) and resolvent (right) analyses. The top row displays the energies (a) and gains (b) for the first six azimuthal wavenumbers. The bottom row (c,d) recasts the above plots (a,b) as a percentage of the sum of energy at each azimuthal wavenumber with contours ranging from 0-40\%.}
\label{fig:LinearGain}
\end{figure}

SPOD and resolvent analyses were performed over a range of frequencies and azimuthal wavenumbers. Figure \ref{fig:LinearGain} (a) shows the SPOD $(E^{SPOD}_{m, St} \equiv \lambda_{m, St})$  energies and  (b) resolvent $(E^{resolvent}_{m, St} \equiv \sigma^2_{m, St})$ amplifications of the leading mode. In figures \ref{fig:LinearGain} (c,d), for SPOD and resolvent respectively, we perform  a supplementary step to highlight the relative contribution of each azimuthal wavenumber at a particular frequency. The data presented in (a,b) are normalized at each frequency with the sum of energy over all azimuthal wavenumbers $(E_{m,St}/\sum_m E_{m, St} \times 100)$ and plotted in the $St-m$ plane. We note that the gains have been normalized in the bottom row (c,d) as a means of visualizing the data, as it is otherwise difficult to identify behaviour occurring at different frequencies as the gain drops off rapidly with increasing frequency. Both plots in figure \ref{fig:LinearGain} (c,d) highlight the dominant wavenumbers at each frequency and facilitate the identification of different mechanisms. The contour maps (c,d) are interpolated for non-integer wavenumbers despite their discrete nature in azimuthal wavenumber, $m$. This representation was chosen as we found it easier to visually interpret the trends, and to more readily compare it to spanwise-streamwise wavenumber contours familiar from boundary-layer \citep{monokrousos2010global} or plane shear \citep{arratia2013transient} flows where wavenumbers are continuous. For reference, the semi-discrete representation of figure \ref{fig:LinearGain} is provided in appendix \ref{sec:SemiDiscrete}.

There is a qualitative agreement between the SPOD energy and resolvent gain. Here, we focus on figure \ref{fig:LinearGain} (c,d) and note three regions in $St-m$ space. i) the energetic region near $St=0.6$ at low azimuthal wavenumbers (particularly $m=0$) is associated with the KH mechanism \citep{garnaud2013preferred,semeraro2016modeling,SchmidtJFM2018,lesshafft2018resolvent}. ii) at low frequencies, meaning $St \rightarrow 0$, and $m=0$ the dominant mode switches from the KH to the Orr mechanism for $St < 0.3$~\citep{SchmidtJFM2018} as the growth of the KH mechanism diminishes. iii) at low frequencies and non-zero azimuthal wavenumbers (mainly $m=1$ and 2), a second energetic region is observed. This region comprises the lift-up mechanism that is described more fully in what follows.  These frequency-wavenumber contour plots present trends similar to those obtained for laminar plane shear \citep{arratia2013transient} and laminar boundary layers~ \citep{monokrousos2010global}, where the lift-up  mechanism dominates as streamwise wavenumbers approach 0 (i.e. streamwise uniform), or as frequency approaches zero for jet flows as documented by local analyses in laminar \citep{boronin2013non,jimenez2017transient} and turbulent jets \citep{Nogueira2019Streaks}. For non-zero azimuthal wavenumbers, where the lift-up mechanism may be permitted due to its three-dimensional characteristics, $m=1$ provides the largest energy across all frequencies, save $m=2$ at $St \approx 0$, followed by a gradual reduction for higher azimuthal wavenumbers.


\subsection{Global characteristics of streaks} \label{sec:Zero_St}

\begin{figure}
\centering
{\small $m=1$ \hspace{25 mm} (a) SPOD \hspace{25 mm} $m=3$ } \\
\includegraphics[width=1\textwidth,trim={1.75cm 0.25cm 1cm 0.95cm},clip]{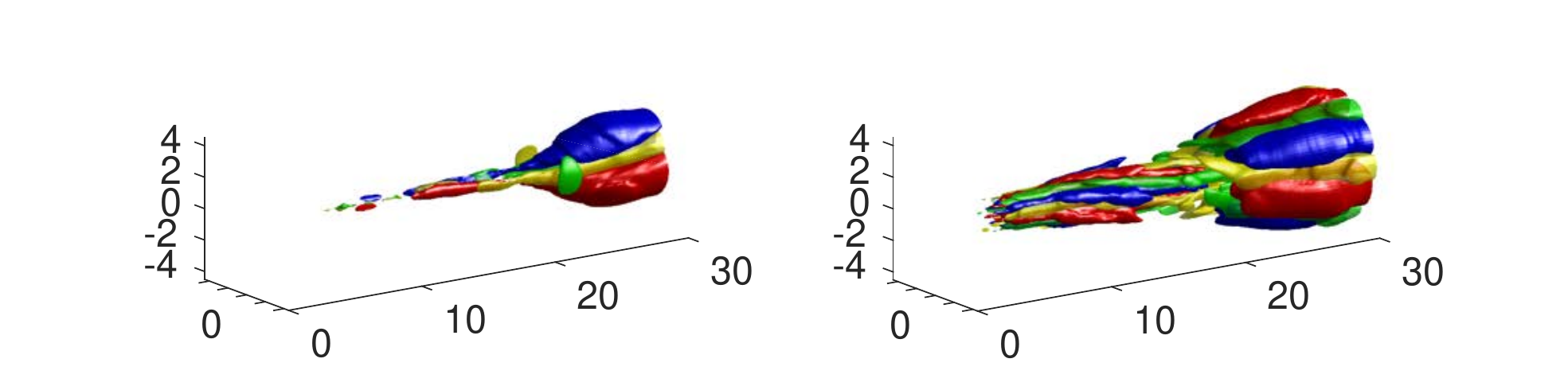}
\\ {\small $m=1$ \hspace{22 mm} (b) Resolvent \hspace{22 mm} $m=3$ } \\
\includegraphics[width=1\textwidth,trim={1.75cm 0.25cm 1cm 0.95cm},clip]{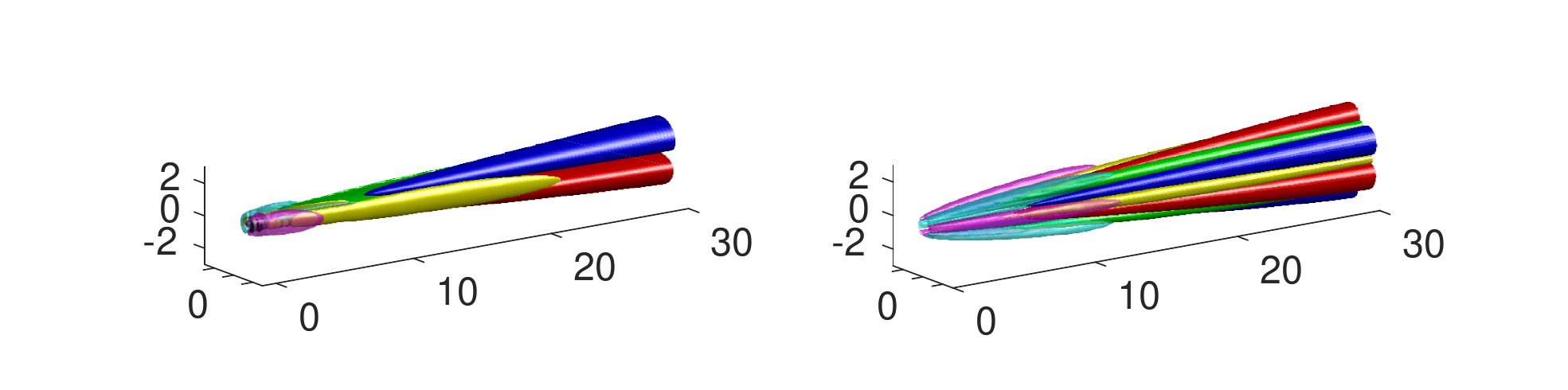}
\caption{Streaks computed via SPOD and resolvent analyses for $m=1$ and 3. (a) Most energetic SPOD mode at $St \rightarrow 0$ for $m=1$ (left) and $m=3$ (right).  Isosurfaces of streamwise velocity ($\bm{\psi_1}:u_x$, red, blue) and streamwise vorticity ($\bm{\psi_1}:\omega_x$, yellow, green), both at $\pm 25\%$ of their maximum value. (b) Global resolvent forcing and response for $m=1$ (left) and $m=3$, (right) at $St=0$. The streamwise forcing vorticity, $\bm{v_1}:\omega_x$ is shown in magenta-cyan with isosurfaces $\pm 0.05 \|\bm{v_1}:\omega_x\|_{\infty}$ for $m=1$ and $\pm 0.15 \|\bm{v_1}:\omega_x\|_{\infty}$ for $m=3$. Streamwise response vorticity, $\bm{u_1}:\omega_x$, is shown in yellow-green with isosurfaces $\pm 0.5 \|\bm{u_1}:\omega_x\|_{\infty}$, and streamwise response velocity, $\bm{u_1}:u_x$, is shown in red-blue with isosurfaces $\pm 0.25 \|\bm{u_1}:u_x\|_{\infty}$.}
\label{fig:M04_Streaks}
\end{figure}

The SPOD modes for $St \rightarrow 0$ are shown in figure \ref{fig:M04_Streaks}. The streaks are visualized in physical space by inverse (azimuthal) Fourier transforming the SPOD mode for a given azimuthal wavenumber. For $m=1$, a well-defined streak ($\bm{\psi_1}:u_x$, red-blue isosurfaces) is observed throughout the domain starting at approximately the end of the potential core. In the $m=3$ plot, streaks are seen upstream and downstream of the potential core, however, at approximately $x/D = [15,20]$ the streaks are less dominant and appear to slightly rotate. The slight rotation is likely an artefact of imperfect SPOD convergence at $St \rightarrow 0 $, as approaching zero frequency still includes small, but non-zero, frequencies. Interestingly, the slight rotation of these modes are rather similar to energetic POD modes found by \cite{freund2009turbulence}.

Further evidence these structures are due to the lift-up mechanism is shown by the yellow-green isosurfaces of streamwise vorticity, $\bm{\psi_1}:\omega_x$, included in figure \ref{fig:M04_Streaks}. The presence of streamwise vorticity, or rolls, and their particular location, situated precisely between positive and negative streamwise velocity contours, are indicative of the lift-up mechanism. The observation of such streamwise vortices in jets has previously been reported by a number of authors, including \cite{bradshaw1964turbulence,liepmann1991streamwise,martin1991numerical,paschereit1992flow,liepmann1992role,arnette1993streamwise,alkislar2007effect,citriniti2000reconstruction,jung2004downstream}, and \cite{caraballo2003application}, among others, each noting the significant impact these vortices imposed upon the mean flow and the accompanying streaks. 

Performing resolvent analysis for $m=1,3$ at $St = 0$ provides a detailed understanding of the forcing mechanisms which give rise to the observed behaviour of the SPOD modes. In figure \ref{fig:M04_Streaks} (b) we see the canonical lift-up progression.  The optimal forcing acts on the cross-stream components upstream at the nozzle, with associated streamwise vorticity $\bm{v_1}:\omega_x$ (magenta-cyan isosurfaces), in order to optimally generate rolls, $\bm{u_1}:\omega_x$. The rolls, in turn, give rise to streamwise velocity responses, $\bm{u_1}:u_x$ , i.e. streaks -- this connection, between these two response structures, has generally been interpreted as lift-up in jets, while the lift-up mechanism we describe refers to the input-output nature of both optimal forcing and response. Further, the responses in both streamwise vorticity and velocity shown by the SPOD modes in figure \ref{fig:M04_Streaks} (a) agree quite well with the resolvent findings for both azimuthal wavenumbers. 

\begin{figure}
\centering
\includegraphics[width=0.75\textwidth,trim={0cm 0cm 0cm 0cm},clip]{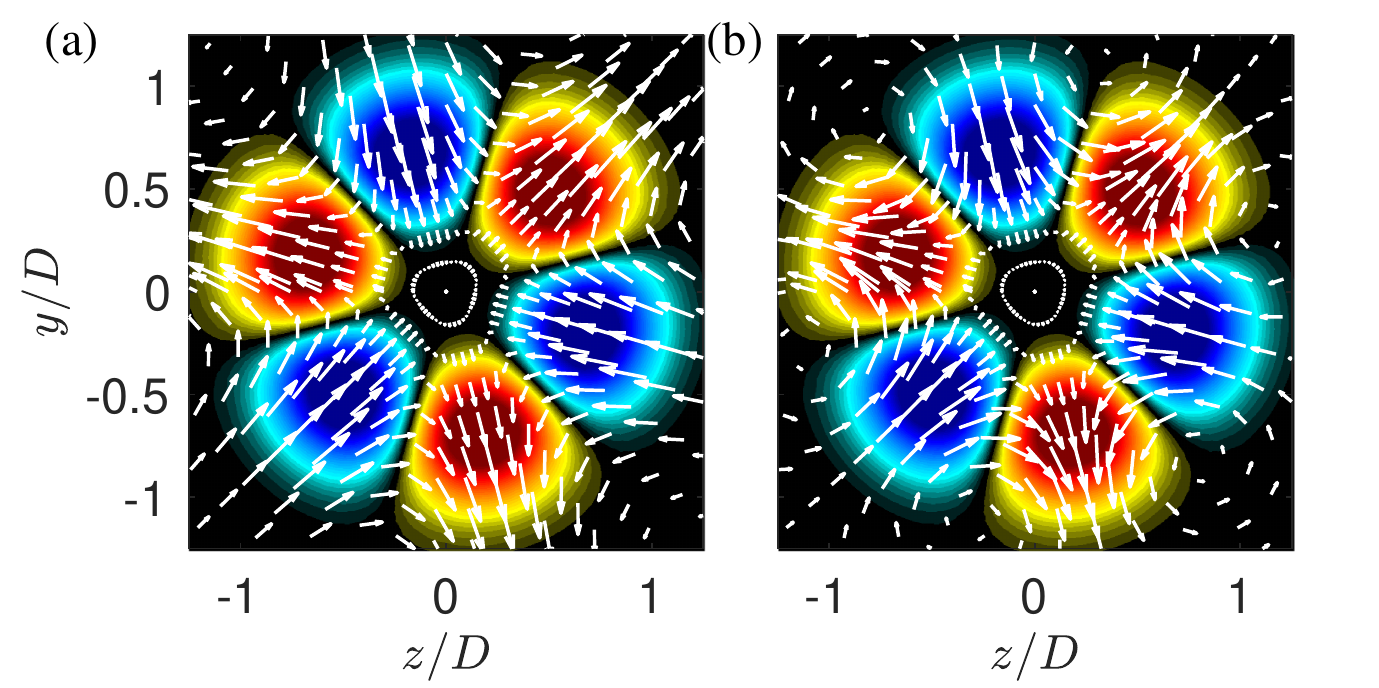}
\caption{Cross plane at $x/D = 5$ of rolls and streaks at $m=3$ computed via resolvent in figure \ref{fig:M04_Streaks}. The red-blue contours present streaks with values set at $\pm 0.75 \|\bm{u_1}:u_x(x/D=5)\|_{\infty}$. The overlaid vectors represent the forcing rolls in (a) and response rolls in (b). }
\label{fig:Cross_Plane}
\end{figure}

Although the three-dimensional (3-D) representation of figure \ref{fig:M04_Streaks} (b) presents many of the salient characteristics of the lift-up mechanism, a 2-D cut at $x/D=5$ shown in figure \ref{fig:Cross_Plane} highlights the radial and azimuthal velocity components of forcing and response rolls with respect to streaks. In figure \ref{fig:Cross_Plane} (a) the forcing velocity vectors show the lifting of high-/low-speed fluid from the centre/outer jet inducing positive/negative streaks, respectively. These mode shape qualities show agreement with previous local analyses of non-modal growth in laminar \citep{boronin2013non,jimenez2017transient} and turbulent jets \citep{Nogueira2019Streaks}. More precisely, the forcing rolls show strong radial velocities coincident with maximum streamwise response for both azimuthal wavenumbers, while strictly azimuthal velocities are located exactly between positive and negative streaks. This latter observation (i.e. location of azimuthal velocities) has implications for results found via the resolvent sensitivity analysis and will be discussed further in \S~\ref{sec:Mod-Low}. 
Additionally, the forcing rolls give rise to response rolls shown in figure \ref{fig:Cross_Plane} (b) presenting similar velocity characteristics as the forcing vectors, however, the response vectors show the vortices are centred near the jet, a quality not shared with the forcing vectors. 

\begin{figure}
\centering
\includegraphics[width=0.65\textwidth]{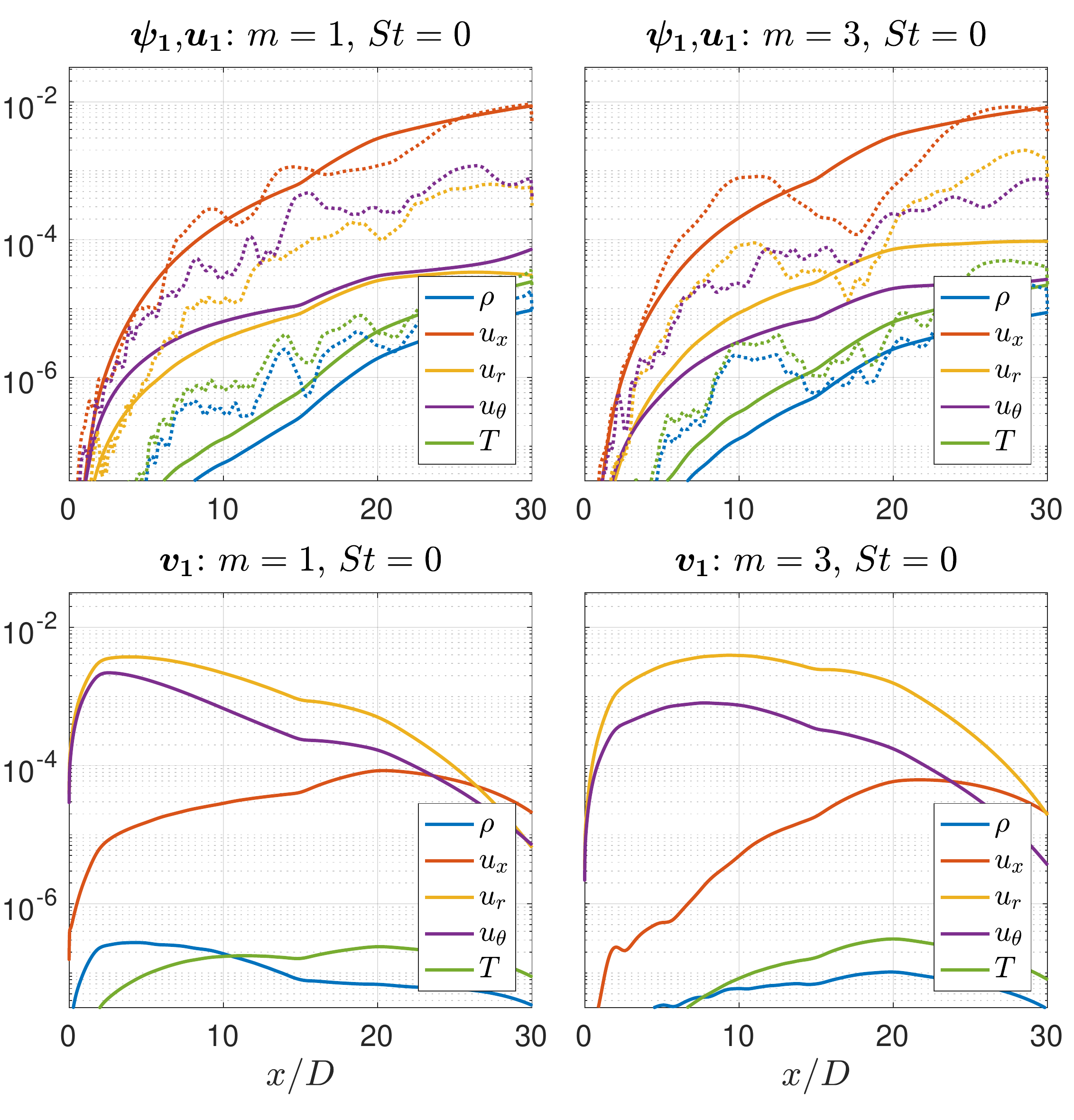}
\caption{Component-wise amplitude curves in the streamwise direction of the streaks for $St=0$ and $m=1$ and 3. Optimal response (top) and forcing (bottom) amplitudes from resolvent (solid) and SPOD (dotted) analyses as a function of streamwise coordinate from the nozzle exit.}
\label{fig:LiftupEnergy}
\end{figure}

To ensure the visualized streamwise components in figure \ref{fig:M04_Streaks} are the dominant variables, we present component-wise amplitudes of each variable. Figure \ref{fig:LiftupEnergy} shows quantitative comparisons of the compressible energy inner product computed at each streamwise position between SPOD ($St \rightarrow 0 $) and resolvent ($St=0$), for $m=1,3$. We emphasize that the full compressible energy inner product of the 5 variables over the streamwise direction is unity by construction. The curves from SPOD (dotted) and resolvent (solid) are trend-wise similar, with noise in the SPOD results due to statistical convergence issues. The streamwise velocity response (i.e. streaks) is clearly the dominant response variable throughout the domain for both the resolvent and SPOD analyses. For the forcing terms, the radial and azimuthal velocities dominate throughout the domain by two orders of magnitude and correspond to the streamwise vorticity (i.e. rolls). Together, the forcing and response amplitudes confirm SPOD and resolvent modes computed at $St = 0$ are dominated by the lift-up mechanism.

The generation of streaks through the lift-up mechanism is also observed for the transonic, $M_j=0.9$, and supersonic, $M_j=1.5$ cases, showing similar trends. SPOD and resolvent analysis results for both jets, providing even clearer agreement, are shown in appendix \ref{sec:TransSuper}.

\section{Interplay of lift-up, Kelvin-Helmholtz, and Orr mechanisms} \label{sec:Mod-Low}

We now discuss the interplay between the three jet mechanisms in the frequency-wavenumber space. 

However, before proceeding, we emphasize that distinct characteristics of each mechanism generally refer to the {\it response}, as, for example, the optimal forcings for both KH and Orr wavepackets are similar.  \cite{garnaud2013preferred} found that the KH preferred jet mode could be more efficiently excited through the Orr mechanism in the pipe upstream of the jet, thus combining algebraic spatial growth in the pipe with the downstream exponential growth associated with KH. This is analogous to the triggering of Tollmien-Schlichting waves via the Orr forcing structures in boundary layers \citep{aakervik2008global}. In a similar vein, the Orr mechanism has also been (for small, but non-zero streamwise wavenumbers) observed to accompany the lift-up mechanism in both the plane shear layer \citep{arratia2013transient} and wall-bounded flows \citep{hack2017algebraic}. Therefore, in this analysis we note that the Orr mechanism will likely be present throughout the frequency-wavenumber space considered. The question we seek to address here is which mechanism is dominant in energy/amplification (i.e. SPOD/resolvent analyses). In most cases, we will find that lift-up (discussed next) or KH are the dominant mechanism, however, when both mechanisms are suppressed, the Orr mechanism becomes dominant by default.

\begin{figure}
\centering
\includegraphics[width=0.8\textwidth,trim={0cm 0cm 0cm 0cm},clip]{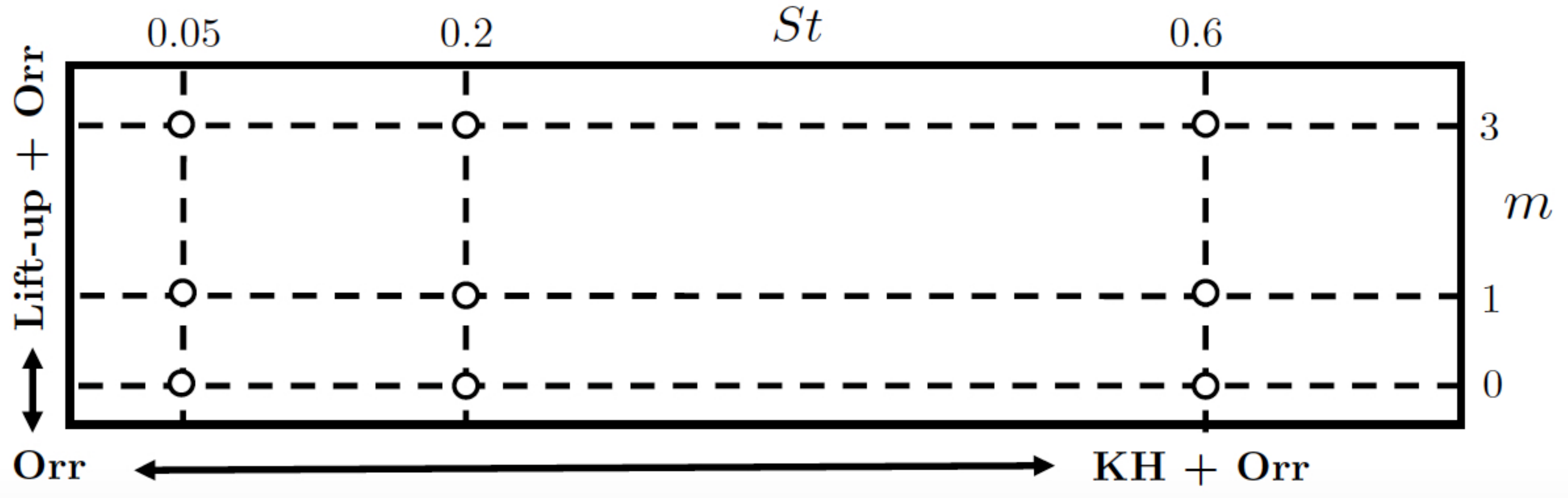}
\caption{ A cartoon mechanism map highlighting the know regions of active mechanisms along the $m=0$ and $St = 0$ axes. The locations of the circles mirror the placement of the subsequent SPOD and resolvent analyses and denote their location in the frequency-wavenumber plane. }
\label{fig:Cartoon}
\end{figure}

Based on previous KH and Orr studies discussed in the introduction and results of the prior section on lift-up, we hypothesize regions of dominance between the mechanisms as a function of frequency and azimuthal mode number in figure \ref{fig:Cartoon}.  This map will be filled in and clarified in the remainder of this section.  The Orr mechanism would be present over all frequencies and wavenumbers, but only dominant where the lift-up and KH are not strongly amplified.  Lift-up would dominant at $St \rightarrow 0$ (but is absent for $m=0$), whereas KH would dominate over a range of intermediate frequencies.  

\begin{figure}
\centering
{\small (a) SPOD}  \\
\includegraphics[width=1\textwidth,trim={0.5cm 0.15cm 0cm 0cm},clip]{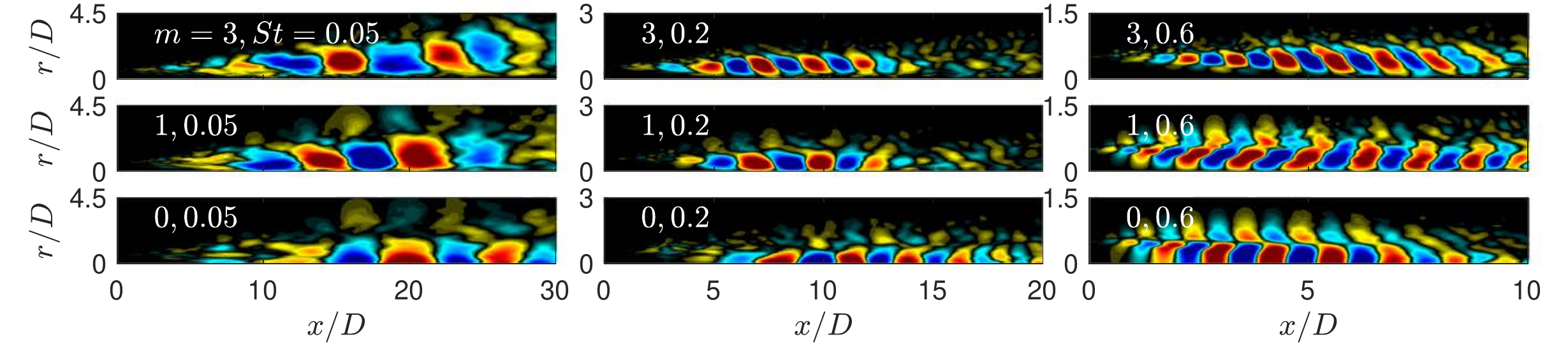}
\\ {\small (b) Resolvent response } \\
\includegraphics[width=1\textwidth,trim={0.5cm 0.15cm 0cm 0cm},clip]{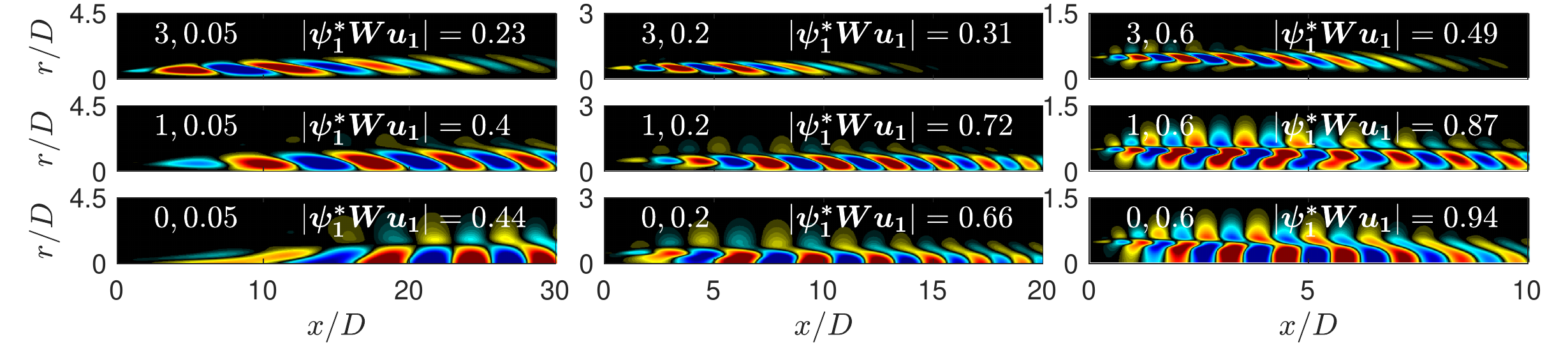}
\\ {\small (c) Resolvent forcing }\\
\includegraphics[width=1\textwidth,trim={0.5cm 0.15cm 0cm 0cm},clip]{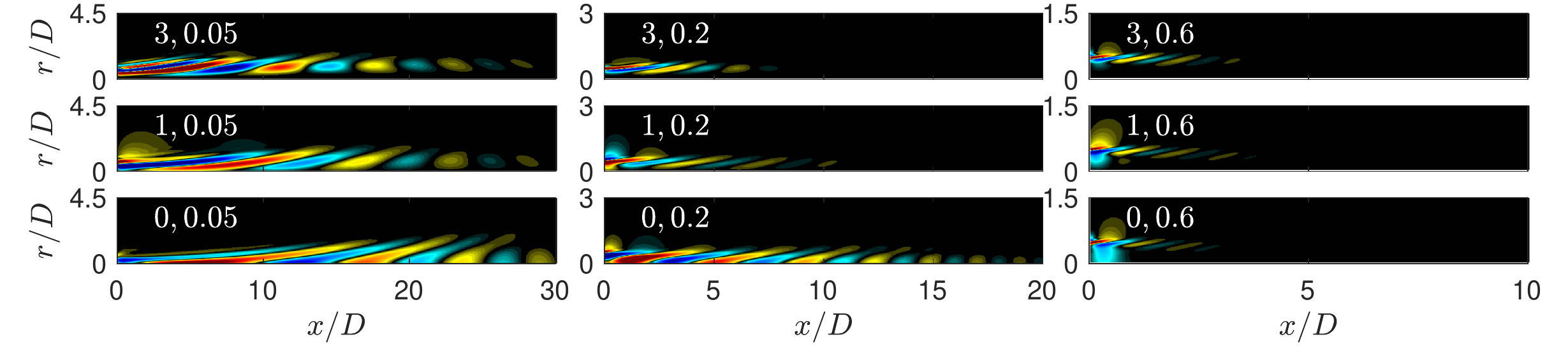}
\caption{ (a) most energetic SPOD, (b) resolvent response, (c) and resolvent forcing modes for azimuthal wavenumbers $m=[0,1,3]$ and frequencies $St=[0.05, 0.2, 0.6]$. Streamwise velocity perturbation, $u_x$, component is shown with contours corresponding to $\pm 0.5 \|u_x\|_\infty$, with projection coefficients, $|\bm{\psi}_1^* \bm{W} \bm{u}_1|$, between the full SPOD responses, $\bm{\psi}_1$, and resolvent responses, $\bm{u}_1$, provided in (b). Note the change in domain limits, $x/D \in [0,30], [0,20]$, and $[0,10]$ for $St = [0.05, 0.2, 0.6]$, respectively. }
\label{fig:Modes}
\end{figure}

\begin{figure}
\centering
\vspace{0.5cm}
\includegraphics[width=0.9\textwidth]{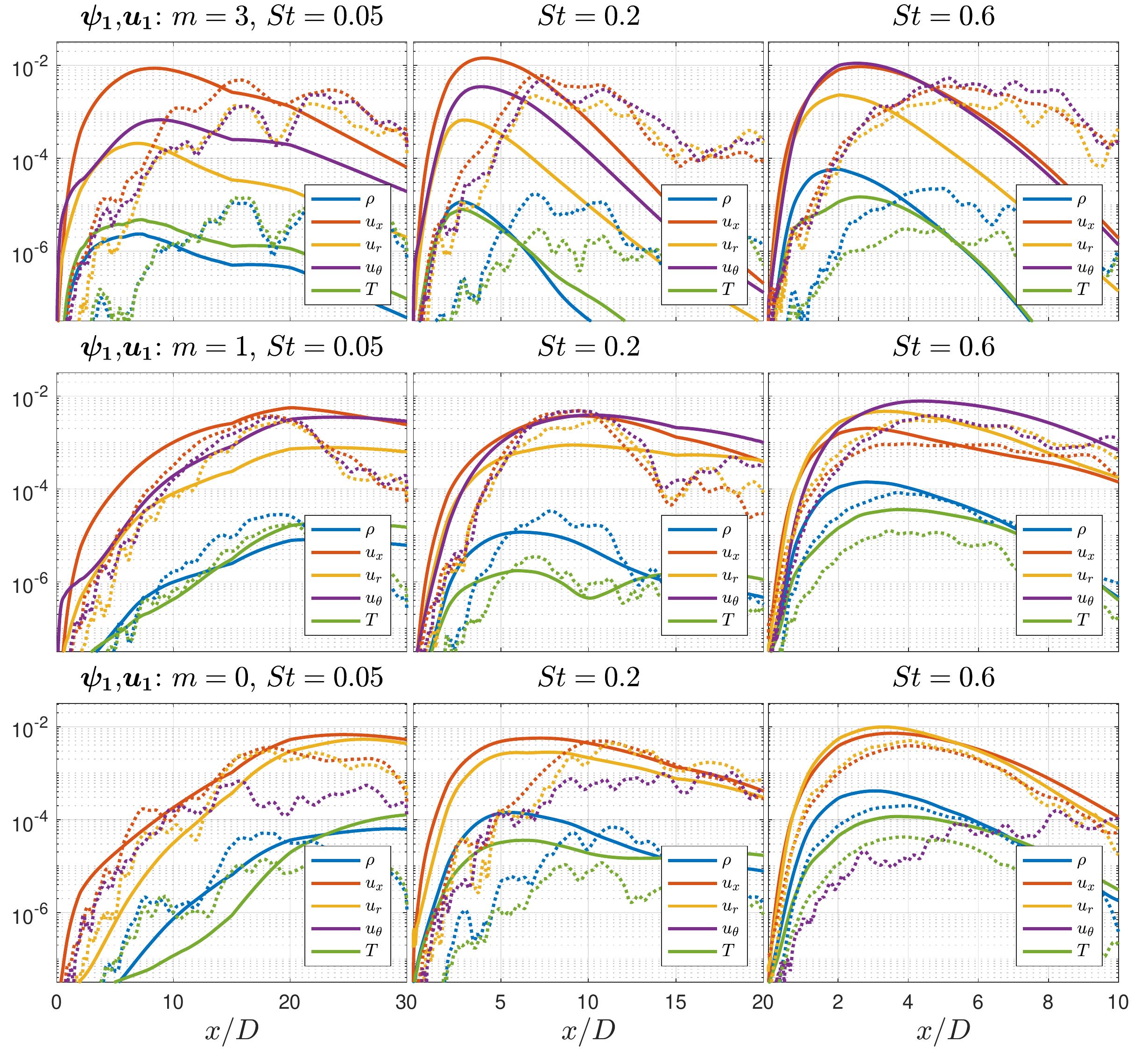}
\caption{Streamwise evolution of component-wise amplitude curves for resolvent (solid lines) and SPOD (dotted lines) analyses. The layout of the figure mirrors that of figure \ref{fig:Modes} with $m=3$ represented in the first row followed by $m=1,0$, and the first column displaying the lowest $St$ increasing with columns to the right. Note that the truncated domains shown in figure \ref{fig:Modes} are maintained for each $St-m$ pair.}
\label{fig:Resolvent_SPOD_Energy}
\end{figure}

\begin{figure}
\centering
\vspace{0.5cm}
\includegraphics[width=1\textwidth]{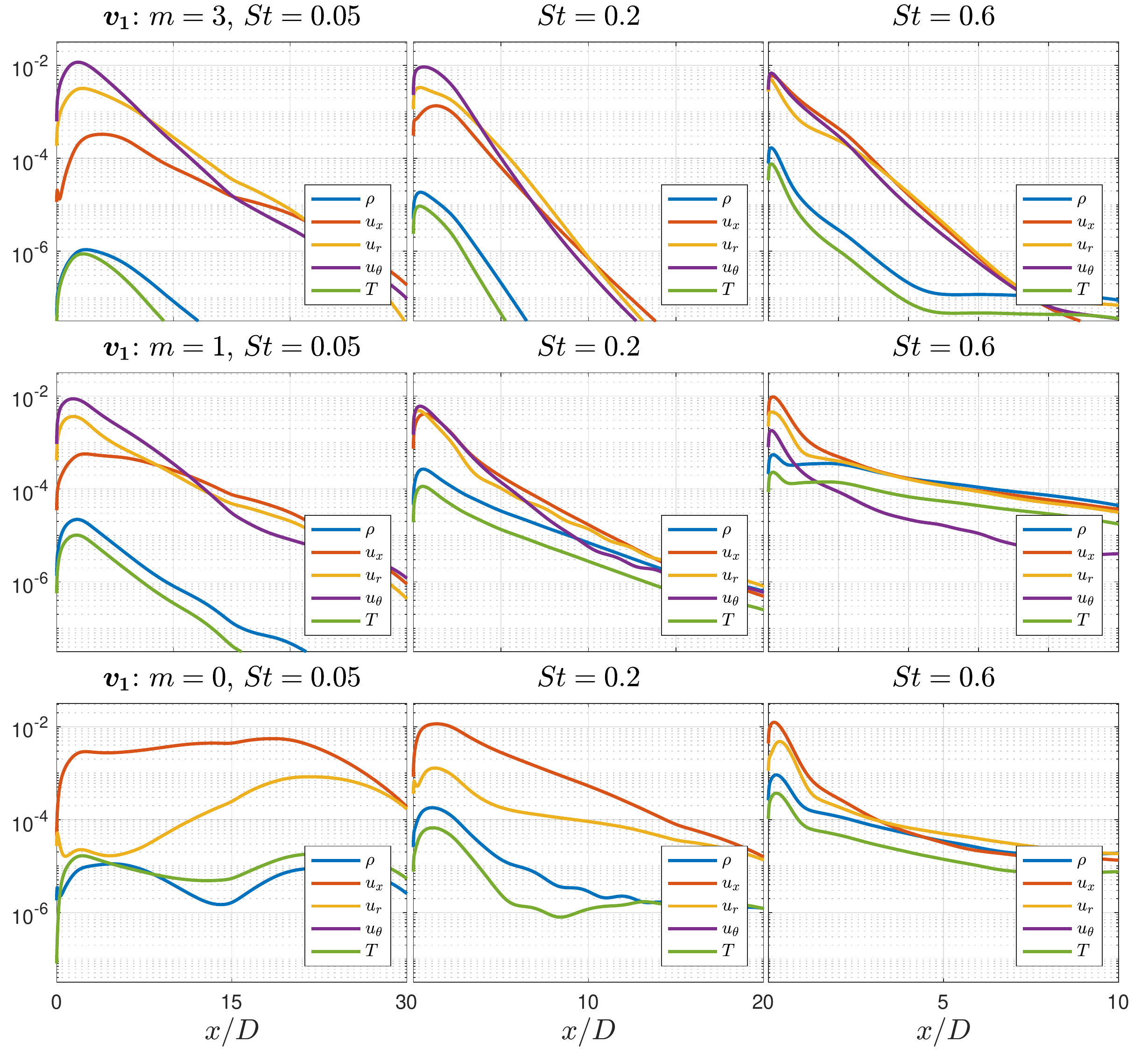}
\caption{Streamwise evolution of component-wise amplitude curves for resolvent forcing modes. The layout and streamwise extent of each plot is identical to figure \ref{fig:Resolvent_SPOD_Energy}.}
\label{fig:Resolvent_Forcing_Energy}
\end{figure}

\begin{figure}
\centering
\vspace{0.5cm}
(a) $m=0$, $St=0.6$. KH-dominated
\includegraphics[width=1\textwidth,trim={0.15cm 0cm 0cm 0cm},clip]{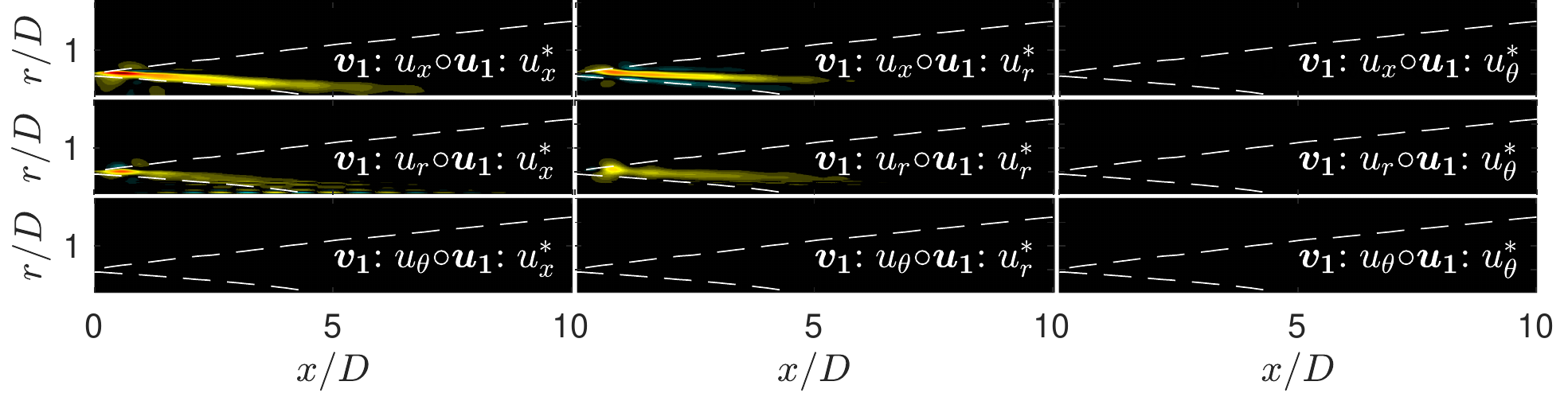}
\\ (b) $m=0$, $St=0.05$. Orr-dominated
\includegraphics[width=1\textwidth,trim={0.15cm 0cm 0cm 0cm},clip]{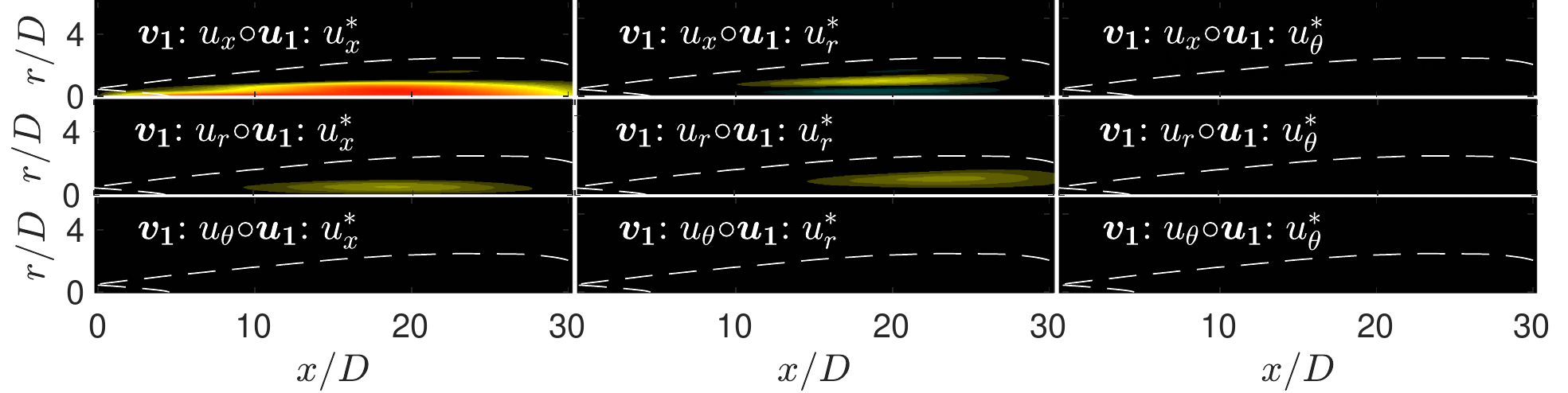}
\\ (c) $m=3$, $St=0$. Lift-up-dominated
\includegraphics[width=1\textwidth,trim={0.15cm 0cm 0cm 0cm},clip]{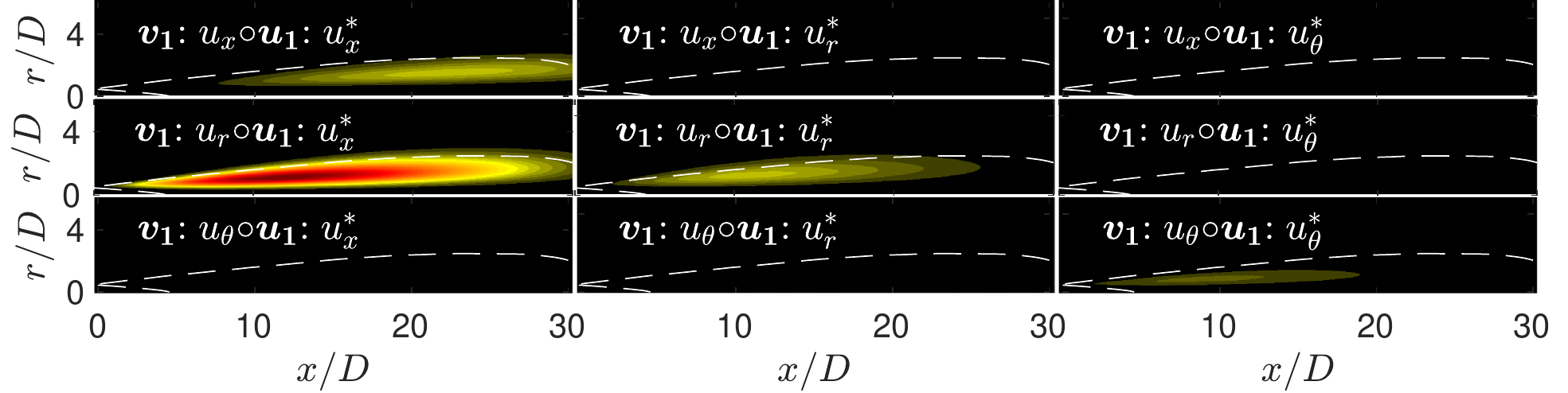}
\caption{Resolvent-based spatial sensitivity plots for all 9 velocity combinations for $[m,St] = [0,0.6]$ (a), $[m,St] = [0,0.05]$ (b), and $[m,St] = [3,0]$ (c). Contours are set for each of the three plots by the maximum sensitivity, $\pm \|\bm{v_1} \circ \bm{u_1}^* \|_\infty$ , across all 9 velocity pairs. Note for the KH case the domain is reduced to $x/D=[0, 10]$ and $r/D=[0, 2]$ to highlight the upstream behaviour, as no sensitivity is observed outside this domain. The white, dotted lines provide reference to the envelope of the jet that is $>10\%$ of the maximum turbulent kinetic energy.}
\label{fig:Sensitivity}
\end{figure}

To fill in the map, we plot in figure~\ref{fig:Modes} the streamwise velocity of the dominant SPOD (a) and resolvent forcing/response (b/c) at the 9 frequency-wavenumber pairs depicted in figure \ref{fig:Cartoon}.  For these same frequency-wavenumber pairs, figure \ref{fig:Resolvent_SPOD_Energy} shows the radially integrated amplitudes for all response flow variables (in both SPOD and resolvent), whereas figure~\ref{fig:Resolvent_Forcing_Energy} shows the radially integrated amplitudes of the forcing flow variables. Finally, the sensitivity of the resolvent gains for selected cases is shown in  figure \ref{fig:Sensitivity}. 

A detailed analysis of these plots is given in the following subsections.  Overall, one can see that the resolvent and SPOD analyses show significant agreement between the mode shapes and their component-wise amplitude content, particularly for the moderate frequencies where KH low-rank behaviour is expected.  The resolvent modes (figure \ref{fig:Modes}b,c) are similar to those presented in \cite{SchmidtJFM2018}, but there are two important differences. The first is that these modes were computed by including an eddy-viscosity model in the resolvent operator, greatly improving the agreement between SPOD and resolvent, particularly at low $St$ \citep{pickering2019eddy}. The second difference is that we plot streamwise velocity rather than pressure, which better allows us to isolate features associated with the differing mechanisms.

\subsection{KH-dominated response ($St=0.6$, $m=0,1$)} \label{sec:KH_dominated}

At $[m,St] = [0,0.6]$, the response is dominated by the KH mechanism \citep{SchmidtJFM2018} and displays the many characteristics described earlier. The SPOD and resolvent modes shown in \ref{fig:Modes} are almost indistinguishable, and similar to what has been previously reported using PSE.  The associated forcing structures (figure \ref{fig:Modes}d) are localized near the lip line of the nozzle and are tilted at an angle $\sim 45^\circ$, representative of Orr-type forcing known to efficiently trigger the KH mechanism \citep{garnaud2013preferred,tissot2017wave,tissot2017sensitivity,lesshafft2018resolvent}). We also find an estimated phase speed of $c_{ph} \approx 0.8$ and a 90 degree phase shift at the critical layer where the apparent phase speed is equal to the mean flow speed \citep{cavalieri2013wavepackets} (phase speed is estimated by taking the Fourier transform in the streamwise direction of the LES pressure along the lip line). Similar behaviour is observed for  $m=1$ wavenumbers at $St=0.6$, for which the KH response is also dominant.


The component-wise amplitude curves for $m \in [0,3]$ and  $St = 0.6$ in figure \ref{fig:Resolvent_SPOD_Energy} also support the KH interpretation with an initially rapid (exponential) growth followed by saturation and decay by $x/D \approx 5$. By contrast with what will be shown for the lowest frequency, the velocity components (streamwise, radial, and azimuthal) have similar amplitudes, apart from the absence of azimuthal velocity when $m=0$.  Forcing modes in figure \ref{fig:Resolvent_Forcing_Energy} for all azimuthal wavenumbers are concentrated upstream with the maximum forcing amplitude occurring within $x/D \leq 0.1$ and along the lip line. The forcing decays by at least two orders of magnitude within the first two jet diameters. This observation is also apparent from the gain sensitivity analysis (figure~\ref{fig:Sensitivity}a), showing sensitivity localized to the inner edge of the shear layer terminating around the end of the potential core, with similar component-wise sensitivity found active for all velocity components (i.e. $\bm{v_1}:u_x,\bm{v_1}:u_r$ and $\bm{u_1}:u_x,\bm{u_1}:{u_r}$ pairs). This localized sensitivity is in stark contrast with the sensitivity contours associated with the Orr and lift-up mechanisms discussed next.

\subsection{Orr-dominated response ($St \rightarrow 0$, $m=0$)}


At $[m,St] = [0,0.05-0.2]$ the Orr mechanism dominates the response as shown previously by \cite{SchmidtJFM2018}. The upstream region of the forcing modes (figure~\ref{fig:Modes}d) are inclined at $-45^\circ$ against the mean shear and the response (figure~\ref{fig:Modes}c) turns from $-45^\circ$ to $+45^\circ$ throughout their envelope.  
The response modes possess a phase speed of approximately 0.4, and although we still observe a phase change across the critical layer, this behaviour is rather weak when comparing to the KH-dominated modes described in \S~\ref{sec:KH_dominated} ($St = 0.6$). In contrast to KH, the response grows gradually with $x/D$, peaking some 20 diameters downstream and is only weakly damped thereafter. We also find both forcing and response modes to peak near the jet centreline, similar to Orr-type modes found by \cite{garnaud2013preferred} and \cite{lesshafft2018resolvent}, rather than at the region of maximum shear in the jets. Although this seems counter-intuitive for the Orr mechanism, by $x/D=15$, well beyond the close of the potential core, the profile is diffuse and the `critical layer' for this phase speed would be relatively close to the jet axis.

Finally, the sensitivity contours (figure~\ref{fig:Sensitivity}b) provide contrast with what was observed for the KH response.  First, the sensitivity is high over a spatially distributed region commensurate with the response region, and reaching far downstream where the flow no longer supports KH amplification. Secondly, there is a large component-wise sensitivity between the streamwise velocity forcing and the streamwise velocity response, a distinguishing characteristic when compared with the KH and lift-up mechanisms.

\subsection{Lift-up-dominated response ($St \rightarrow 0$, $m>0$)}

For $m \ne 0$, and approaching zero frequency, the wavepackets in figure \ref{fig:Modes} (a,b) show a different structure from the KH-dominated regimes. Although the response and forcing mode structures are inclined at $45^\circ$ (this response behaviour in jets was also observed by \cite{bradshaw1964turbulence}), and seem to indicate Orr mechanism behaviour (again, likely present), it is the component-wise amplitude curves that provide evidence of lift-up.  For $[m,St] = [3,0.05]$  in figure \ref{fig:Resolvent_SPOD_Energy}, the streamwise velocity is approximately an order of magnitude larger than the other components, whereas for $[m,St] = [0,0.05]$ (i.e. Orr only) both streamwise and radial velocities contribute similarly to the overall mode. The presence of streaks is directly responsible for this higher streamwise velocity. The forcing in figure \ref{fig:Resolvent_Forcing_Energy}, by contrast, shows a dominance of radial and azimuthal forcing components, producing streamwise vortical forcing, $\bm{v_1}:\omega_x$. In tandem, the response and forcing amplitude curves describe the lifting of fluid, via cross-plane rolls, from high- and low-speed regions of the jet to the fluctuating streamwise velocity response. These results show that the lift-up mechanism, demonstrated at zero frequency in \S~\ref{sec:Zero_St}, persists at small, non-zero frequencies.  In this case the streaks are not stationary, and slowly rotate about the jet.  We find evidence of streaks for all non-zero wavenumbers up to frequencies of $St \approx 0.2$.  

The sensitivity analysis (figure~\ref{fig:Sensitivity}c), after being described for KH and Orr, gives further evidence that the lift-up mechanism may be uniquely identified in turbulent jets.  Unlike KH and Orr, we find the largest sensitivity in radial forcing to streamwise response, and unlike KH, but similar to Orr, the sensitivity is spatially distributed throughout the domain. The critical difference between Orr and lift-up is the shift of sensitivity to the radial forcing, streamwise response pair. We expect this for the lift-up mechanism as the radial forcing component lifts high- and low-speed regions of the jet to the streamwise response. 

One puzzling aspect of the results is the lack of sensitivity of the streamwise velocity to the azimuthal velocity forcing (figure \ref{fig:Sensitivity}c).  In general, streamwise vorticity is associated with either variation of azimuthal velocity with radius, or radial velocity with azimuth, and it seems surprising that the sensitivity identifies changes to the latter, but not the former, as a means to amplify the streaks.  This is explained by the phase between radial/azimuthal forcing and the corresponding response.  As is observed in figure \ref{fig:Cross_Plane} (a,c) in \S~\ref{sec:Zero_St}, the largest regions of radial forcing are coincident with the most energetic regions of the streaks (i.e. large sensitivity), but the largest azimuthal forcing occurs where streaks have low amplitude, resulting in low sensitivity. The above interpretation is limited to the sensitivity map as defined by \cite{qadri2017frequency}, which examines the structural sensitivity to spatially localized (delta-function) changes to the structure of the linear system defining the resolvent.  In reality, any finite-sized perturbation would convolve the response and forcing (rather than multiply them), and likely lead to larger sensitivity to the azimuthal velocity than reported here.

\subsection{Linear mechanism map}

\begin{figure}
\centering
\vspace{0.5cm}
\includegraphics[width=0.9\textwidth]{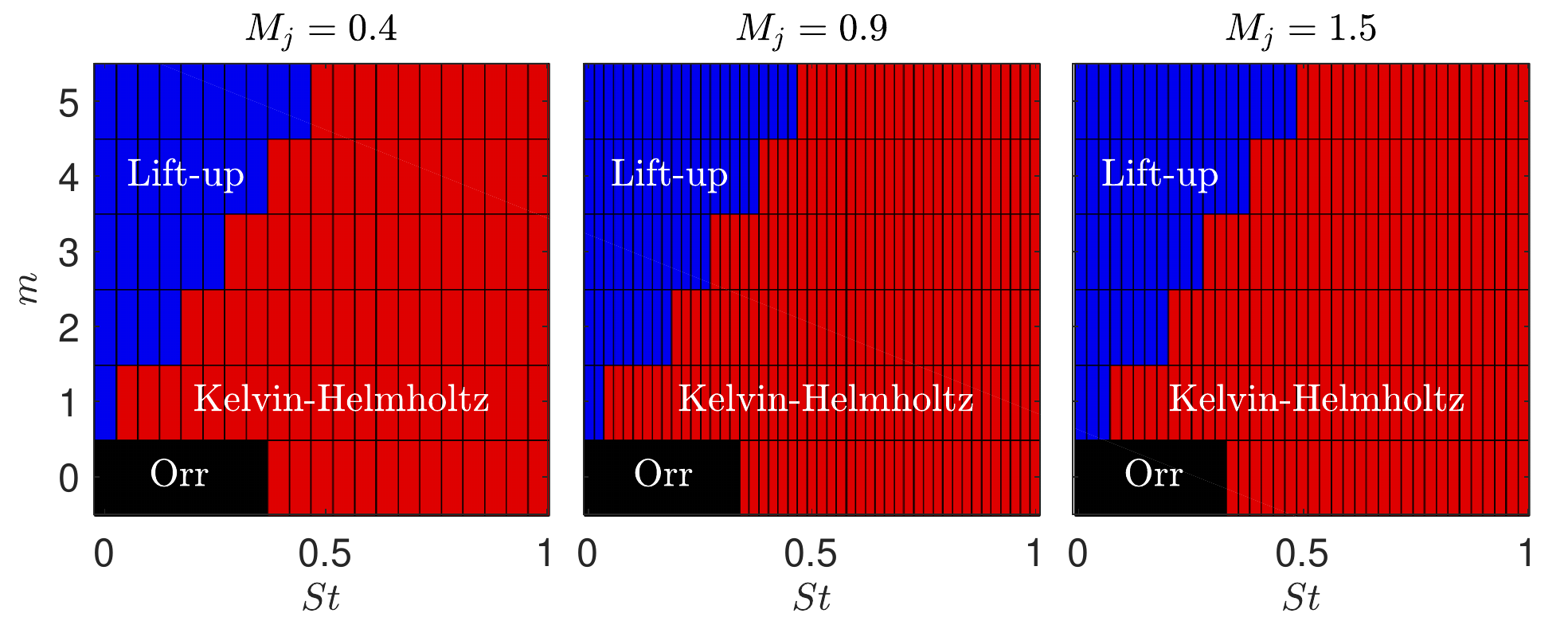}
\caption{Linear mechanism map estimating the dominant linear mechanisms in the frequency-wavenumber space of the most amplified resolvent response for turbulent jets. Red, black, and blue represent the KH, Orr, and lift-up mechanisms, respectively.}
\label{fig:Mechanism_Map}
\end{figure}

Following identification of the salient properties of each turbulent jet mechanism, we propose the mechanism map shown in figure \ref{fig:Mechanism_Map} estimating the regions of mechanism dominance in the frequency-wavenumber space for turbulent jets. We determine the mechanism map through inspection of individual frequency/azimuthal modes and identify the nature of the dominant mode by applying the criteria discussed earlier in this section. For $m=0$ the map reflects the shift in dominance of the KH at moderate Strouhal numbers to Orr at low frequencies, with mechanism shifts occurring at $St \approx 0.3$ for $M_j=0.4,0.9,1.5$. For the lift-up mechanism, the region of dominance may be estimated as $St/m \leq 0.1$ in figure \ref{fig:Mechanism_Map}. Although, the lift-up mechanism continues to become more dominant as $St/m \rightarrow 0$, similar to the findings of \cite{arratia2013transient} for a plane shear layer as lift-up dominates when $k_x/k_z \rightarrow 0 $, where $k_x,k_y$ are the streamwise and spanwise wavenumbers, respectively. It is important to stress that this map is an estimate of mechanism dominance and mechanisms are simultaneously present at the boundaries of dominant mechanism regions.

\section{Local analysis for higher azimuthal modes}\label{sec:Domain}

Up until now, the SPOD and resolvent analyses have used the full computational domain extending 30 diameters downstream.  The associated Chu compressible energy norm, which is integrated over this entire region, is biased toward the lower azimuthal modes that have long wavelengths and dominate the jet far downstream.  For example, the dominant azimuthal wavenumber for much of the frequency range is $m=1$, and as $St \rightarrow 0$ the dominant azimuthal wavenumber is $m=2$.  When one inspects the results for the higher azimuthal modes in the global domain large noise downstream prevents the identification of coherent structures likely near the nozzle.  The question arises as to what is the dominant azimuthal wavenumber when smaller regions closer to the nozzle exit are considered, particularly as $St \rightarrow 0$.

\begin{figure}
\centering
\includegraphics[width=0.6\textwidth]{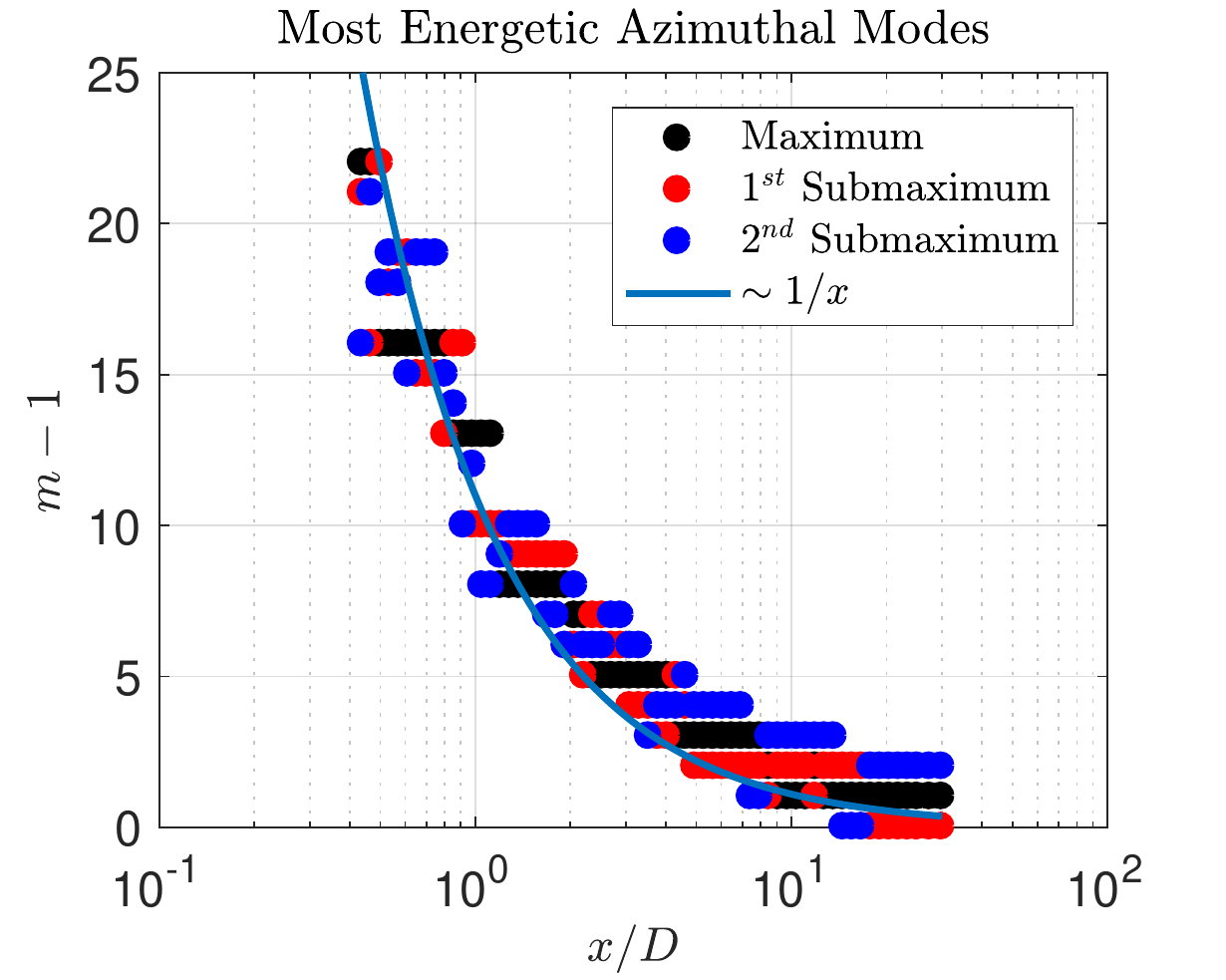}
\caption{Wavenumbers of the three largest SPOD energies from the $St \rightarrow 0$ SPOD bin as a function of streamwise distance. Here SPOD was performed locally on 2-D streamwise cross sections and we present a scaling of maximum azimuthal wavenumber as $\sim 1/x + 1 $.}
\label{fig:SPOD_Planes}
\end{figure}

To determine an overall picture of which azimuthal wavenumbers dominate at different streamwise positions in the jet, we first performed SPOD (using blocks of 256 snapshots and a 50\% overlap) in 2-D cross-stream planes throughout the range $x/D = [0.5 - 30]$.   The three azimuthal wavenumbers exhibiting the largest energy at each $x/D$ are plotted in figure \ref{fig:SPOD_Planes}.  We see that the maximum energy for $x/D = 30$ is in line with the global analysis result, at azimuthal wavenumber $m=2$.  However, as the plane moves upstream we see a trend towards higher azimuthal wavenumbers. In fact, this trend scales as $m_{max} \sim 1/x + 1$ (i.e. the maximum wavenumber approaches 1). This scaling is inversely proportional to the linear scaling of the shear layer of the jet, $\sim x$ \citep{SchmidtJFM2018}, and has also been reported from experimental observations of \cite{jung2004downstream} and \cite{Nogueira2019Streaks} for axial stations $x/D \leq 6$ and $x/D\leq 8$, respectively. This suggests that the width of the shear layer determines support for particular azimuthal wavenumbers as $St \rightarrow 0$. 

\begin{figure}
\centering
{\small (a) SPOD Spectra } \\
\includegraphics[width=1\textwidth,trim={0cm 0cm 0cm 0cm},clip]{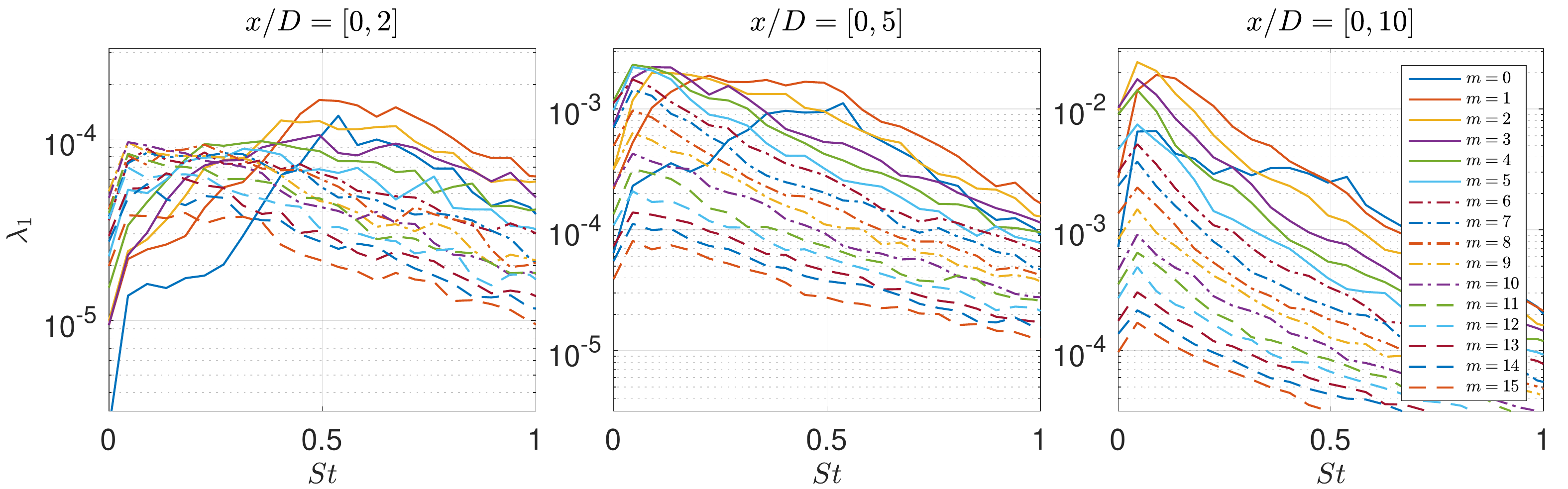}
\\ {\small (b) SPOD frequency-wavenumber maps }\\
\includegraphics[width=1\textwidth,trim={0cm 0cm 0cm 0cm},clip]{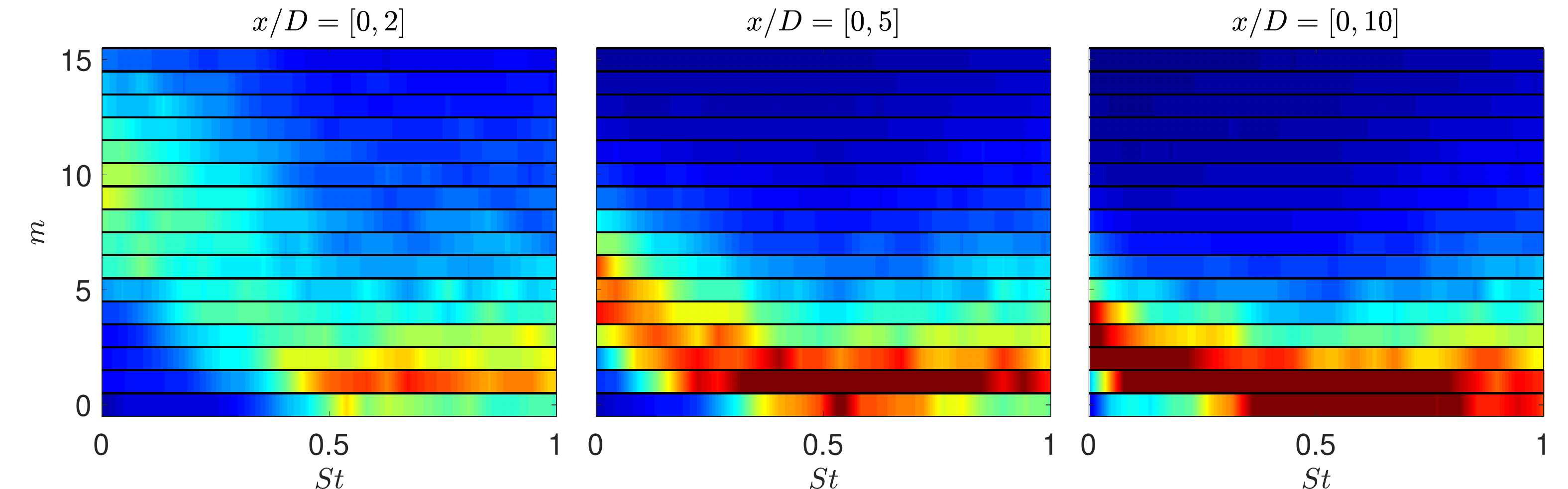}
\caption{SPOD spectra, $\lambda_1(St,m)$, for truncated domains of $x/D \in [0,2], [0,5], [0,10]$ left to right respectively, for azimuthal wavenumbers $m\in[0,15]$. (a) SPOD spectrum curves for all wavenumbers. (b) Semi-discrete frequency-wavenumber maps showing the distribution of energy (as a percentage) among wavenumbers at each frequency with contour levels at 0-20\%.}
\label{fig:Domains}
\end{figure}

\begin{figure}
\centering
{\small (a) Resolvent Spectra } \\
\includegraphics[width=1\textwidth,trim={0cm 0cm 0cm 0cm},clip]{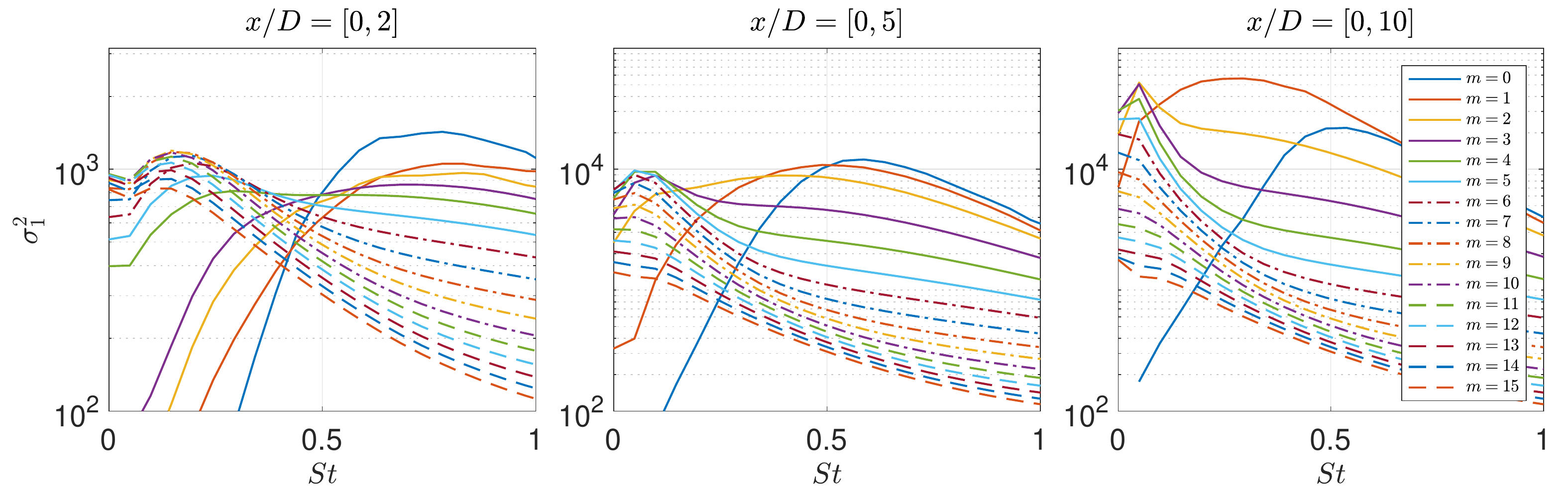}
\\ {\small (b) Resolvent frequency-wavenumber maps }\\
\includegraphics[width=1\textwidth,trim={0cm 0cm 0cm 0cm},clip]{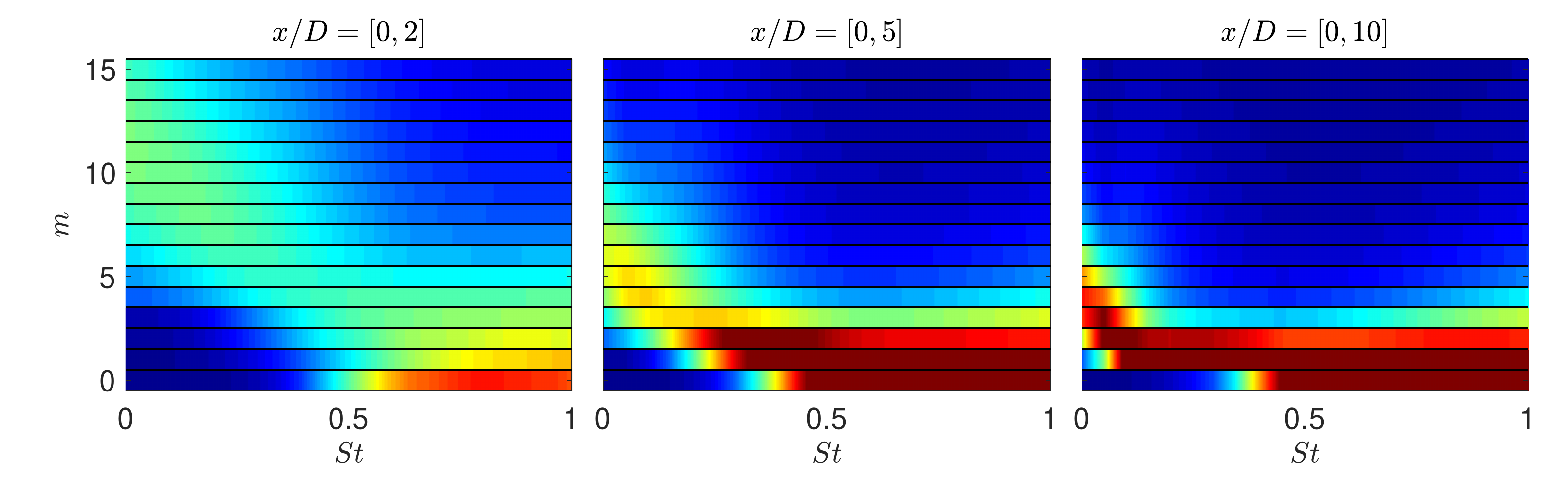}
\caption{Resolvent spectra, $\sigma^2_1(St,m)$, for truncated domains of $x/D \in [0,2], [0,5], [0,10]$ left to right respectively, for azimuthal wavenumbers $m\in[0,15]$. (a) Resolvent spectrum curves for all wavenumbers. (b) Semi-discrete frequency-wavenumber maps showing the distribution of energy (as a percentage) among wavenumbers at each frequency with contour levels at 0-20\%.}
\label{fig:DomainsResolvent}
\end{figure}

Next, to better isolate streaks related to higher azimuthal wavenumbers, the truncated domains (including the 2-D cross-stream planes above) $x/D \in [0,2], [0,5], [0,10]$ are examined for $m \in [0,15]$ using SPOD and resolvent analyses. For these truncated domains, blocks of 256 snapshots and a 50\% overlap suffice even at low frequencies and are used for this final analysis. Figure~\ref{fig:Domains} shows the SPOD energy spectra and normalized azimuthal frequency-wavenumber maps similar to the ones reported in figure~\ref{fig:LinearGain} for the restricted domains, and figure~\ref{fig:DomainsResolvent} shows the associated resolvent results.

\begin{figure}
\centering
\includegraphics[width=\textwidth,trim={1cm 0cm 0cm 0cm},clip]{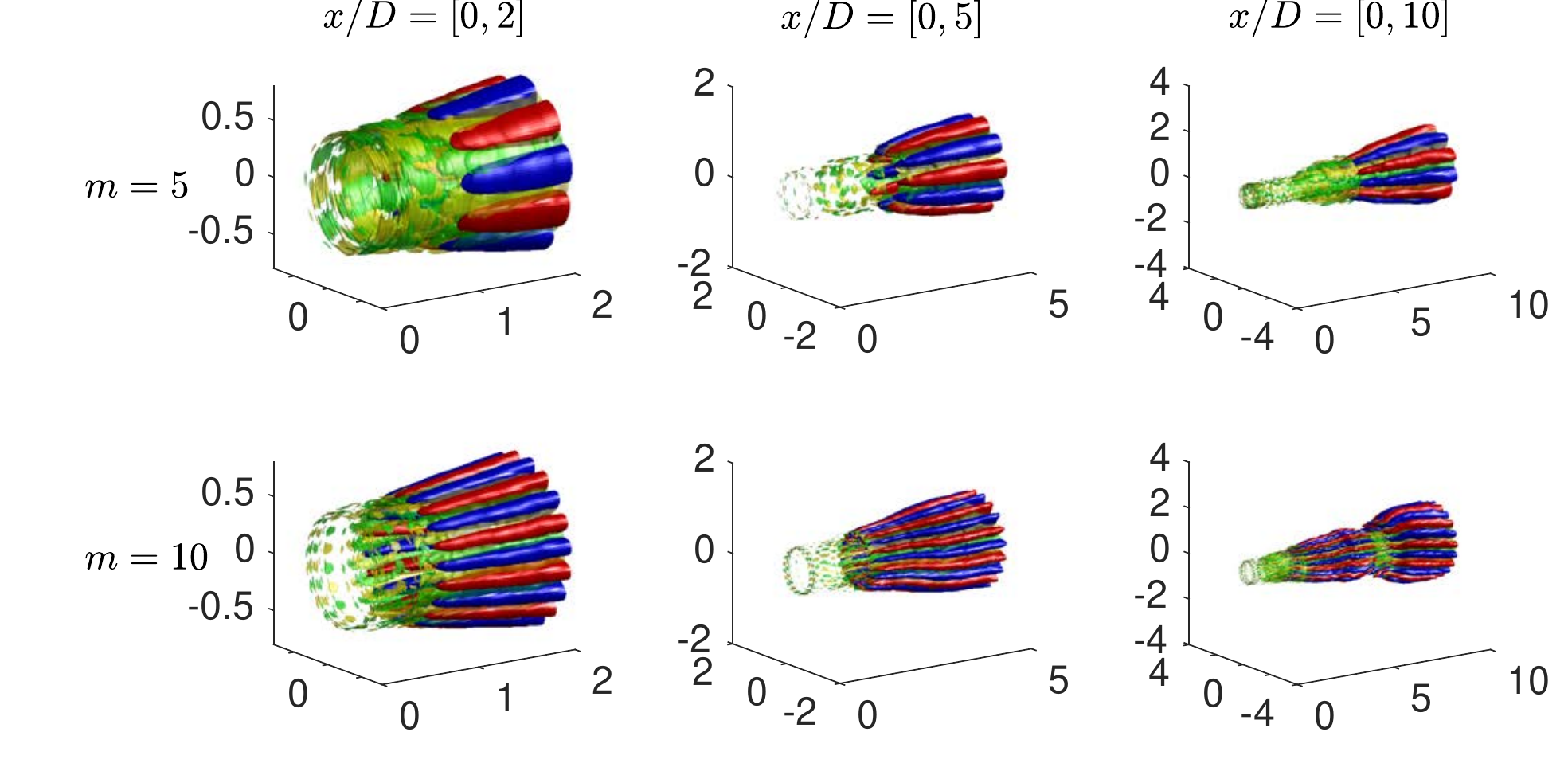}
\caption{Three-dimensional reconstruction of the first SPOD mode for $m = [5,10]$ (top and bottom, respectively) at three truncated domains $x/D \in [0,2], [0,5], [0,10]$ (left to right, respectively) for  $St \rightarrow 0$. Streamwise velocity, $\bm{\psi_1}:u_x$, is denoted as red-blue with isosurfaces $\pm 25\%$ the maximum streamwise velocity and streamwise vorticity, $\bm{\psi_1}:\omega_x$, is shown as yellow-green with isosurfaces as $\pm 50\%$ of the maximum streamwise vorticity.}
\label{fig:Domain_SPOD_3D}
\end{figure}

\begin{figure}
\centering
\includegraphics[width=\textwidth,trim={1cm 0cm 0cm 0cm},clip]{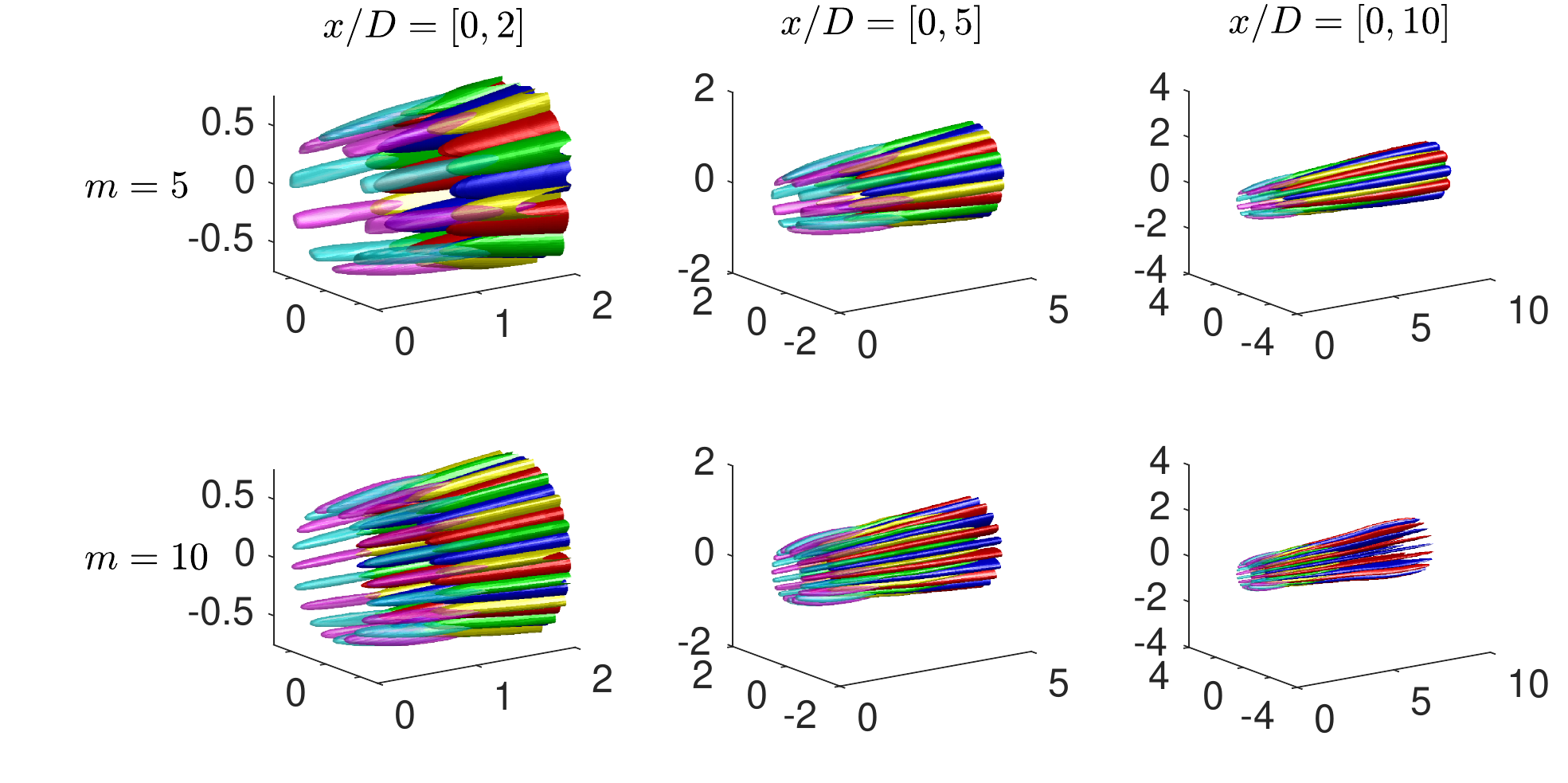}
\caption{Three-dimensional reconstruction of the first resolvent mode for $m = [5,10]$ (top and bottom, respectively) at three truncated domains $x/D \in [0,2], [0,5], [0,10]$ (left to right, respectively) for  $St = 0$. Streamwise velocity, $\bm{u_1}:u_x$, is denoted as red-blue with isosurfaces $\pm 25\%$ the maximum streamwise velocity, streamwise vorticity, $\bm{u_1}:\omega_x$, is shown as yellow-green with isosurfaces as $\pm 50\%$ of the maximum streamwise vorticity, and magenta-cyan represents isosurfaces of streamwise forcing vortices, $\bm{v_1}:\omega_x$, at $\pm 50\%$ of the maximum streamwise forcing vorticity.}
\label{fig:Domain_Resolvent_3D}
\end{figure}

The restricted SPOD and resolvent spectra and frequency-wavenumber maps are in striking agreement, and provide insight into both the total energy and the overall influence of each wavenumber as the domain is truncated towards the nozzle. The largest domain, $x/D \in [0,10]$, shows energy peaking at low frequencies and dominated by low azimuthal wavenumbers, similar to the global spectrum in figure \ref{fig:LinearGain}. We also see that the peak azimuthal wavenumber at $St \rightarrow 0$ has increased, yet the moderate-frequency behaviour associated with KH remains active for low azimuthal wavenumbers, similar to the full domain analysis. Truncating the domain to $x/D \in [0,5]$ presents a relative shift in total energy towards higher frequencies and the KH content maintains its strong $m=[0,1,2]$ behaviour at these frequencies, while the distribution of energy at low frequencies is now centred at $m=6$. The final truncation, $x/D \in [0,2]$, shows a significant drop in energy at low frequencies and higher frequencies dominate the energy spectrum, however, the peak maximum azimuthal wavenumber at low frequencies has shifted much higher, centred at $m=10$ near zero frequency. 

The figures also show a clear separation between the lift-up mechanism's influence in the low-frequency portion of the energy map versus the KH mechanism's influence in moderate-frequency regions. Interestingly, both of these observations, despite truncation of the jet domain, still fall in line with the estimated mechanism map, figure \ref{fig:Mechanism_Map}, with lift-up increasingly apparent as $St/m \rightarrow 0$.

To further show that the $St \rightarrow 0$ SPOD/resolvent energies/amplifications in this truncated domain analysis correspond to streaks, we show 3-D reconstructions of streamwise velocity and streamwise vorticity for $m=5,10$ for both SPOD and resolvent in figures \ref{fig:Domain_SPOD_3D} and \ref{fig:Domain_Resolvent_3D}, respectively. The truncated SPOD and resolvent modes show close alignment with each plot for $m=5$ and 10 presenting smooth streaks of streamwise velocity ($\bm{u_1}:u_x$), accompanied with streamwise rolls ($\bm{u_1}:\omega_x$) indicative of the lift-up mechanism, for all domains considered. Azimuthal wavenumber $m=5$ presents a case in which streaks are present with significant energy for each truncated domain, and as such, give rise to well-defined streaks that occupy the entire domain. For $m=10$, streamwise velocity streaks and streamwise vorticity rolls are easily identified in the truncated domains $x/D \in [0,2]$ and $ [0,5]$. However, the $x/D \in [0,10]$ domain gives a differing behaviour for both SPOD and resolvent modes. In the resolvent case, highly amplified streaks are observed until $x/D \approx 5$ where response rolls disappear and the streaks begin to decay, showing that steady streaks are no longer supported downstream at this azimuthal wavenumber. This is also seen in the SPOD results as the original streaks decay at $x/D \approx 5$ and a low-frequency, slowly rotating streak enters towards the end of the domain. This can be attributed to the various domain and convergence issues described earlier, where a larger domain introduces additional energetic structures and noise which may be aliased into the $St \rightarrow 0$ bin. Again, for the global results presented earlier in the paper, this was averted by increasing the number of snapshots in the discrete Fourier transform (DFT) and reducing the bin sizes. Nevertheless, these results show that although high azimuthal wavenumbers do not appear to provide significant energy to the full flow field energy (e.g. figure \ref{fig:LinearGain}), their energetic impact is significant in the near nozzle region and is a result of the lift-up mechanism.

\section{Conclusions \& Outlook} \label{sec:Conclusions}

We have extended the linear resolvent and data-driven SPOD analyses of turbulent jet mean flow fields to the zero-frequency limit.  The main result is a confirmation and extension of the local analysis of \cite{Nogueira2019Streaks}, namely the identification of the lift-up mechanism as an important linear amplifier of disturbances in turbulent jets. 

We found lift-up responsible for the generation of streamwise elongated structures, known as streaks, at low-frequency, non-zero azimuthal wavenumbers for turbulent round jets at Mach numbers 0.4, 0.9, and 1.5. At moderate frequencies, KH becomes the globally dominant mechanism, and the Orr mechanism is active over all frequencies but plays a subdominant role at those frequencies and azimuthal wavenumbers where lift-up and KH are active.  The behaviour of the $m=0$ response is unique, as axisymmetric streaks cannot exist; rather, the Orr response is dominant for low and high frequencies, with the KH response dominating over an intermediate-frequency regime centred on $St=0.6$.  For non-axisymmetric modes, the lift-up mechanism, and resulting streak response, is dominant as $St/m \rightarrow 0$, although with progressively higher wavenumber, lift-up responses are limited in spatial extent to nearer the nozzle exit.  We find that the azimuthal wavenumber of dominant streak responses is inversely proportional to shear layer width and scales with $\sim 1/x$.

For all regimes, there is a reasonable agreement between the resolvent analysis and SPOD modes from the associated LES database, confirming that the theoretical mechanisms are active (and dominant) in the turbulent regime. This agreement is predicated on the addition of an eddy-viscosity model in the linear resolvent \citep{morra2019relevance,pickering2019eddy}.  While the simple model we employed suffices to establish the link between theory and observation, further refinements to the model would be required to establish a resolvent analysis that is {\it predictive} of turbulence structure.

Streaks observed using SPOD show similar structure to (space-only) POD modes reported by \cite{freund2009turbulence} and SPOD modes with limited spatial extent by \cite{citriniti2000reconstruction} and \cite{jung2004downstream}, and predicted by resolvent analyses for $m>0$ at $St = 0$. Both SPOD and resolvent modes provide significant qualitative agreement in both streamwise vorticity and streamwise velocity, which are related to rolls and streaks, respectively. The resolvent results show optimal forcing in the form of streamwise vortices (rolls, $\bm{v_1}:\omega_x$) from the nozzle exit decaying slowly downstream, followed by a response of streamwise vortices  ($\bm{u_1}:\omega_x$), and, finally, further downstream there appears a response of streamwise velocity ($\bm{u_1}:u_x$), or streaks. These characterizations of the flow, from both resolvent and SPOD, now link multiple previous experimental and numerical observations in transitional and turbulent jets of streamwise vortices \citep{bradshaw1964turbulence,liepmann1991streamwise,martin1991numerical,paschereit1992flow,liepmann1992role,arnette1993streamwise} and streaks \citep{citriniti2000reconstruction,caraballo2003application,jung2004downstream,freund2009turbulence,cavalieri2013wavepackets} to the lift-up mechanism, forced upstream by a separate set of streamwise vortices.

Our results show that the lift-up mechanism, like the KH and Orr mechanisms, may be modelled as a direct amplification of disturbances to the turbulent jet mean flow. However, this work does not refute the idea, extrapolated from transitional free shear flows, that streaks arise via secondary instability of the KH rollers.  The resolvent forcing that generates lift-up could involve, through triadic interactions, braid vortices.  Indeed, the optimal forcing computed via the resolvent is non-zero in the region where the KH response is active. An explicit analysis of the triadic interactions involved in specific realizations of the turbulent flow would be needed to answer this question, a topic left for future research.  Nevertheless, even though a nonlinear analysis would provide deeper insight into turbulence, the simplicity and ability of the linear resolvent framework to capture observed mode shapes and qualitatively capture their relative amplitudes across the wavenumber-frequency domain is remarkable.

The presence of the lift-up mechanism in turbulent jets also suggests further investigation of the resulting dynamics, and its potential impact for control of quantities such as jet noise. Considering streaks are highly energetic structures it is likely that their behaviour (i.e. breakdown and regeneration) have significant impact on the mean flow and other structures in turbulent jets, such as Orr and KH wavepackets and qualities associated with these structures' intermittency. In fact, noise reduction has been accomplished via the introduction of streamwise vortices at the nozzle exit using tabs \citep{samimy1993effect,zaman1994control,zaman1999spreading}, chevrons \citep{bridges2003control,bridges2004parametric,saiyed2003acoustics,callender2005far,alkislar2007effect,violato2011three} and microjets \citep{arakeri2003use, greska2005effects, yang2016turbulent}. However, these changes have, in certain configurations, increased noise, leaving fundamental questions regarding the dynamic impact of imposed streamwise vortices and streaks. Recently, \cite{rigas2019streaks} showed the presence of both chevron induced streamwise vortices and their accompanying streaks in the base flow of a turbulent, chevron jet, while \cite{marant2018influence} have reported the ability of streaks, at particular wavenumbers, to stabilize the KH instability for a planar laminar shear flow. Considering past computational work of \cite{sinha2016linear} (using PSE) demonstrated the ability of chevrons to `stabilize' (i.e. weaken) KH wavepackets, the dominant mechanism in jet noise \citep{jordan2013wave}, it is likely that lift-up presents a fundamental mechanism of turbulent jet flow that may be tuned appropriately for optimal jet noise reduction techniques. 

\section*{Acknowledgments}
This research was supported by a grant from the Office of Naval Research (grant No. N00014-16-1-2445) with Dr. Steven Martens as program manager. E.P. was supported by the Department of Defense (DoD) through the National Defense Science \& Engineering Graduate Fellowship (NDSEG) Program. The LES study was performed at Cascade Technologies, with support from ONR and NAVAIR SBIR project, under the supervision of Dr. John T. Spyropoulos. The main LES calculations were carried out on DoD HPC systems in ERDC DSRC.

\section*{Declaration of interests}
The authors report no conflict of interest.

\appendix

\section{Transonic and supersonic jets} \label{sec:TransSuper}

\begin{figure}
\centering
\includegraphics[width=0.9\textwidth]{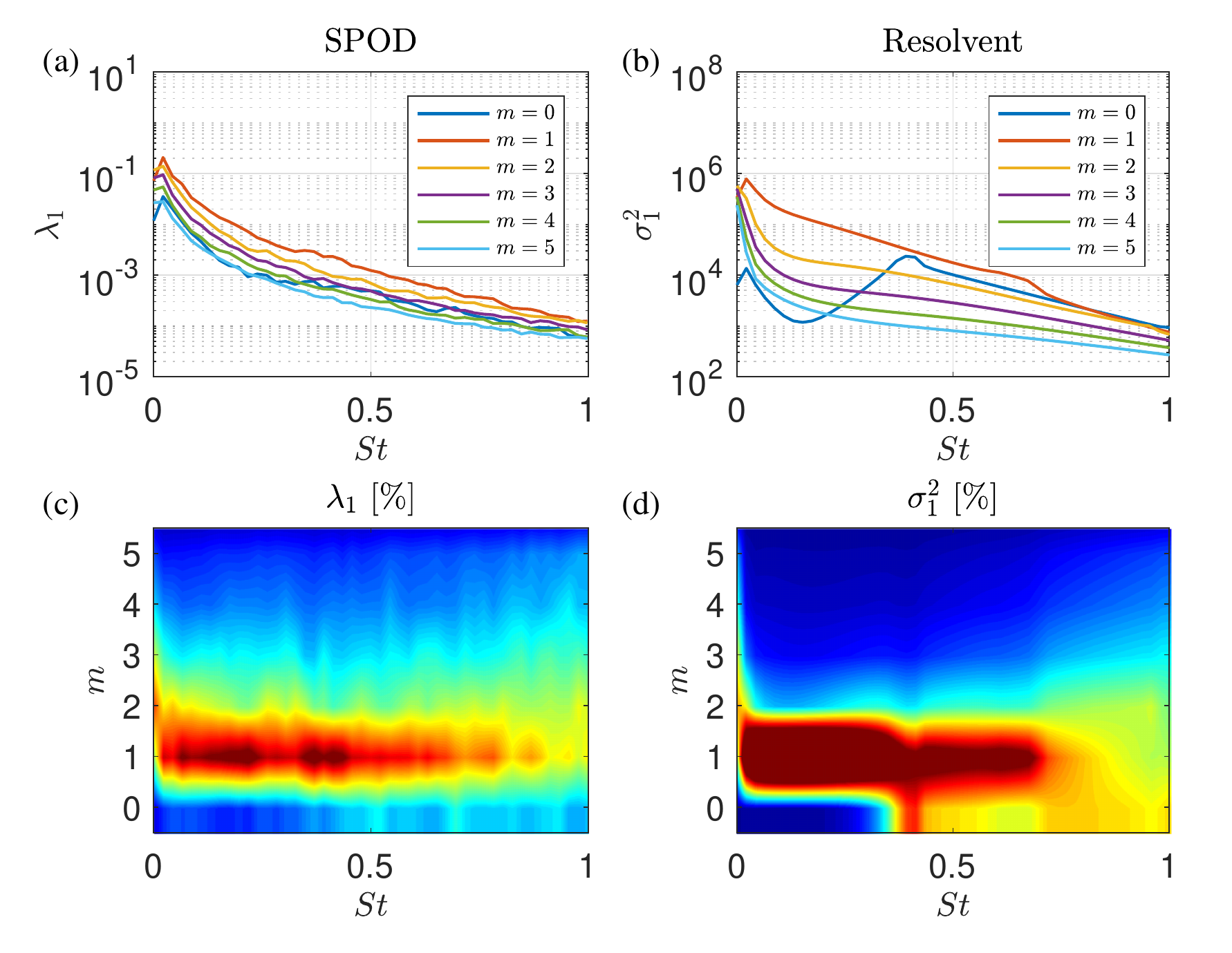}
\caption{Modal energy from SPOD and resolvent analyses of the $M_j = 0.9$ round jet.}

\label{fig:M09_Map}
\end{figure}
\begin{figure}
\centering
\includegraphics[width=0.9\textwidth]{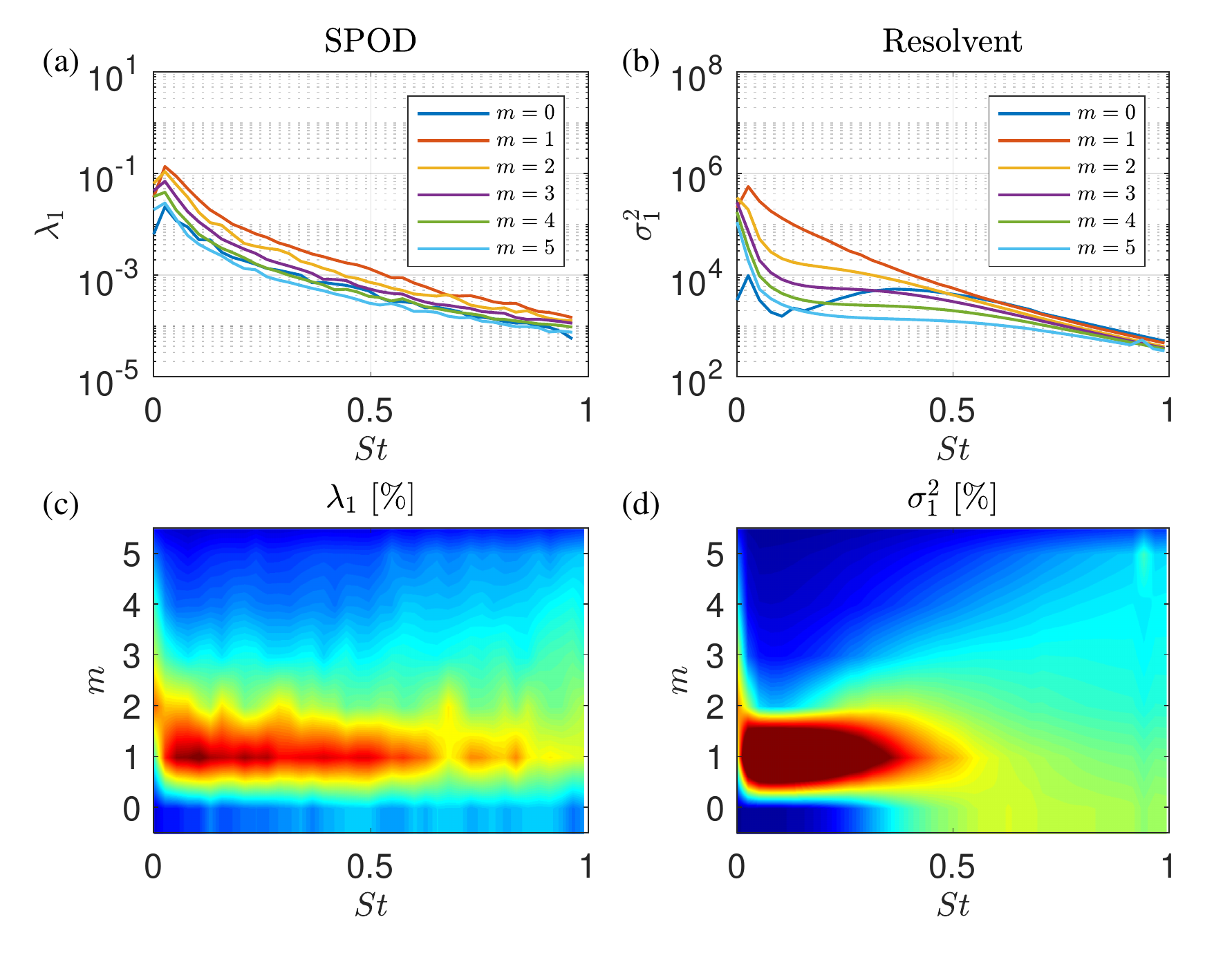}
\caption{Modal energy from SPOD and resolvent analyses of the $M_j = 1.5$ round jet.}
\label{fig:M15_Map}
\end{figure}

This section presents similar results supporting the presence of the lift-up mechanisms and streaks for $M_j = 0.9$ and $M_j = 1.5$ turbulent round jets. We first show the azimuthal frequency-wavenumber maps for each of the jets in figure \ref{fig:M09_Map} and figure \ref{fig:M15_Map}. For both figures, there is significant qualitative agreement between the SPOD energies and the resolvent gains. 

In the $M_j = 0.9$ jet we see a small spike at $St \sim 0.4$ for the resolvent analysis, due to trapped acoustic modes, that is slightly over predicted when compared to the SPOD energies. Another small discrepancy between SPOD and resolvent is the additional influence of the $m=0$ mode in the resolvent analysis when compared to SPOD. However, both show large energies for $m=1$ and at low frequencies, displaying behaviour similar to the $M_j = 0.4$ jet with energy peaking at $m=2$ as $St \rightarrow 0$.

For the $M_j = 1.5$ jet there is a similar over-prediction by resolvent analysis for $m=0$ when compared to SPOD. Despite this, the remainder of the resolvent map follows SPOD characteristics quite well. For higher frequencies, energy is distributed across multiple azimuthal wavenumbers and at low frequencies both analyses display high-energy behaviour similar to the $M_j = 0.4$ jet with energy peaking at $m=2$ as $St \rightarrow 0$. 

\begin{figure}
\centering
{\small $m=1$ \hspace{25 mm} (a) $M_j = 0.9$ \hspace{25 mm} $m=3$ } \\
\includegraphics[width=1\textwidth,trim={1.75cm 0.25cm 1cm 0.95cm},clip]{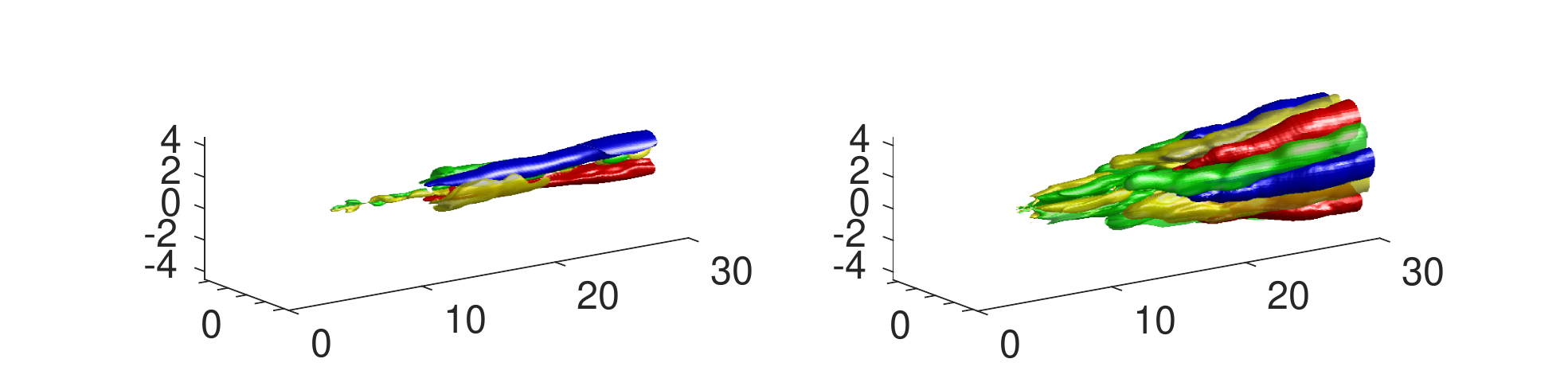}
\\ {\small $m=1$ \hspace{25 mm} (b) $M_j = 1.5$ \hspace{25 mm} $m=3$ } \\
\includegraphics[width=1\textwidth,trim={1.75cm 0.25cm 1cm 0.95cm},clip]{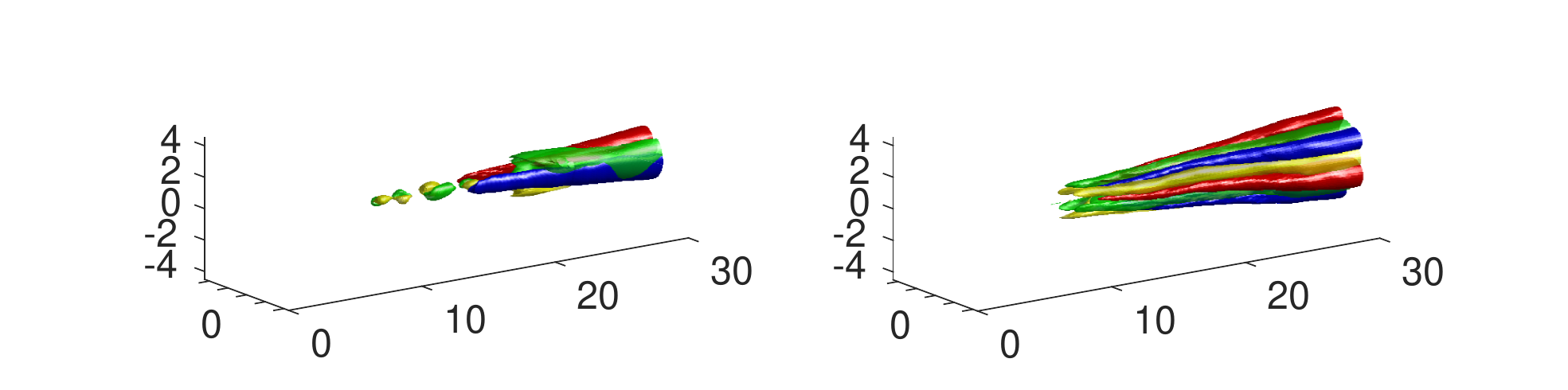}
\caption{Three-dimensional reconstruction of the first SPOD mode (streamwise velocity, $\bm{\psi_1}:u_x$, red-blue, streamwise vorticity, $\bm{\psi_1}:\omega_x$, yellow-green) as $St \rightarrow 0$ for $m=1$ (left column) and $m=3$ (right column) using isosurfaces of $\pm 50\%$ of the maximum streamwise velocity and isosurfaces of $\pm 25\%$ of the maximum streamwise vorticity, with the exception of the $M_j = 1.5$, $m=1$, case where red-blue isosurfaces are instead $\pm 30\%$ of the maximum streamwise velocity.}
\label{fig:M09_M15_SPOD}
\end{figure}

\begin{figure}
\centering
{\small $m=1$ \hspace{25 mm} (a) $M_j = 0.9$ \hspace{25 mm} $m=3$ } \\
\includegraphics[width=1\textwidth,trim={1.75cm 0.25cm 1cm 0.95cm},clip]{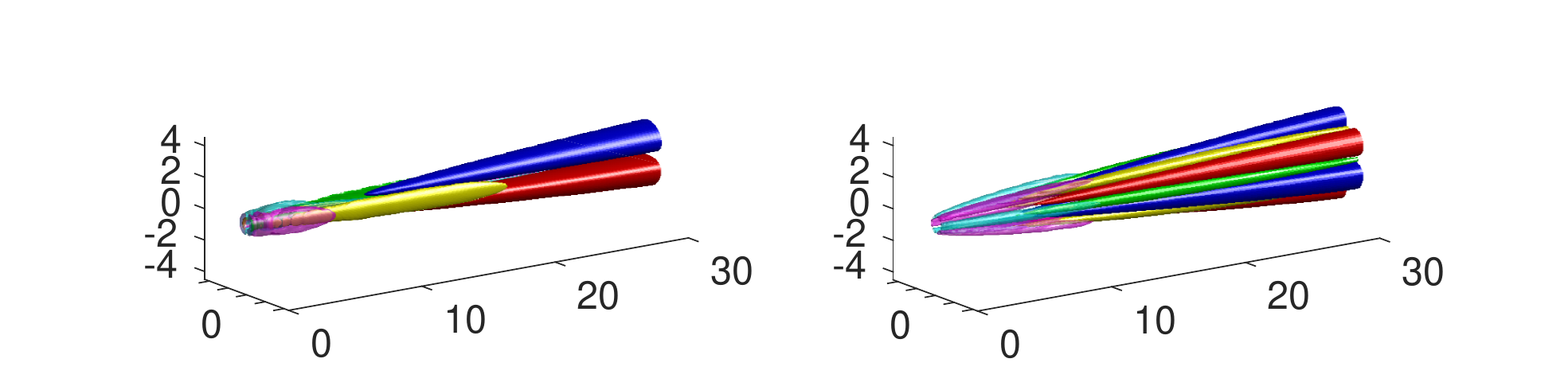}
\\ {\small $m=1$ \hspace{25 mm} (b) $M_j = 1.5$ \hspace{25 mm} $m=3$ } \\
\includegraphics[width=1\textwidth,trim={1.75cm 0.25cm 1cm 0.95cm},clip]{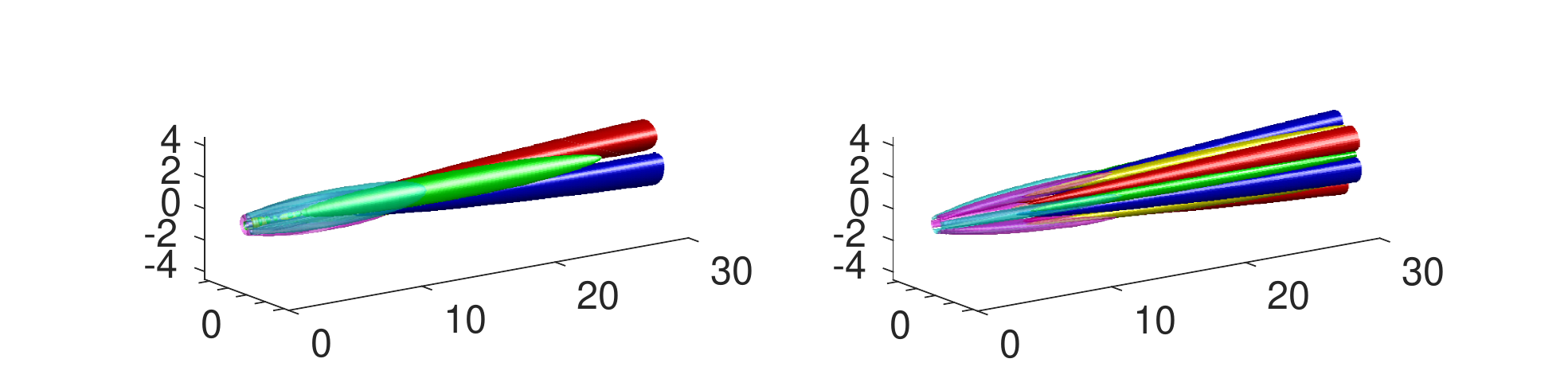}
\caption{Global resolvent forcing and response for $m=1$ (left) and $m=3$, (right) at $St=0$. The streamwise forcing vorticity is shown in magenta-cyan with isosurfaces $\pm 0.05 \|\bm{v_1}:\omega_x\|_{\infty}$ for $m=1$ and $\pm 0.2 \|\bm{v_1}:\omega_x\|_{\infty}$ for $m=3$, streamwise response vorticity is shown in yellow-green with isosurfaces $\pm 0.5 \|\bm{u_1}:\omega_x\|_{\infty}$, and streamwise response velocity is shown in red-blue with isosurfaces $\pm 0.25 \|\bm{u_1}:u_x\|_{\infty}$.}
\label{fig:M09_M15_Resolvent}
\end{figure}

We also show the 3-D reconstructions of both the $M_j = 0.9$ and $ 1.5$ jet SPOD modes at $St=0$ and $m=1,3$ shown in figure \ref{fig:M09_M15_SPOD} using 1024 snapshots and 75\% overlap. All four plots show streaks of streamwise velocity response along with the associated streamwise vorticity, with the clearest descriptions shown for the $m=3$ cases. The $m=3$ case for both jets presents streaky structures paired with streamwise vorticity rolls placed perfectly between each streak. The $m=1$ plots are not quite as appealing, but both plots present streaks in streamwise velocity and are also paired with streamwise vorticity placed between each streak.

Finally, we show the resolvent analysis at $St=0$ for both jets and wavenumbers $m=1,3$ in figure \ref{fig:M09_M15_Resolvent}. Here, we show the same behaviour as previously described for the $M_j = 0.4$ jet. Forcing in the form of streamwise vorticity begins upstream near the nozzle, followed by responses of streamwise rolls, that then lead to lift-up and large streamwise velocity responses. Considering all of the presented evidence for $M_j = 0.9$ and 1.5 jets, we conclude that the lift-up mechanism is present in all turbulent jets.

\section{SPOD and resolvent semi-discrete energy maps for Mach 0.4}\label{sec:SemiDiscrete}

\begin{figure}
\centering
\vspace{0.5cm}
\includegraphics[width=1\textwidth]{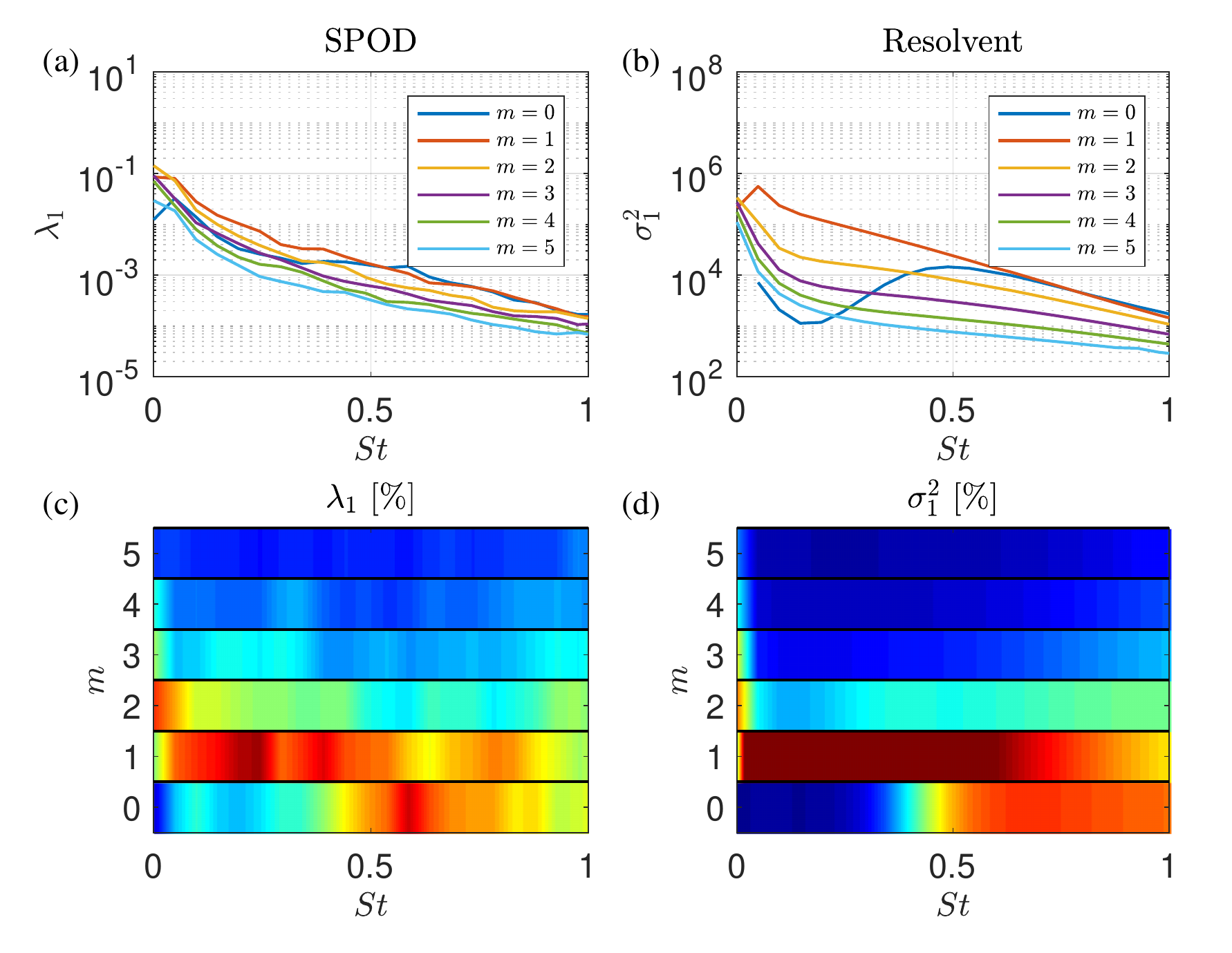}
\caption{Modal energy from SPOD and resolvent analyses of a Mach 0.4 round jet, shown here in semi-discrete form.}
\label{fig:M04_Map_Discrete}
\end{figure}

Figure \ref{fig:M04_Map_Discrete} gives the semi-discrete, continuous in frequency and discrete in azimuthal wavenumber, representation of figure \ref{fig:LinearGain}. Through both representations of SPOD and resolvent analyses we can make similar conclusions as the interpolated results of figure \ref{fig:LinearGain}. We can clearly see the shift from the KH-dominated space at $m=0$ as frequency is decreased to the Orr dominated region. For non-zero azimuthal wavenumbers we see significant energy for frequencies approaching 0 and find $m=2$ has the largest energetic contribution for $St \approx 0$.

\bibliographystyle{jfm}
\bibliography{jfmbib}

\begin{thebibliography}{115}
\expandafter\ifx\csname natexlab\endcsname\relax\def\natexlab#1{#1}\fi
\def\au#1{#1} \def\ed#1{#1} \def\yr#1{#1}\def\at#1{#1}\def\jt#1{\textit{#1}}
  \def\bt#1{#1}\def\bvol#1{\textbf{#1}} \def\vol#1{#1} \def\pg#1{#1}
  \def\publ#1{#1}\def\arxiv#1{#1}\def\org#1{#1}\def\st#1{\textit{#1}}

\bibitem[Abreu {\em et~al.\/}(2019)Abreu, Cavalieri, Schlatter, Vinuesa \&
  Henningson]{abreureduced2019}
{\sc \au{Abreu, L.~I.}, \au{Cavalieri, A.~V.~G.}, \au{Schlatter, P.},
  \au{Vinuesa, R.} \& \au{Henningson, D.}} \yr{2019} Reduced-order models to
  analyse coherent structures in turbulent pipe flow.  \bt{In {\em 11th
  International Symposium on Turbulence and Shear Flow Phenomena\/}}.
  \publ{University of Southampton}.

\bibitem[Ag{\"u}{\'\i} \& Hesselink(1988)]{agui1988flow}
{\sc \au{Ag{\"u}{\'\i}, J.~C.} \& \au{Hesselink, L.}} \yr{1988}  \at{Flow
  visualization and numerical analysis of a coflowing jet: a three-dimensional
  approach}.  \jt{J. Fluid Mech.}  \bvol{191},  \pg{19--45}.

\bibitem[{\AA}kervik {\em et~al.\/}(2008){\AA}kervik, Ehrenstein, Gallaire \&
  Henningson]{aakervik2008global}
{\sc \au{{\AA}kervik, E.}, \au{Ehrenstein, U.}, \au{Gallaire, F.} \&
  \au{Henningson, D.~S.}} \yr{2008}  \at{Global two-dimensional stability
  measures of the flat plate boundary-layer flow}.  \jt{European Journal of
  Mechanics-B/Fluids}  \bvol{27}~(5),  \pg{501--513}.

\bibitem[Alkislar {\em et~al.\/}(2007)Alkislar, Krothapalli \&
  Butler]{alkislar2007effect}
{\sc \au{Alkislar, M.~B.}, \au{Krothapalli, A.} \& \au{Butler, G.~W.}}
  \yr{2007}  \at{The effect of streamwise vortices on the aeroacoustics of a
  mach 0.9 jet}.  \jt{J. Fluid Mech.}  \bvol{578},  \pg{139--169}.

\bibitem[Arakeri {\em et~al.\/}(2003)Arakeri, Krothapalli, Siddavaram, Alkislar
  \& Lourenco]{arakeri2003use}
{\sc \au{Arakeri, V.~H.}, \au{Krothapalli, A.}, \au{Siddavaram, V.},
  \au{Alkislar, M.~B.} \& \au{Lourenco, L.~M.}} \yr{2003}  \at{On the use of
  microjets to suppress turbulence in a mach 0.9 axisymmetric jet}.  \jt{J.
  Fluid Mech.}  \bvol{490},  \pg{75--98}.

\bibitem[Arnette {\em et~al.\/}(1993)Arnette, Samimy \&
  Elliott]{arnette1993streamwise}
{\sc \au{Arnette, S.~A.}, \au{Samimy, M.} \& \au{Elliott, G.~S.}} \yr{1993}
  \at{On streamwise vortices in high {R}eynolds number supersonic axisymmetric
  jets}.  \jt{Physics of Fluids A: Fluid Dynamics}  \bvol{5}~(1),
  \pg{187--202}.

\bibitem[Arratia {\em et~al.\/}(2013)Arratia, Caulfield \&
  Chomaz]{arratia2013transient}
{\sc \au{Arratia, C.}, \au{Caulfield, C.~P.} \& \au{Chomaz, J.~M.}} \yr{2013}
  \at{Transient perturbation growth in time-dependent mixing layers}.  \jt{J.
  Fluid Mech.}  \bvol{717},  \pg{90--133}.

\bibitem[Becker \& Massaro(1968)]{becker1968vortex}
{\sc \au{Becker, H.~A.} \& \au{Massaro, T.~A.}} \yr{1968}  \at{Vortex evolution
  in a round jet}.  \jt{J. Fluid Mech.}  \bvol{31}~(3),  \pg{435--448}.

\bibitem[Benney(1961)]{benney1961non}
{\sc \au{Benney, D.~J.}} \yr{1961}  \at{A non-linear theory for oscillations in
  a parallel flow}.  \jt{J. Fluid Mech.}  \bvol{10}~(2),  \pg{209--236}.

\bibitem[Benney \& Lin(1960)]{Benney1960}
{\sc \au{Benney, D.~J.} \& \au{Lin, C.~C.}} \yr{1960}  \at{On the secondary
  motion induced by oscillations in a shear flow}.  \jt{The Physics of Fluids}
  \bvol{3}~(4),  \pg{656--657}.

\bibitem[Bernal \& Roshko(1986)]{bernal1986streamwise}
{\sc \au{Bernal, L.P.} \& \au{Roshko, A.}} \yr{1986}  \at{Streamwise vortex
  structure in plane mixing layers}.  \jt{J. Fluid Mech.}  \bvol{170},
  \pg{499--525}.

\bibitem[Bernal(1981)]{bernal1981coherent}
{\sc \au{Bernal, L.~P.}} \yr{1981}  \at{The coherent structure of turbulent
  mixing layers}. PhD thesis, Ph.D. Thesis, California Institute of Technology.

\bibitem[Bernal {\em et~al.\/}(1979)Bernal, Breidenthal, Brown, Konrad \&
  Roshko]{bernal1979development}
{\sc \au{Bernal, L.~P.}, \au{Breidenthal, R.~E.}, \au{Brown, G.~L.},
  \au{Konrad, J.~H.} \& \au{Roshko, A.}} \yr{1979} On the development of three
  dimensional small scales in turbulent mixing layers.  \bt{In {\em Proc. 2nd
  Int. Symp. on Turbulent Shear Flows, Imperial College, London\/}},  \pg{pp.
  8.1--8.6}.

\bibitem[Boronin {\em et~al.\/}(2013)Boronin, Healey \& Sazhin]{boronin2013non}
{\sc \au{Boronin, S.~A.}, \au{Healey, J.~J.} \& \au{Sazhin, S.~S.}} \yr{2013}
  \at{Non-modal stability of round viscous jets}.  \jt{J. Fluid Mech.}
  \bvol{716},  \pg{96--119}.

\bibitem[Bradshaw {\em et~al.\/}(1964)Bradshaw, Ferriss \&
  Johnson]{bradshaw1964turbulence}
{\sc \au{Bradshaw, P.}, \au{Ferriss, D.~H.} \& \au{Johnson, R.~F.}} \yr{1964}
  \at{Turbulence in the noise-producing region of a circular jet}.  \jt{J.
  Fluid Mech.}  \bvol{19}~(4),  \pg{591--624}.

\bibitem[Brandt(2014)]{brandt2014lift}
{\sc \au{Brandt, L.}} \yr{2014}  \at{The lift-up effect: the linear mechanism
  behind transition and turbulence in shear flows}.  \jt{European Journal of
  Mechanics-B/Fluids}  \bvol{47},  \pg{80--96}.

\bibitem[Breidenthal(1981)]{breidenthal1981structure}
{\sc \au{Breidenthal, R.}} \yr{1981}  \at{Structure in turbulent mixing layers
  and wakes using a chemical reaction}.  \jt{J. Fluid Mech.}  \bvol{109},
  \pg{1--24}.

\bibitem[Breidenthal(1978)]{breidenthal1978chemically}
{\sc \au{Breidenthal, R.~E.}} \yr{1978}  \at{A chemically reacting shear
  layer}. PhD thesis, Ph.D. Thesis, California Institute of Technology.

\bibitem[Br{\`e}s {\em et~al.\/}(2017)Br{\`e}s, Ham, Nichols \&
  Lele]{bres2017unstructured}
{\sc \au{Br{\`e}s, G.~A.}, \au{Ham, F.~E.}, \au{Nichols, J.~W.} \& \au{Lele,
  S.~K.}} \yr{2017}  \at{Unstructured large-eddy simulations of supersonic
  jets}.  \jt{AIAA Journal}  \bvol{55}~(4),  \pg{1164--1184}.

\bibitem[Br{\`e}s {\em et~al.\/}(2018)Br{\`e}s, Jordan, Jaunet, Le~Rallic,
  Cavalieri, Towne, Lele, Colonius \& Schmidt]{bres2018importance}
{\sc \au{Br{\`e}s, G.~A.}, \au{Jordan, P.}, \au{Jaunet, V.}, \au{Le~Rallic,
  M.}, \au{Cavalieri, A.~V.~G.}, \au{Towne, A.}, \au{Lele, S.~K.},
  \au{Colonius, T.} \& \au{Schmidt, O.~T.}} \yr{2018}  \at{Importance of the
  nozzle-exit boundary-layer state in subsonic turbulent jets}.  \jt{J. Fluid
  Mech.}  \bvol{851},  \pg{83--124}.

\bibitem[Bridges \& Brown(2004)]{bridges2004parametric}
{\sc \au{Bridges, J.} \& \au{Brown, C.}} \yr{2004} Parametric testing of
  chevrons on single flow hot jets.  \bt{In {\em 10th AIAA/CEAS Aeroacoustics
  Conference\/}},  \pg{p. 2824}.

\bibitem[Bridges {\em et~al.\/}(2003)Bridges, Wernet \&
  Brown]{bridges2003control}
{\sc \au{Bridges, J.}, \au{Wernet, M.} \& \au{Brown, C.}} \yr{2003}
  \at{Control of jet noise through mixing enhancement}.  \jt{NASA Rep. No.
  NASA/TM 2003-212335} .

\bibitem[Browand \& Laufer(1975)]{browand1975roles}
{\sc \au{Browand, F.~K.} \& \au{Laufer, J.}} \yr{1975} The roles of large scale
  structures in the initial development of circular jets.  \bt{In {\em Symposia
  on Turbulence in Liquids\/}},  \pg{p.~35}.  \publ{University of
  Missouri--Rolla}.

\bibitem[Brown \& Roshko(1974)]{brown1974density}
{\sc \au{Brown, G.~L.} \& \au{Roshko, A.}} \yr{1974}  \at{On density effects
  and large structure in turbulent mixing layers}.  \jt{J. Fluid Mech.}
  \bvol{64}~(4),  \pg{775--816}.

\bibitem[Butler \& Farrell(1992)]{butler1992three}
{\sc \au{Butler, K.~M.} \& \au{Farrell, B.~F.}} \yr{1992}
  \at{Three-dimensional optimal perturbations in viscous shear flow}.
  \jt{Physics of Fluids A: Fluid Dynamics}  \bvol{4}~(8),  \pg{1637--1650}.

\bibitem[Callender {\em et~al.\/}(2005)Callender, Gutmark \&
  Martens]{callender2005far}
{\sc \au{Callender, B.}, \au{Gutmark, E.~J.} \& \au{Martens, S.}} \yr{2005}
  \at{Far-field acoustic investigation into chevron nozzle mechanisms and
  trends}.  \jt{AIAA journal}  \bvol{43}~(1),  \pg{87--95}.

\bibitem[Caraballo {\em et~al.\/}(2003)Caraballo, Samimy, Scott, Narayanan \&
  DeBonis]{caraballo2003application}
{\sc \au{Caraballo, E.}, \au{Samimy, M.}, \au{Scott, J.}, \au{Narayanan, S.} \&
  \au{DeBonis, J.}} \yr{2003}  \at{Application of proper orthogonal
  decomposition to a supersonic axisymmetric jet}.  \jt{AIAA journal}
  \bvol{41}~(5),  \pg{866--877}.

\bibitem[Cavalieri {\em et~al.\/}(2013)Cavalieri, Rodr{\'\i}guez, Jordan,
  Colonius \& Gervais]{cavalieri2013wavepackets}
{\sc \au{Cavalieri, A. V.~G.}, \au{Rodr{\'\i}guez, D.}, \au{Jordan, P.},
  \au{Colonius, T.} \& \au{Gervais, Y.}} \yr{2013}  \at{Wavepackets in the
  velocity field of turbulent jets}.  \jt{J. Fluid Mech.}  \bvol{730},
  \pg{559--592}.

\bibitem[Chantry {\em et~al.\/}(2016)Chantry, Tuckerman \&
  Barkley]{chantry2016turbulent}
{\sc \au{Chantry, M.}, \au{Tuckerman, L.~S.} \& \au{Barkley, D.}} \yr{2016}
  \at{Turbulent--laminar patterns in shear flows without walls}.  \jt{J. Fluid
  Mech.}  \bvol{791},  \pg{R8}.

\bibitem[Chu(1965)]{chu1965energy}
{\sc \au{Chu, B.-T.}} \yr{1965}  \at{On the energy transfer to small
  disturbances in fluid flow ({P}art {I})}.  \jt{Acta Mechanica}  \bvol{1}~(3),
   \pg{215--234}.

\bibitem[Citriniti \& George(2000)]{citriniti2000reconstruction}
{\sc \au{Citriniti, J.~H.} \& \au{George, W.~K.}} \yr{2000}  \at{Reconstruction
  of the global velocity field in the axisymmetric mixing layer utilizing the
  proper orthogonal decomposition}.  \jt{J. Fluid Mech.}  \bvol{418},
  \pg{137--166}.

\bibitem[Crighton \& Gaster(1976)]{crighton1976stability}
{\sc \au{Crighton, D.~G.} \& \au{Gaster, M.}} \yr{1976}  \at{Stability of
  slowly diverging jet flow}.  \jt{J. Fluid Mech.}  \bvol{77}~(2),
  \pg{397--413}.

\bibitem[Crow \& Champagne(1971)]{crow1971orderly}
{\sc \au{Crow, S.~C.} \& \au{Champagne, F.~H.}} \yr{1971}  \at{Orderly
  structure in jet turbulence}.  \jt{J. Fluid Mech.}  \bvol{48}~(3),
  \pg{547--591}.

\bibitem[Davoust {\em et~al.\/}(2012)Davoust, Jacquin \&
  Leclaire]{davoust2012Dynamics}
{\sc \au{Davoust, S.}, \au{Jacquin, L.} \& \au{Leclaire, B.}} \yr{2012}
  \at{Dynamics of m= 0 and m= 1 modes and of streamwise vortices in a turbulent
  axisymmetric mixing layer}.  \jt{J. Fluid Mech.}  \bvol{709},  \pg{408--444}.

\bibitem[Dergham {\em et~al.\/}(2013)Dergham, Sipp \&
  Robinet]{dergham2013stochastic}
{\sc \au{Dergham, G.}, \au{Sipp, D.} \& \au{Robinet, J.~C.}} \yr{2013}
  \at{Stochastic dynamics and model reduction of amplifier flows: the backward
  facing step flow}.  \jt{J. Fluid Mech.}  \bvol{719},  \pg{406--430}.

\bibitem[Dimotakis {\em et~al.\/}(1983)Dimotakis, Miake-Lye \&
  Papantoniou]{dimotakis1983structure}
{\sc \au{Dimotakis, P.~E.}, \au{Miake-Lye, R.~C.} \& \au{Papantoniou, D.~A.}}
  \yr{1983}  \at{Structure and dynamics of round turbulent jets}.  \jt{The
  Physics of fluids}  \bvol{26}~(11),  \pg{3185--3192}.

\bibitem[Eitel-Amor {\em et~al.\/}(2014)Eitel-Amor, {\"O}rl{\"u} \&
  Schlatter]{eitel2014simulation}
{\sc \au{Eitel-Amor, G.}, \au{{\"O}rl{\"u}, R.} \& \au{Schlatter, P.}}
  \yr{2014}  \at{{Simulation and validation of a spatially evolving turbulent
  boundary layer up to Re$\theta$= 8300}}.  \jt{International Journal of Heat
  and Fluid Flow}  \bvol{47},  \pg{57--69}.

\bibitem[Ellingsen \& Palm(1975)]{ellingsen1975stability}
{\sc \au{Ellingsen, T.} \& \au{Palm, E.}} \yr{1975}  \at{Stability of linear
  flow}.  \jt{Phys. Fluids}  \bvol{18}~(4),  \pg{487--488}.

\bibitem[Farrell(1988)]{farrell1988optimal}
{\sc \au{Farrell, B.~F.}} \yr{1988}  \at{Optimal excitation of perturbations in
  viscous shear flow}.  \jt{The Physics of fluids}  \bvol{31}~(8),
  \pg{2093--2102}.

\bibitem[Farrell \& Ioannou(1993)]{farrell1993optimal}
{\sc \au{Farrell, B.~F.} \& \au{Ioannou, P.~J.}} \yr{1993}  \at{Optimal
  excitation of three-dimensional perturbations in viscous constant shear
  flow}.  \jt{Physics of Fluids A: Fluid Dynamics}  \bvol{5}~(6),
  \pg{1390--1400}.

\bibitem[Freund \& Colonius(2009)]{freund2009turbulence}
{\sc \au{Freund, J.~B.} \& \au{Colonius, T.}} \yr{2009}  \at{Turbulence and
  sound-field pod analysis of a turbulent jet}.  \jt{Int. J. Aeroacoust.}
  \bvol{8}~(4),  \pg{337--354}.

\bibitem[Garnaud {\em et~al.\/}(2013{\natexlab{{\em a\/}}})Garnaud, Lesshafft,
  Schmid \& Huerre]{garnaud2013modal}
{\sc \au{Garnaud, X.}, \au{Lesshafft, L.}, \au{Schmid, P.~J.} \& \au{Huerre,
  P.}} \yr{2013{\natexlab{{\em a\/}}}}  \at{Modal and transient dynamics of jet
  flows}.  \jt{Physics of Fluids}  \bvol{25}~(4),  \pg{044103}.

\bibitem[Garnaud {\em et~al.\/}(2013{\natexlab{{\em b\/}}})Garnaud, Lesshafft,
  Schmid \& Huerre]{garnaud2013preferred}
{\sc \au{Garnaud, X.}, \au{Lesshafft, L.}, \au{Schmid, P.~J.} \& \au{Huerre,
  P.}} \yr{2013{\natexlab{{\em b\/}}}}  \at{The preferred mode of
  incompressible jets: linear frequency response analysis}.  \jt{J. Fluid
  Mech.}  \bvol{716},  \pg{189--202}.

\bibitem[Greska {\em et~al.\/}(2005)Greska, Krothapalli, Seiner, Jansen \&
  Ukeiley]{greska2005effects}
{\sc \au{Greska, B.}, \au{Krothapalli, A.}, \au{Seiner, J.}, \au{Jansen, B.} \&
  \au{Ukeiley, L.}} \yr{2005} The effects of microjet injection on an {F}404
  jet engine.  \bt{In {\em 11th AIAA/CEAS Aeroacoustics Conference\/}},  \pg{p.
  3047}.

\bibitem[Gudmundsson \& Colonius(2011)]{gudmundsson2011instability}
{\sc \au{Gudmundsson, K.} \& \au{Colonius, T.}} \yr{2011}  \at{Instability wave
  models for the near-field fluctuations of turbulent jets}.  \jt{J. Fluid
  Mech.}  \bvol{689},  \pg{97--128}.

\bibitem[Hack \& Moin(2017)]{hack2017algebraic}
{\sc \au{Hack, M.~J.~P.} \& \au{Moin, P.}} \yr{2017}  \at{Algebraic disturbance
  growth by interaction of {O}rr and lift-up mechanisms}.  \jt{J. Fluid Mech.}
  \bvol{829},  \pg{112--126}.

\bibitem[Hellstr{\"o}m {\em et~al.\/}(2011)Hellstr{\"o}m, Sinha \&
  Smits]{hellstrom2011visualizing}
{\sc \au{Hellstr{\"o}m, L.~H.~O.}, \au{Sinha, A.} \& \au{Smits, A.~J.}}
  \yr{2011}  \at{Visualizing the very-large-scale motions in turbulent pipe
  flow}.  \jt{Phys. Fluids}  \bvol{23}~(1),  \pg{011703}.

\bibitem[Ho \& Huerre(1984)]{ho1984perturbed}
{\sc \au{Ho, C.-M.} \& \au{Huerre, P.}} \yr{1984}  \at{Perturbed free shear
  layers}.  \jt{Annual review of fluid mechanics}  \bvol{16}~(1),
  \pg{365--422}.

\bibitem[Hutchins \& Marusic(2007)]{hutchins2007evidence}
{\sc \au{Hutchins, N.} \& \au{Marusic, I.}} \yr{2007}  \at{Evidence of very
  long meandering features in the logarithmic region of turbulent boundary
  layers}.  \jt{J. Fluid Mech.}  \bvol{579},  \pg{1--28}.

\bibitem[Hwang \& Cossu(2010)]{hwang2010amplification}
{\sc \au{Hwang, Y.} \& \au{Cossu, C.}} \yr{2010}  \at{Amplification of coherent
  streaks in the turbulent couette flow: an input--output analysis at low
  {R}eynolds number}.  \jt{J. Fluid Mech.}  \bvol{643},  \pg{333--348}.

\bibitem[Jaunet {\em et~al.\/}(2017)Jaunet, Jordan \& Cavalieri]{jaunet2017two}
{\sc \au{Jaunet, V.}, \au{Jordan, P.} \& \au{Cavalieri, A.~V.~G.}} \yr{2017}
  \at{Two-point coherence of wave packets in turbulent jets}.  \jt{Physical
  Review Fluids}  \bvol{2}~(2),  \pg{024604}.

\bibitem[Jeun {\em et~al.\/}(2016)Jeun, Nichols \&
  Jovanovi{\'c}]{jeun2016input}
{\sc \au{Jeun, J.}, \au{Nichols, J.~W.} \& \au{Jovanovi{\'c}, M.~R.}} \yr{2016}
   \at{Input-output analysis of high-speed axisymmetric isothermal jet noise}.
  \jt{Phys. Fluids}  \bvol{28}~(4),  \pg{047101}.

\bibitem[Jim{\'e}nez(2013)]{jimenez2013linear}
{\sc \au{Jim{\'e}nez, J.}} \yr{2013}  \at{How linear is wall-bounded
  turbulence?}  \jt{Physics of Fluids}  \bvol{25}~(11),  \pg{110814}.

\bibitem[Jim{\'e}nez(2018)]{jimenez2018coherent}
{\sc \au{Jim{\'e}nez, J.}} \yr{2018}  \at{Coherent structures in wall-bounded
  turbulence}.  \jt{J. Fluid Mech.}  \bvol{842},  \pg{P1}.

\bibitem[Jimenez {\em et~al.\/}(1985)Jimenez, Cogollos \&
  Bernal]{jimenez1985perspective}
{\sc \au{Jimenez, J.}, \au{Cogollos, M.} \& \au{Bernal, L.~P.}} \yr{1985}
  \at{A perspective view of the plane mixing layer}.  \jt{J. Fluid Mech.}
  \bvol{152},  \pg{125--143}.

\bibitem[Jim{\'e}nez \& Pinelli(1999)]{jimenez1999autonomous}
{\sc \au{Jim{\'e}nez, J.} \& \au{Pinelli, A.}} \yr{1999}  \at{The autonomous
  cycle of near-wall turbulence}.  \jt{J. Fluid Mech.}  \bvol{389},
  \pg{335--359}.

\bibitem[Jimenez-Gonzalez \& Brancher(2017)]{jimenez2017transient}
{\sc \au{Jimenez-Gonzalez, J.~I.} \& \au{Brancher, P.}} \yr{2017}
  \at{Transient energy growth of optimal streaks in parallel round jets}.
  \jt{Physics of Fluids}  \bvol{29}~(11),  \pg{114101}.

\bibitem[Jordan \& Colonius(2013)]{jordan2013wave}
{\sc \au{Jordan, P.} \& \au{Colonius, T.}} \yr{2013}  \at{Wave packets and
  turbulent jet noise}.  \jt{Annu. Rev. Fluid Mech.}  \bvol{45},
  \pg{173--195}.

\bibitem[Jung {\em et~al.\/}(2004)Jung, Gamard \& George]{jung2004downstream}
{\sc \au{Jung, D.}, \au{Gamard, S.} \& \au{George, W.~K.}} \yr{2004}
  \at{Downstream evolution of the most energetic modes in a turbulent
  axisymmetric jet at high {R}eynolds number. {P}art 1. the near-field region}.
   \jt{J. Fluid Mech.}  \bvol{514},  \pg{173--204}.

\bibitem[Kantharaju {\em et~al.\/}(2020)Kantharaju, Courtier, Leclaire \&
  Jacquin]{kantharaju2020interactions}
{\sc \au{Kantharaju, J.}, \au{Courtier, R.}, \au{Leclaire, B.} \& \au{Jacquin,
  L.}} \yr{2020}  \at{Interactions of large-scale structures in the near field
  of round jets at high {R}eynolds numbers}.  \jt{J. Fluid Mech.}  \bvol{888},
  \pg{A8}.

\bibitem[Kim {\em et~al.\/}(1971)Kim, Kline \& {R}eynolds]{kim1971production}
{\sc \au{Kim, H.~T.}, \au{Kline, S.~J.} \& \au{{R}eynolds, W.~C.}} \yr{1971}
  \at{The production of turbulence near a smooth wall in a turbulent boundary
  layer}.  \jt{J. Fluid Mech.}  \bvol{50}~(1),  \pg{133--160}.

\bibitem[Klebanoff(1971)]{klebanoff1971effect}
{\sc \au{Klebanoff, P.~S.}} \yr{1971}  \at{Effect of freestream turbulence on
  the laminar boundary layer}.  \jt{Bulletin of the American Physical Society}
  \bvol{16},  \pg{1321}.

\bibitem[Kline {\em et~al.\/}(1967)Kline, {R}eynolds, Schraub \&
  Runstadler]{kline1967structure}
{\sc \au{Kline, S.~J.}, \au{{R}eynolds, W.~C.}, \au{Schraub, F.~A.} \&
  \au{Runstadler, P.~W.}} \yr{1967}  \at{The structure of turbulent boundary
  layers}.  \jt{J. Fluid Mech.}  \bvol{30}~(4),  \pg{741--773}.

\bibitem[Konrad(1976)]{konrad_1976}
{\sc \au{Konrad, J.~H.}} \yr{1976}  \at{An experimental investigation of mixing
  in two-dimensional turbulent shear flows with applications to
  diffusion-limited chemical reactions}. PhD thesis, Ph.D. Thesis, California
  Institute of Technology.

\bibitem[Landahl(1980)]{landahl1980note}
{\sc \au{Landahl, M.~T.}} \yr{1980}  \at{A note on an algebraic instability of
  inviscid parallel shear flows}.  \jt{J. Fluid Mech.}  \bvol{98}~(2),
  \pg{243--251}.

\bibitem[Lesshafft {\em et~al.\/}(2019)Lesshafft, Semeraro, Jaunet, Cavalieri
  \& Jordan]{lesshafft2018resolvent}
{\sc \au{Lesshafft, L.}, \au{Semeraro, O.}, \au{Jaunet, V.}, \au{Cavalieri,
  A.~V.~G.} \& \au{Jordan, P.}} \yr{2019}  \at{Resolvent-based modelling of
  coherent wavepackets in a turbulent jet}.  \jt{Phys. Rev. Fluids}
  \bvol{4}~(6),  \pg{063901}.

\bibitem[Liepmann(1991)]{liepmann1991streamwise}
{\sc \au{Liepmann, D.}} \yr{1991}  \at{Streamwise vorticity and entrainment in
  the near field of a round jet}.  \jt{Physics of Fluids A: Fluid Dynamics}
  \bvol{3}~(5),  \pg{1179--1185}.

\bibitem[Liepmann \& Gharib(1992)]{liepmann1992role}
{\sc \au{Liepmann, D.} \& \au{Gharib, M.}} \yr{1992}  \at{The role of
  streamwise vorticity in the near-field entrainment of round jets}.  \jt{J.
  Fluid Mech.}  \bvol{245},  \pg{643--668}.

\bibitem[Lin(1981)]{lin1981}
{\sc \au{Lin, C.~C.}} \yr{1981}  \at{The evolution of streamwise vorticity in
  the free shear layer}. PhD thesis, Ph.D. Thesis, Univ. Calif., Berkeley.

\bibitem[Lin \& Corcos(1984)]{lin1984mixing}
{\sc \au{Lin, S.~J.} \& \au{Corcos, G.~M.}} \yr{1984}  \at{The mixing layer:
  deterministic models of a turbulent flow. part 3. the effect of plane strain
  on the dynamics of streamwise vortices}.  \jt{J. Fluid Mech.}  \bvol{141},
  \pg{139--178}.

\bibitem[Lumley(1967)]{lumley1967}
{\sc \au{Lumley, J.~L.}} \yr{1967}  \at{The structure of inhomogeneous
  turbulent flows}.  \jt{Atmospheric turbulence and radio propagation}  \pg{pp.
  166--178}.

\bibitem[Lumley(1970)]{lumley1970}
{\sc \au{Lumley, J.~L.}} \yr{1970} {\em Stochastic tools in turbulence\/}.
  \publ{New York: Academic Press}.

\bibitem[Malik \& Chang(1997)]{malik1997pse}
{\sc \au{Malik, M.} \& \au{Chang, C.~L.}} \yr{1997} {PSE} applied to supersonic
  jet instability.  \bt{In {\em 35th Aerospace Sciences Meeting and
  Exhibit\/}},  \pg{p. 758}.

\bibitem[Marant \& Cossu(2018)]{marant2018influence}
{\sc \au{Marant, M.} \& \au{Cossu, C.}} \yr{2018}  \at{Influence of optimally
  amplified streamwise streaks on the {K}elvin--{H}elmholtz instability}.
  \jt{J. Fluid Mech.}  \bvol{838},  \pg{478--500}.

\bibitem[Martin \& Meiburg(1991)]{martin1991numerical}
{\sc \au{Martin, J.~E.} \& \au{Meiburg, E.}} \yr{1991}  \at{Numerical
  investigation of three-dimensionally evolving jets subject to axisymmetric
  and azimuthal perturbations}.  \jt{J. Fluid Mech.}  \bvol{230},
  \pg{271--318}.

\bibitem[Mattsson \& Nordstr{\"o}m(2004)]{mattsson2004summation}
{\sc \au{Mattsson, K.} \& \au{Nordstr{\"o}m, J.}} \yr{2004}  \at{Summation by
  parts operators for finite difference approximations of second derivatives}.
  \jt{J. Computat. Phys.}  \bvol{199}~(2),  \pg{503--540}.

\bibitem[McKeon \& Sharma(2010)]{mckeon2010critical}
{\sc \au{McKeon, B.~J.} \& \au{Sharma, A.~S.}} \yr{2010}  \at{A critical-layer
  framework for turbulent pipe flow}.  \jt{J. Fluid Mech.}  \bvol{658},
  \pg{336--382}.

\bibitem[Metcalfe {\em et~al.\/}(1987)Metcalfe, Orszag, Brachet, Menon \&
  Riley]{metcalfe1987secondary}
{\sc \au{Metcalfe, R.~W.}, \au{Orszag, S.~A.}, \au{Brachet, M.~E.}, \au{Menon,
  S.} \& \au{Riley, J.~J.}} \yr{1987}  \at{Secondary instability of a
  temporally growing mixing layer}.  \jt{J. Fluid Mech.}  \bvol{184},
  \pg{207--243}.

\bibitem[Michalke(1984)]{michalke1984survey}
{\sc \au{Michalke, A.}} \yr{1984}  \at{Survey on jet instability theory}.
  \jt{Progress in Aerospace Sciences}  \bvol{21},  \pg{159--199}.

\bibitem[Miksad(1972)]{miksad1972experiments}
{\sc \au{Miksad, R.~W.}} \yr{1972}  \at{Experiments on the nonlinear stages of
  free-shear-layer transition}.  \jt{J. Fluid Mech.}  \bvol{56}~(4),
  \pg{695--719}.

\bibitem[Mizuno \& Jim{\'e}nez(2013)]{mizuno2013wall}
{\sc \au{Mizuno, Y.} \& \au{Jim{\'e}nez, J.}} \yr{2013}  \at{Wall turbulence
  without walls}.  \jt{J. Fluid Mech.}  \bvol{723},  \pg{429--455}.

\bibitem[Moffatt(1965)]{moffatt1965interaction}
{\sc \au{Moffatt, H.~K.}} \yr{1965}  \at{The interaction of turbulence with
  strong wind shear}.  \jt{Atmospheric Turbulence and Radio Waves Propagation,
  Proc. Intern. Collq. Moscow, 1965}  \pg{pp. 139--156}.

\bibitem[Mohseni \& Colonius(2000)]{mohseni2000numerical}
{\sc \au{Mohseni, K.} \& \au{Colonius, T.}} \yr{2000}  \at{Numerical treatment
  of polar coordinate singularities}.  \jt{J. Computat. Phys.}  \bvol{157}~(2),
   \pg{787--795}.

\bibitem[Mollo-Christensen(1967)]{mollo1967jet}
{\sc \au{Mollo-Christensen, E.}} \yr{1967}  \at{Jet noise and shear flow
  instability seen from an experimenter's viewpoint}.  \jt{J. Appl. Mech.}
  \bvol{34},  \pg{1--7}.

\bibitem[Monokrousos {\em et~al.\/}(2010)Monokrousos, {\AA}kervik, Brandt \&
  Henningson]{monokrousos2010global}
{\sc \au{Monokrousos, A.}, \au{{\AA}kervik, E.}, \au{Brandt, L.} \&
  \au{Henningson, D.~S.}} \yr{2010}  \at{Global three-dimensional optimal
  disturbances in the blasius boundary-layer flow using time-steppers}.  \jt{J.
  Fluid Mech.}  \bvol{650},  \pg{181--214}.

\bibitem[Monty {\em et~al.\/}(2007)Monty, Stewart, Williams \&
  Chong]{monty2007large}
{\sc \au{Monty, J.~P.}, \au{Stewart, J.~A.}, \au{Williams, R.~C.} \& \au{Chong,
  M.~S.}} \yr{2007}  \at{Large-scale features in turbulent pipe and channel
  flows}.  \jt{J. Fluid Mech.}  \bvol{589},  \pg{147--156}.

\bibitem[Morra {\em et~al.\/}(2019)Morra, Semeraro, Henningson \&
  Cossu]{morra2019relevance}
{\sc \au{Morra, P.}, \au{Semeraro, O.}, \au{Henningson, D.~S.} \& \au{Cossu,
  C.}} \yr{2019}  \at{On the relevance of {R}eynolds stresses in resolvent
  analyses of turbulent wall-bounded flows}.  \jt{J. Fluid Mech.}  \bvol{867},
  \pg{969--984}.

\bibitem[Morris(1976)]{morris1976spatial}
{\sc \au{Morris, P.~J.}} \yr{1976}  \at{The spatial viscous instability of
  axisymmetric jets}.  \jt{J. Fluid Mech.}  \bvol{77}~(3),  \pg{511--529}.

\bibitem[Neu(1984)]{neu1984dynamics}
{\sc \au{Neu, J.~C.}} \yr{1984}  \at{The dynamics of stretched vortices}.
  \jt{J. Fluid Mech.}  \bvol{143},  \pg{253--276}.

\bibitem[Nichols \& Lele(2011)]{nichols2011global}
{\sc \au{Nichols, J.~W.} \& \au{Lele, S.~K.}} \yr{2011}  \at{Global modes and
  transient response of a cold supersonic jet}.  \jt{J. Fluid Mech.}
  \bvol{669},  \pg{225--241}.

\bibitem[Nogueira {\em et~al.\/}(2019)Nogueira, Cavalieri, Jordan \&
  Jaunet]{Nogueira2019Streaks}
{\sc \au{Nogueira, P.~A.~S.}, \au{Cavalieri, A.~V.~G.}, \au{Jordan, P.} \&
  \au{Jaunet, V.}} \yr{2019}  \at{Large-scale, streaky structures in turbulent
  jets}.  \jt{J. Fluid Mech.}  \bvol{873},  \pg{211--237}.

\bibitem[Paschereit {\em et~al.\/}(1992)Paschereit, Oster, Long, Fiedler \&
  Wygnanski]{paschereit1992flow}
{\sc \au{Paschereit, C.~O.}, \au{Oster, D.}, \au{Long, T.~A.}, \au{Fiedler,
  H.~E.} \& \au{Wygnanski, I.}} \yr{1992}  \at{Flow visualization of
  interactions among large coherent structures in an axisymmetric jet}.
  \jt{Experiments in Fluids}  \bvol{12}~(3),  \pg{189--199}.

\bibitem[Pickering {\em et~al.\/}(2019)Pickering, Rigas, Sipp, Schmidt \&
  Colonius]{pickering2019eddy}
{\sc \au{Pickering, E.}, \au{Rigas, G.}, \au{Sipp, D.}, \au{Schmidt, O.~T.} \&
  \au{Colonius, T.}} \yr{2019} Eddy viscosity for resolvent-based jet noise
  models.  \bt{In {\em 25th AIAA/CEAS Aeroacoustics Conference\/}},  \pg{p.
  2454}.

\bibitem[Pierrehumbert \& Widnall(1982)]{pierrehumbert1982two}
{\sc \au{Pierrehumbert, R.~T.} \& \au{Widnall, S.~E.}} \yr{1982}  \at{The
  two-and three-dimensional instabilities of a spatially periodic shear layer}.
   \jt{J. Fluid Mech.}  \bvol{114},  \pg{59--82}.

\bibitem[Qadri \& Schmid(2017)]{qadri2017frequency}
{\sc \au{Qadri, U.~A.} \& \au{Schmid, P.~J.}} \yr{2017}  \at{Frequency
  selection mechanisms in the flow of a laminar boundary layer over a shallow
  cavity}.  \jt{Phys. Rev. Fluids}  \bvol{2},  \pg{013902}.

\bibitem[Rigas {\em et~al.\/}(2019)Rigas, Pickering, Schmidt, Nogueira,
  Cavalieri, Br{\`e}s \& Colonius]{rigas2019streaks}
{\sc \au{Rigas, G.}, \au{Pickering, E.}, \au{Schmidt, O.~T.}, \au{Nogueira,
  P.~A.}, \au{Cavalieri, A.~V.}, \au{Br{\`e}s, G.~A.} \& \au{Colonius, T.}}
  \yr{2019} Streaks and coherent structures in jets from round and serrated
  nozzles.  \bt{In {\em 25th AIAA/CEAS Aeroacoustics Conference\/}},  \pg{p.
  2597}.

\bibitem[Rodr{\'\i}guez {\em et~al.\/}(2015)Rodr{\'\i}guez, Cavalieri, Colonius
  \& Jordan]{rodriguez2015study}
{\sc \au{Rodr{\'\i}guez, D.}, \au{Cavalieri, A. V.~G.}, \au{Colonius, T.} \&
  \au{Jordan, P.}} \yr{2015}  \at{A study of linear wavepacket models for
  subsonic turbulent jets using local eigenmode decomposition of piv data}.
  \jt{European Journal of Mechanics-B/Fluids}  \bvol{49},  \pg{308--321}.

\bibitem[Rogers \& Moser(1992)]{rogers1992three}
{\sc \au{Rogers, M.~M.} \& \au{Moser, R.~D.}} \yr{1992}  \at{The
  three-dimensional evolution of a plane mixing layer: the
  {K}elvin--{H}elmholtz rollup}.  \jt{J. Fluid Mech.}  \bvol{243},
  \pg{183--226}.

\bibitem[Saiyed {\em et~al.\/}(2003)Saiyed, Mikkelsen \&
  Bridges]{saiyed2003acoustics}
{\sc \au{Saiyed, N.~H.}, \au{Mikkelsen, K.~L.} \& \au{Bridges, J.~E.}}
  \yr{2003}  \at{Acoustics and thrust of quiet separate-flow high-bypass-ratio
  nozzles}.  \jt{AIAA journal}  \bvol{41}~(3),  \pg{372--378}.

\bibitem[Samimy {\em et~al.\/}(1993)Samimy, Zaman \& Reeder]{samimy1993effect}
{\sc \au{Samimy, M.}, \au{Zaman, K.} \& \au{Reeder, M.F.}} \yr{1993}
  \at{Effect of tabs on the flow and noise field of an axisymmetric jet}.
  \jt{AIAA journal}  \bvol{31}~(4),  \pg{609--619}.

\bibitem[{Schmidt} {\em et~al.\/}(2018){Schmidt}, {Towne}, Rigas, {Colonius} \&
  {Br{\`e}s}]{SchmidtJFM2018}
{\sc \au{{Schmidt}, O.~T.}, \au{{Towne}, A.}, \au{Rigas, G.}, \au{{Colonius},
  T.} \& \au{{Br{\`e}s}, G.~A.}} \yr{2018}  \at{Spectral analysis of jet
  turbulence}.  \jt{J. Fluid Mech.}  \bvol{855},  \pg{953--982}.

\bibitem[Semeraro {\em et~al.\/}(2016)Semeraro, Lesshafft, Jaunet \&
  Jordan]{semeraro2016modeling}
{\sc \au{Semeraro, O.}, \au{Lesshafft, L.}, \au{Jaunet, V.} \& \au{Jordan, P.}}
  \yr{2016}  \at{Modeling of coherent structures in a turbulent jet as global
  linear instability wavepackets: theory and experiment}.  \jt{International
  Journal of Heat and Fluid Flow}  \bvol{62},  \pg{24--32}.

\bibitem[Sinha {\em et~al.\/}(2016)Sinha, Rajagopalan \&
  Singla]{sinha2016linear}
{\sc \au{Sinha, A.}, \au{Rajagopalan, A.} \& \au{Singla, S.}} \yr{2016} Linear
  stability implications of chevron geometry modifications for turbulent jets.
  \bt{In {\em 22nd AIAA/CEAS Aeroacoustics Conference\/}},  \pg{p. 3053}.

\bibitem[Sipp \& Marquet(2013)]{sipp2013characterization}
{\sc \au{Sipp, D.} \& \au{Marquet, O.}} \yr{2013}  \at{Characterization of
  noise amplifiers with global singular modes: the case of the leading-edge
  flat-plate boundary layer}.  \jt{Theoretical and Computational Fluid
  Dynamics}  \bvol{27}~(5),  \pg{617--635}.

\bibitem[Swearingen \& Blackwelder(1987)]{swearingen1987growth}
{\sc \au{Swearingen, J.~D.} \& \au{Blackwelder, R.~F.}} \yr{1987}  \at{The
  growth and breakdown of streamwise vortices in the presence of a wall}.
  \jt{J. Fluid Mech.}  \bvol{182},  \pg{255--290}.

\bibitem[Tam \& Hu(1989)]{tam1989three}
{\sc \au{Tam, C. K.~W.} \& \au{Hu, F.~Q.}} \yr{1989}  \at{On the three families
  of instability waves of high-speed jets}.  \jt{J. Fluid Mech.}  \bvol{201},
  \pg{447--483}.

\bibitem[Tissot {\em et~al.\/}(2017{\natexlab{{\em a\/}}})Tissot, Laj{\'u}s~Jr,
  Cavalieri \& Jordan]{tissot2017wave}
{\sc \au{Tissot, G.}, \au{Laj{\'u}s~Jr, F.~C.}, \au{Cavalieri, A.~V.~G.} \&
  \au{Jordan, P.}} \yr{2017{\natexlab{{\em a\/}}}}  \at{Wave packets and {O}rr
  mechanism in turbulent jets}.  \jt{Phys. Rev. Fluids}  \bvol{2}~(9),
  \pg{093901}.

\bibitem[Tissot {\em et~al.\/}(2017{\natexlab{{\em b\/}}})Tissot, Zhang,
  Laj{\'u}s, Cavalieri \& Jordan]{tissot2017sensitivity}
{\sc \au{Tissot, G.}, \au{Zhang, M.}, \au{Laj{\'u}s, F.~C.}, \au{Cavalieri, A.
  V.~G.} \& \au{Jordan, P.}} \yr{2017{\natexlab{{\em b\/}}}}  \at{Sensitivity
  of wavepackets in jets to nonlinear effects: the role of the critical layer}.
   \jt{J. Fluid Mech.}  \bvol{811},  \pg{95--137}.

\bibitem[Towne {\em et~al.\/}(2018)Towne, Schmidt \&
  Colonius]{towne2018spectral}
{\sc \au{Towne, A.}, \au{Schmidt, O.~T.} \& \au{Colonius, T.}} \yr{2018}
  \at{Spectral proper orthogonal decomposition and its relationship to dynamic
  mode decomposition and resolvent analysis}.  \jt{J. Fluid Mech.}  \bvol{847},
   \pg{821--867}.

\bibitem[Violato \& Scarano(2011)]{violato2011three}
{\sc \au{Violato, D.} \& \au{Scarano, F.}} \yr{2011}  \at{Three-dimensional
  evolution of flow structures in transitional circular and chevron jets}.
  \jt{Physics of Fluids}  \bvol{23}~(12),  \pg{124104}.

\bibitem[Widnall {\em et~al.\/}(1974)Widnall, Bliss \&
  Tsai]{widnall1974instability}
{\sc \au{Widnall, S.~E.}, \au{Bliss, D.~B.} \& \au{Tsai, C.-Y.}} \yr{1974}
  \at{The instability of short waves on a vortex ring}.  \jt{J. Fluid Mech.}
  \bvol{66}~(1),  \pg{35--47}.

\bibitem[Yang {\em et~al.\/}(2016)Yang, Zhou, So \& Liu]{yang2016turbulent}
{\sc \au{Yang, H.}, \au{Zhou, Y.}, \au{So, R. M.~C.} \& \au{Liu, Y.}} \yr{2016}
   \at{Turbulent jet manipulation using two unsteady azimuthally separated
  radial minijets}.  \jt{Proceedings of the Royal Society A: Mathematical,
  Physical and Engineering Sciences}  \bvol{472}~(2191),  \pg{20160417}.

\bibitem[Yule(1978)]{yule1978large}
{\sc \au{Yule, A.~J.}} \yr{1978}  \at{Large-scale structure in the mixing layer
  of a round jet}.  \jt{J. Fluid Mech.}  \bvol{89}~(3),  \pg{413--432}.

\bibitem[Zaman(1999)]{zaman1999spreading}
{\sc \au{Zaman, K.}} \yr{1999}  \at{Spreading characteristics of compressible
  jets from nozzles of various geometries}.  \jt{J. Fluid Mech.}  \bvol{383},
  \pg{197--228}.

\bibitem[Zaman {\em et~al.\/}(1994)Zaman, Reeder \& Samimy]{zaman1994control}
{\sc \au{Zaman, K.}, \au{Reeder, M.~F.} \& \au{Samimy, M.}} \yr{1994}
  \at{Control of an axisymmetric jet using vortex generators}.  \jt{Physics of
  Fluids}  \bvol{6}~(2),  \pg{778--793}.

\end{thebibliography}
\end{document}